\documentclass[12pt]{article}

\usepackage{newtxtext}
\usepackage[varvw]{newtxmath}

\usepackage{graphicx}

\usepackage{braket}
\usepackage{multirow}
\usepackage{graphicx} 
\usepackage{array}
\usepackage{amsmath}

\usepackage[letterpaper,margin=1in]{geometry}

\renewenvironment{abstract}
	{\quotation}
	{\endquotation}

\date{}

\makeatletter
\renewcommand{\fnum@figure}{\textbf{Figure \thefigure}}
\renewcommand{\fnum@table}{\textbf{Table \thetable}}
\makeatother

\usepackage{scicite}
\usepackage{url}
\usepackage{hyperref}

\def\scititle{
Morphological computational capacity of \textit{Physarum polycephalum}
}
\title{\bfseries \boldmath \scititle}

\author{
	S. Bajpai$^{\dagger}$,
        A. Lucas-DeMott$^{\ddagger}$,
        N. J. Murugan$^{\&}$,
        M. Levin$^{\ddagger}$, and
	P. Kurian$^{\ast \dagger}$ \and
	\small$^{\ddagger}$Allen Discovery Center, Tufts University, 200 Boston Avenue, Medford, MA 02155\and
	\small$^{\&}$Department of Health Science, Wilfrid Laurier University, Waterloo, Canada, N2L 3C5\and
    \small$^{\dagger}$Quantum Biology Laboratory, Howard University, 2041 Georgia Avenue NW, Washington, DC 20060\and
	\small$^\ast$Corresponding author. Email: pkurian@howard.edu\and
}

\begin{document} 

\maketitle

\begin{abstract} \bfseries \boldmath
While computational capacity limits of the universe and carbon-based life have been estimated, a stricter bound for aneural organisms has not been established. \textit{Physarum polycephalum}, a unicellular, multinucleated amoeba, is capable of complex problem-solving despite lacking neurons. 
\textcolor{black}{By analyzing growth dynamics from high-throughput imaging data—more than 500 samples tracked every 30 minutes for one to three days—across two distinct \textit{Physarum} strains under diverse biological conditions, we map morphological evolution to information processing.} \textcolor{black}{
We revisit the Margolus–Levitin bound that constrains maximum computation rates, deriving a correction for these driven-dissipative living systems. Time-series data of \textit{Physarum}'s morphology—-quantified through area, perimeter, circularity, and fractal dimension—enable the determination of upper bounds on the number of logical operations achievable through its hydromechanical, chemical, kinetic, and quantum-optical degrees of freedom. \textcolor{black}{The time-series trends of these morphological quantities remain statistically robust across 100,000 bootstrap resamples at both 95$\%$ and 99.5$\%$ confidence intervals. Based on the organism's explored areas and characteristic values of ATP available for work, the computational capacity of \textit{Physarum} is upper bounded by ${\sim}10^{33}$ logical operations in 24 hours, scaling linearly in the non-equilibrium steady state.}}
This framework enables comparison of the computational capacities of life, exploiting either classical or quantum degrees of freedom.

\end{abstract}

\section{Introduction}

Morphological computation refers to how the physical structure or morphology of a system contributes to its computational capabilities or behavior \cite{pfeifer2005new,muller2017morphological, nowakowski2017bodily,zahedi2013quantifying}. It can be broadly understood as information processing that occurs through the body of a system via its dynamic interaction with the environment, offloading some of the computational cost. A classic example is the functioning of bird wings, where flight stability and lift are not only governed by neural control, but emerge from the physical interaction between the structure of the wing and the surrounding air currents \cite{wootton1992functional}. This idea has been explored in areas such as self-assembly and DNA computing, where information is encoded directly into DNA sequences \cite{adleman1994molecular, kari1997dna}. 

In a pioneering study, Adleman (1994) demonstrated that DNA molecules can be used to solve a combinatorial problem—the Hamiltonian path problem—thus inaugurating the field of molecular-scale computation \cite{adleman1994molecular}, as originally envisaged by Feynman in his visionary 1959 lecture “There’s Plenty of Room at the Bottom” \cite{feynman2018there}. This breakthrough showed that biological molecules could be harnessed to perform complex information-processing tasks. Building on this foundation, researchers have shown that DNA strands can be precisely engineered to self-assemble into algorithmically defined nanostructures, tiling patterns, and origami, effectively embedding logic and computation into the physical geometry of molecular assemblies \cite{winfree1998design, rothemund2000program, stephanopoulos2020hybrid, han2011dna}. Parallel to these biochemical approaches, other groups have shown that even non-biochemical colloidal systems can exhibit computationally relevant behavior: rod-like colloids self-assemble into membrane-like monolayers purely through entropy-driven interactions, and variations in chirality can reconfigure these structures into morphologically distinct states \cite{barry2010entropy, gibaud2012reconfigurable}. These systems—ranging from DNA nanostructures to colloidal membranes—underscore a shared principle: that physical form, shaped by local rules and global constraints, can serve as a substrate for distributed computation. Such efforts have pushed the boundaries of computation beyond silicon-based systems to include self-organizing biological systems capable of adapting, optimizing, and solving problems dynamically in response to their environment \cite{nakagaki2000maze, ben2009learning, zhu2018remarkable}. Understanding the ultimate limits on computational capabilities, set by physical laws and embedded in biological morphology, is crucial for both biological insights and the development of bio-inspired computational platforms, as described by two of the authors (SB, PK) in our companion work \cite{bajpai2025tracking}.

At the fundamental level, the computational limits of nature have been explored. Lloyd \cite{lloyd2002computational} proposed that the total number of operations that can have been performed by the universe since the big bang is on the order of $10^{120}$. The senior author of this work (PK) recently conjectured that the maximum number of operations that can have been performed by carbon-based life on Earth is approximately the square root of that value, or $10^{60}$ \cite{kurian2025computational}. This revised upper bound for the maximum number of operations for carbon-based life takes into account the phenomenon of single-photon superradiance from tryptophan networks in cytoskeletal protein fibers \cite{babcock2024ultraviolet, patwa2024quantum} found across aneural and neural living systems. Such superradiant states are able to compute about a billion times faster than the speed of Hodgkin–Huxley neurons in animal species \cite{kurian2025computational}. 

The potential astrophysical origin of tryptophan—via the UV photodegradation of so-called astronomical polycyclic aromatic hydrocarbons (PAHs) in interstellar and circumstellar environments—invites a broader narrative, in which the molecular scaffolds for quantum-enhanced computation may have been seeded prior to the emergence of life on Earth \cite{kurian2025computational}. \textit{Physarum polycephalum} belongs to the Amoebozoa, an ancient evolutionary lineage that originated over one billion years ago \cite{tekle2022new}, with plasmodial slime molds evolving as part of this eukaryotic supergroup's later diversification about 500 million years ago. Notably, it contains tryptophan-rich networks in its actomyosin cortex, which is critical in the formation of vein networks for cytosolic ``shuttle-streaming'' that have been strongly linked to the organism's ability to compute. 

The presence of microtubules in the intranuclear spindle during mitosis has been demonstrated in synchronous plasmodia of \textit{Physarum} wild-type strains \cite{goodman1969plasmodial, guttes1968electron, sakai1972electron, wille1979fine}. In particular, their occurrence in the mitotic nucleus has been confirmed in a Japanese strain \cite{sakai1972electron}, as well as in Colonia and diploid wild-type strains \cite{wille1979fine}. In addition, they have been reported in CLd-AXE (an axenic derivative of Colonia) and M3C laboratory strains \cite{havercroft1983demonstration}. However, their existence in the cytoplasm has long been disputed. Early electron microscopy and staining studies \cite{dugas1961electron, rhea1966electron} did not detect cytoplasmic microtubules during the coenocytic stage; subsequent works drew on these observations to establish the prevailing view that the mature plasmodium lacks them \cite{ havercroft1983demonstration, rhea1966electron, burland1983cell, burland1988gene, green1987developmental, gull1974ultrastructural, paul1987patterns, roobol1984identification, sauer1982developmental, solnica1990variable}.

In contrast, the Salles-Passador group reported single and bundled cytoplasmic microtubules forming a cold- and drug-sensitive three-dimensional meshwork in \textit{Physarum} (strains Colonia, M3CIV, TU291, CH713\(\times\)CH957, and CH713\(\times\)LU860), arguing that earlier failures to detect them were due to high background fluorescence \cite{salles1991physarum}. These results were indirectly supported by subsequent work, which showed that purified microtubules \textit{in vitro} mirrored the organism’s physiological response to inhibitors \textit{in vivo} \cite{quinlan1981correlation}. The Salles-Passador group later demonstrated distinct organizing systems for mitosis and for nucleating interphase microtubules \cite{salles1992intranuclear}. However, the presence of cytoplasmic microtubules has not been independently confirmed in the \textit{Physarum} strains we consider in this work, leaving unresolved whether they represent stable cytoplasmic structures or appear only in specific strains or at particular growth stages.

Together, these cytoskeletal and intranuclear networks raise the possibility that quantum mechanical effects such as superradiance may underlie \textit{Physarum}’s remarkable capacity for decentralized computation. These observations lead to a natural question: can similar, yet more specific, computational limits be derived at the scale of such an aneural organism—one that performs computation through its morphology?

It is often suggested that the primary purpose of morphological computation is to reduce the load on a central control unit \cite{paul2004morphology}.  However, our model organism, \textit{Physarum polycephalum}—a unicellular, multinucleated slime mold known for its remarkable problem-solving abilities \cite{boussard2021adaptive,zhu2018remarkable,reid2016decision, nakagaki2000maze}—lacks a centralized nervous system. \textit{Physarum} does not reduce the computational load through its morphology; rather, its entire body performs the computation through complex, spatially distributed oscillatory dynamics \cite{siriwardana2012fast, bajpai2025tracking}. In this sense, the organism's morphology serves as the computer. Its body acts as a living substrate for computation, akin to a reservoir computer, dynamically adapting its growth in response to environmental constraints to solve pathfinding and optimization problems. 

For example, \textit{Physarum} optimizes its morphology to maximize nutrient absorption when food particles are spatially distributed \cite{nakagaki2000maze,nakagaki2007minimum, tero2010rules, dussutour2010amoeboid, jones2014computation}. Moreover, \textit{Physarum} is able to integrate mechanosensory information across its body and redistribute this information internally to guide directional growth decisions \cite{murugan2021mechanosensation}, highlighting its capacity to compute and coordinate behavior through whole-organism information flow. It can also retain the memory of previously applied stimuli and reproduce specific behavioral responses upon re-exposure at later times \cite{saigusa2008amoebae}.  The organism has been shown to construct spatial graphs \cite{baumgarten2010plasmodial} and solve constraint satisfaction problems \cite{aono2007amoeba}. Notably, it has demonstrated the ability to tackle NP-hard problems such as the traveling salesperson problem and the Boolean satisfiability problem \cite{aono2007amoeba, aono2009amoeba, zhu2013amoeba}.

To understand the constraints that govern these behaviors, it is essential to consider the physical scaling laws that shape the structure and function of the biological system. Kleiber observed that the total basal metabolic rate in animals scales as $M^{3/4}$, where $M$ is body mass, establishing a classic allometric law \cite{kleiber1947body}. Building on this, West et al. \cite{west2002allometric} demonstrated that this scaling extends over 27 orders of magnitude --- from enzyme molecules to multicellular organisms --- and arises from hierarchical, fractal-like transport networks optimized to minimize energy dissipation. Since the number of cells scales linearly with $M$, the average per-cell metabolic rate \textit{in vivo} decreases with body size as $M^{-1/4}$. In contrast, isolated cells \textit{in vitro} exhibit an invariant metabolic rate, since they are no longer constrained by the organism’s transport networks. Analogous to the metabolic constraints identified by West et al.\ across hierarchical levels of biological organization, our analysis seeks to establish the limits on the computational capacity of \textit{Physarum} by deriving bounds across distinct biophysical processes, including hydrodynamical oscillations, chemical (adenosine triphosphate, or ATP) conversion, kinetic, and quantum optical processes.

Theoretical limits on the computational capacity of an individual aneural organism—based on its morphological dynamics—have not been addressed before. While prior studies have explored behavioral responsiveness, adaptive growth, and the ability of such organisms to solve complex problems \cite{vallverdu2018slime, beekman2015brainless,nakagaki2008intelligent}, they have not quantified the upper bounds of their computational rates in physical terms. In particular, no existing framework estimates upper limits on the number of logical operations that an organism like \textit{Physarum polycephalum}  can perform purely through morphological transformations.

In this work, we investigate the growth dynamics of \textit{Physarum polycephalum} under varying physiological conditions, including strain, age, biomass, and initial vein network. \textcolor{black}{Using high-throughput time-lapse imaging of more than 500 samples, tracked every 30 minutes for one to three days, we extracted key morphological features—area, perimeter, circularity, and fractal dimension—using custom-developed analysis scripts. The resulting temporal trends in these morphological quantities remain statistically robust under bootstrap resampling with 100,000 resamples at both 95$\%$ and 99.5$\%$ confidence intervals.}
 
\begin{figure}[h!]
    \centering
    \includegraphics[width=1\linewidth]{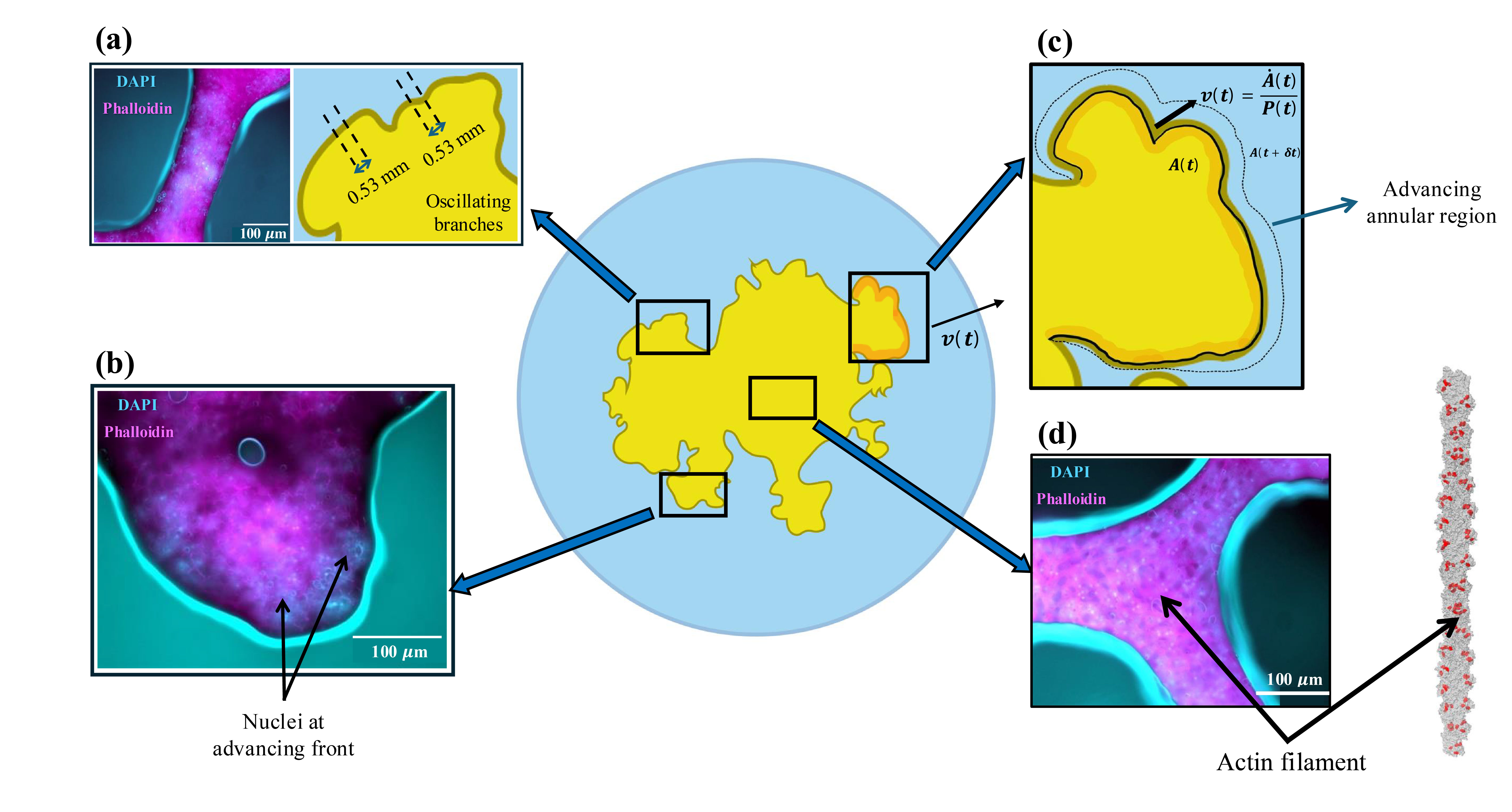}  
    \caption{\textbf{Hydrodynamic cytosol oscillations, distribution of ATP across \textit{Physarum}’s body, kinetic motion of the advancing perimeter, and theoretically predicted superradiant states in actin bundles provide the organism with distinct computational capabilities, enabling determination of their respective upper bounds---a first-ever quantification in an aneural organism.} 
    The figure illustrates with histological stains (DAPI for DNA, phalloidin for actin filaments) the biophysical processes that confer on \textit{Physarum} distinct zones of computational power: 
    (a) Oscillating branches (width $\sim$0.53 mm) act as individual hydrodynamic oscillators.
    (b) Dividing nuclei localized at the advancing front mirror the spatial distribution of ATP, as shown by experiments in ~\cite{hirose1980changes, ueda1987patterns}. (c) The rate of radial expansion, $v(t)=\dot{A}(t)/P(t)$, characterizes the growth of the organism’s advancing front. 
    (d) Actin filament bundles with organized tryptophan networks (shown in red) have been theoretically predicted to maintain photoexcited superradiant states \cite{patwa2024quantum}, with lifetimes of tens of picoseconds for ultrafast information processing.}
    \label{fig:Schematic}
\end{figure}

We present estimates of four distinct upper bounds on the computational capacity of the slime mold, limited by the accessible energies of the relevant physical degrees of freedom used to process information. These bounds are each derived from the Margolus-Levitin theorem \cite{margolus1998maximum} and include: (1) a hydrodynamical cytosolic bound, obtained from the organism’s well-documented macroscopic streaming patterns; (2) a chemical ATP consumption bound, obtained from prior experimental measurements across the organism body; (3) a kinetic energy bound, obtained from the kinetic motion of the thin, advancing annulus of \textit{Physarum} near its boundary, which represents the actively moving region of the organism; and (4) a quantum optical bound, which estimates the maximum computing speed from experimental measurements of actin fibril densities and numerical predictions of the characteristic lifetime of a single-photon superradiant state in an actin filament bundle. A schematic overview of the above-mentioned bounds, rooted in distinct aspects of the respective microphysics, is provided in Fig.~\ref{fig:Schematic}.

After analyzing time-series data of \textit{Physarum}’s morphological growth, we fit these morphological quantities with the appropriate models. These models are then used to obtain analytical expressions for the computational bounds described above. The fitting parameters are then substituted into these expressions to estimate the corresponding upper bounds on the number of logical operations that the organism can perform within a given time interval. These estimates provide a concrete, physically grounded measure of its information-processing capacity. We show that these upper bounds vary across experimental conditions, including strain, age, initial biomass, vein network, and feeding condition. To complement this analysis, we estimate the time at which the organism transitions to a nonequilibrium steady state (NESS), the point at which morphological growth is stabilized. The transition to the NESS time scale is defined as the first time point beyond the beginning of the experiment at which the growth rate of the organism is significantly slowed, below a small threshold value, marking the transition to a stable area region. \textcolor{black}{Together, this framework enables quantitative estimates of the information-processing capacity of two distinct, widely studied \textit{Physarum} strains examined here and provides a basis for future comparisons across additional strains and other living systems with differing morphological, biomolecular, and computational characteristics.}

The text in this paper is organized as follows. Section~2 introduces the morphological indices used to characterize \textit{Physarum}'s growth. Section~3 discusses the application of the Margolus--Levitin bound to \textit{Physarum} in the macroscopic limit. The computational framework and derivations of distinct bounds on \textit{Physarum}'s computational capacity are presented in Section~4. Section~5 reports our results on morphological growth and the corresponding values of distinct bounds for various experimental subgroups, disaggregated across strain, age, initial biomass, vein network, and feeding condition. Section~6 provides a broader discussion of these findings. Finally, the materials and methods utilized in this study are described in Section~7.

\section{Morphological growth of \textit{\textit{Physarum}}}

\textit{Physarum}’s unique information-processing capabilities arise from its adaptive growth morphology, which depends on factors such as strain, age, biomass, and feeding conditions. These morphological characteristics can be quantitatively assessed using indices such as area, perimeter, circularity, and fractal dimension of the macroplasmodia boundary, which provide insights into the structure and dynamical evolution of \textit{Physarum}’s growth. For an organism like \textit{Physarum}, the macroplasmodial area and perimeter reflecting its spatial expansion over time are proportional to several biological observables, which in turn serve as the physical bases for various computational capacities. Simultaneous changes in area and perimeter can be quantified using circularity, which characterizes the equiradial growth of the organism, distinguishing at extreme values between uniform, radial growth and more irregular, protuberant extensions. The fractal dimension captures the geometric complexity of the network and can be associated with abrupt perimeter changes or burst-like growth patterns, offering additional insight into the adaptive strategies of the organism. Together, these morphological indices serve as useful metrics for assessing internal processing and decision making, and they directly inform our estimates of the number of logical operations in the model. Circularity and fractal dimension are described in the following section.

\subsection{Circularity}
The area and perimeter explored by \textit{Physarum} can be quantified using circularity, a dimensionless index that characterizes the degree of roundness of the organism’s growth. It is defined as
\begin{equation}
\text{Circularity} = 4\pi  \frac{\text{Area}}{\text{(Perimeter)}^2}
\end{equation}

A circularity value of 1 indicates exactly circular growth, whereas lower values correspond to more anisotropic or protuberant morphologies. Circularity can also be used to detect transitions in growth behavior, with higher values typically associated with radial, symmetric expansion, and lower values reflecting irregular or exploratory growth.

\subsection{Fractal dimension}
Fractal dimension is related to how quantities, such as perimeter, increase as the scale of measurement decreases. For example, when measuring the length of a coastline, the measured length increases as the size of the measuring stick gets smaller. The fractal dimension of curves in two dimensions can range from 1 (for a smooth, line-like curve) to 2 (for highly jagged or irregular curves that fill the two-dimensional area). If the side length of the bounding boxes is $r$, and $N(r)$ is the number of boxes needed to cover the shape, the relationship can be expressed as:
\begin{equation}
\log N(r)=-d_f \log (r)+\log (k)
\end{equation}
where $d_f$ is the fractal dimension and $k$ is a constant. If we plot the logarithm of the number of boxes, $\log N(r)$, against the logarithm of the box size, $\log (r)$, we obtain a straight line. The absolute value of the slope of this line, $d_f$, is the fractal dimension of the shape, and the intercept is $\log (k)$.

\section{Application of macroscopic Margolus-Levitin limit to \textit{\textit{Physarum}}}
\label{sec:macMarg-LevinPhys}
The Margolus-Levitin theorem provides a lower bound on the time required to distinguish between two orthogonal states (e.g., $\ket{0}$ and $\ket{1}$). This lower bound is given by 
\begin{equation} \label{marglev}
    \tau \geq \frac{\pi \hbar}{2 \langle \mathcal{E} \rangle},
\end{equation}
where $\langle \mathcal{E} \rangle$ is the average energy, the expectation value of a time-independent Hamiltonian where the ground state energy is set to zero. Similar bounds have been demonstrated to hold in classical systems, and across the quantum-to-classical transition \cite{okuyama_quantum_2018,shanahan_quantum_2018}. 
 
The maximum number of operations per unit time that can be performed by such a physical system, according to the Margolus-Levitin theorem, is then given by
\begin{equation} \label{Nmax}
N_{\max }=\left(\tau_{\min }\right)^{-1}=\frac{2\langle\mathcal{E}\rangle}{\pi \hbar}
\end{equation}
Starting from Eq.~\ref{Nmax}, we can derive bounds on the computational capacity of \textit{Physarum polycephalum}, a single-cell, multinucleate amoeboid syncytium. 

The mean energy $\langle\mathcal{E}\rangle$ of a single harmonic oscillator in the macroscopic limit (i.e., when the number of the maximum energy eigenstate $N\gg 1$) 
approaches one-half of the maximum accessible energy, provided that the system has non-degenerate energy levels. \textcolor{black}{However, in degenerate systems, such as coupled oscillator systems, the mean energy shifts toward the maximum energy due to the increasing number of degenerate states at higher energy levels (see Methods and Supplementary Material for further details and proofs)}. 

Since \textit{Physarum} can be modeled effectively as a system of coupled 3D harmonic oscillators, we have also analyzed the mean energy for such coupled systems. These examples demonstrate that, in the macroscopic limit, increasing the overall number of oscillators drives the mean energy progressively closer to the system's maximum energy. This behavior arises due to the increasing degeneracy of energy states at higher energies, which skews the average energy upward in the statistical distribution.

In general, for a system of harmonic oscillators each of $d$-dimensions with a maximum degeneracy $G$, the mean energy can be approximated as
\begin{equation}
\langle \mathcal{E} \rangle_{Gd}\approx \frac{G d}{G d+1} E_{\max },
\end{equation}
as proven in the Methods and Supplementary Material. Since \textit{Physarum} can be treated effectively as a system composed of a very large number of coupled three-dimensional oscillators, we can conveniently approximate its mean energy as being very close to the system's maximum accessible energy, $\langle \mathcal{E}\rangle_{Gd} \approx E_{\max }$. 

\textcolor{black}{Using this estimate for the mean energy in \textit{Physarum}, we can rewrite the expression for the maximum number of operations per unit time for a macroscopic closed cycle of many mutually orthogonal states as}
\begin{equation}\label{eq:ML_Energy}
N_{\max }=\frac{E_{\max }}{\pi \hbar},
\end{equation}
and we obtain the number of operations for a given time interval by integration:
\begin{equation}\label{eq:slimeops}
\mathcal{N}(t)=\int_0^t d t^{\prime} N_{\max }\left(t^{\prime}\right)=\int_0^t d t^{\prime} \frac{E_{\max }\left(t^{\prime}\right)}{\pi \hbar}.
\end{equation}
We obtain distinct bounds on the computational capacities of \textit{Physarum} by substituting morphological estimates for the maximum energies accessible from certain degrees of freedom into the expression for $\mathcal{N}(t)$, depending on the type of microphysics considered (see Fig. \ref{fig:Schematic}). Specifically, we estimate the hydromechanical, chemical ATP, kinetic energy, and quantum optical upper bounds on computational capacity using Eq.~\ref{eq:slimeops}.

\textcolor{black}{Since \textit{Physarum} is inherently an open, driven, and dissipative biological system, we additionally consider phenomenological corrections to the Margolus-Levitin bounds in Eq.~\ref{eq:slimeops}, which are derived for unitarily evolving closed systems. By incorporating drive and dissipation into open-system quantum speed-limit formulation for non-unitary dynamics, and in the limit of weak system-environment coupling consistent with the Markovian approximation, we obtain a representative enhancement factor
\begin{equation}
\widetilde{\mathcal N}
=
\frac{\pi}{\sqrt2}\frac{N_{MT}}{N_{ML}},
\end{equation}
where $N_{MT}$ and $N_{ML}$ represent the computational rates associated with the Mandelstam–Tamm (MT) and Margolus–Levitin (ML) bounds, respectively. In the regime where the ML and MT computational rates are comparable, i.e., $N_{MT}\approx N_{ML}$, this yields the representative enhancement factor $\widetilde{\mathcal{N}}\approx \pi/\sqrt{2}$. This resulting driven-dissipative correction is incorporated into the corresponding computational bounds presented below. More details on the enhancement factor for single harmonic oscillator and coupled-oscillator systems are provided in the Methods section and Section S11 of the Supplementary Material. Applying this driven-dissipative correction, the corresponding upper bound on the number of operations performed over a time interval $t$ may be estimated as
\begin{equation} \label{MLopen}
\mathcal{N}_{\mathrm{open}}(t)
\approx
\frac{\pi}{\sqrt2}\mathcal N(t)
=
\frac{\pi}{\sqrt2}
\int_{0}^{t}
 dt^\prime
\frac{E_{\max}(t^\prime)}{\pi\hbar},
\end{equation}
where $\mathcal{N}(t)$ denotes the corresponding Margolus-Levitin estimate for the closed-system case.}

\subsection{\textcolor{black}{Definition of \textit{operations} and interpretation of distinguishable states}}
\textcolor{black}{A logical operation is traditionally defined as a Boolean logic-gate operation in the conventional digital-computing sense. However, an operation in our case is defined as a transition between distinguishable physical states associated with different physical modalities in \textit{Physarum}. For the hydrodynamic bound, these states correspond to collective cytosolic oscillatory configurations; for the chemical ATP bound, to biochemical energy states based on ATP distribution and hydrolysis; for the kinetic energy bound, to collective position and momentum states of the advancing annular region; and for the quantum-optical bound, to the absorption and decay dynamics of predicted superradiant states in actin fibers. 
\textcolor{black}{The distinguishability between such states can be quantified through appropriate measures of overlap between classical quasi-probability distributions (Wigner functions), analogous to the Bures angle used in quantum systems to determine the fidelity (overlap) between states \cite{shanahan_quantum_2018}, with the fidelity reducing to the overlap for pure states. See Section S9 of the Supplementary Material for further details.}}

\textcolor{black}{The relevant energy states, microphysical observables, characteristic timescales, and resulting orders of magnitude differ across the distinct degrees of freedom considered here (see Section S8 of the Supplementary Material for further details). Therefore, operations counted within one physical modality are not counted again in another. However, these estimates should be interpreted as effective upper bounds on logical operations derived from coarse-grained microphysical descriptions, observed at the resolution of our imaging experiments, rather than exact counts of microscopic computational events.}

\textcolor{black}{Any logical operation implemented by a physical system must ultimately be instantiated through transitions between distinguishable states of its underlying physical degrees of freedom. Consequently, a physical limit on the number of distinguishable-state transitions also sets an upper limit on the number of logical operations that can be physically realized. The resulting limits form a hierarchy: the fundamental physical bounds derived here establish an upper ceiling on information-processing capacity, while additional system-specific physical and biological constraints can progressively tighten these bounds toward the physically realizable capacity. For example, for the chemical ATP bound, additional kinetic information from ATP hydrolysis ($k_\text{hyd}$) and steady-state ATPase turnover ($k_{\text{basal}}$) rates can further tighten the coarse-grained energetic estimate toward a physically realizable bound; $k_\text{hyd}$ can exceed $k_\text{basal}$ by factors ranging from a few to $10^4$ for relevant ATPases (see Table S7 in the Supplementary Material). Further incorporation of local and spatially resolved information specific to the underlying physical processes can tighten these estimates toward the actual information-processing capacity of the organism. Such a determination would require a comprehensive analysis of the local physical and biological processes throughout the organism, which is beyond the scope of the present work. Thus, the bounds presented here constrain logical information processing without implying a one-to-one correspondence between every distinguishable-state transition and an actual logical operation.}

\textcolor{black}{However, an independent benchmark for the lower bound on the actual logical operations performed by \textit{Physarum} can be obtained from the companion study by two of the authors (SB, PK) of \textit{Physarum} solving traveling salesperson problem (TSP) instances \cite{bajpai2025tracking}. The \textit{Physarum}-based TSP solver successfully finds high-quality solutions for the eight-city TSP, which consists of a search space of $8!=40,320$ tours. Computing the total length of each tour requires eight logical additions of intercity distances counting return to the starting city, while comparison with the current minimum requires one logical comparison operation ($<?$). Thus, evaluating the minimum-length tour requires $8!\times 9=9!=362,880$ logical operations. 
Since \textit{Physarum} consistently solves the TSP within an hour on average, frequently yielding solutions within $30 \%$ of the exact minimum tour length, we estimate that, for such optimized exploratory search over a 24-hr period, the organism is capable of performing at least 70\% of the $8.7 \times 10^6$ logical ops, corresponding to ${\sim}6.1$ megaflops in one day.
This value exceeds in all strains and conditions considered the maximum allowed by strictly hydrodynamic (cytosolic) processes (see Section ~\ref{subsec:Hyd_bound_values}), strongly suggesting the involvement of other physical degrees of freedom in solving the problem.} 

\textcolor{black}{
Moreover, this estimate captures only the logical operations associated with solving the TSP and therefore constitutes a lower benchmark for the organism's total computational activity. \textit{Physarum} concurrently performs \textit{bio}-logical operations associated with physical degrees of freedom throughout the lanes and central hub of the stellate chip used in TSP experiments, including kinetic exploration of the agar substrate, spatiotemporal thickness wave formation and reorganization in the central hub, photoresponsiveness to TSP-encoding optical stimuli, and maintenance of internal order and metabolic coherence. This contrasts sharply with conventional silicon-based computation and motivates our analysis of the computational capacities associated with the organism's multiple physical degrees of freedom. See Section S8.1 of the Supplementary Material for further details.}

\section{Upper bounds on computational capacities in \textit{Physarum polycephalum}}

\subsection{Hydrodynamical bound}

\textit{Physarum} exhibits microscopic cytosol oscillations driven by peristaltic contraction and relaxation of its tubular vein network. These oscillations typically have a period ranging from 70 to 140 seconds. To estimate the maximum number of such operations performed by the pseudopod-like extensions of \textit{Physarum} along its perimeter per unit time, we divide the organism's advancing perimeter into discrete segments of the minimal pseudopod width $l_d$, with each pseudopod able to function as an independent mesoscopic oscillator. 

\textcolor{black}{For our calculations, we adopt an oscillation period of 100 seconds, corresponding to a frequency of 0.01 Hz, which was the dominant frequency peak in the power spectral density during the steady state in the companion work \cite{bajpai2025tracking} by two of the authors (SB, PK).} \textcolor{black}{Thus, each segment performs approximately 0.01 operations per second. It should be noted that the experimentally observed oscillation periods in this work range from ${\sim}50-100$ seconds, corresponding to representative local operation rates of approximately $0.01-0.02$ operations per second. The number of such independently oscillating segments at time $t$ is given by $P(t) / l_d$, where $P(t)$ is the length of the advancing perimeter of \textit{Physarum} at time $t$, and $l_d$ is the characteristic width of each oscillating pseudopod-like segment. We quantified branch-width distributions from the experimental images by measuring more than 300 branches from representative samples belonging to both the Carolina and Japanese subsets. In both cases, the mean branch width was found to be approximately $0.53$ mm, which we adopt for the main hydrodynamic bound calculations presented here. The smallest observed branch widths were approximately 0.2 mm, while the largest observed widths approached 2 mm for both strains. Sensitivity analyses of the hydrodynamic bounds using representative local operation rates of 0.01 and 0.02 operations per second, together with multiple branch-width choices spanning the experimentally observed range $0.2-2 \mathrm{~mm}$, are provided in Section S15 of the Supplementary Material.}

Multiplying the number of segments by the local operation rate per segment, denoted by $n_{\text {hydro }}^{\text{local}}$, gives the hydrodynamical bound on the maximum number of operations per second from these physical degrees of freedom:
\begin{equation}\label{eq:local_Hyd_2}
N_{\text{hydro}}(t)=n_{\text{hydro}}^{\text{local}} \times \frac{P(t)}{l_d}
\end{equation}
where $n_{\text{hydro}}^{\text{local}}=0.01$ operations per second per segment. By comparing the hydrodynamical bound on the maximum number of operations in Eq.~\ref{eq:local_Hyd_2} with the upper bound on number of operations computed in the macroscopic Margolus-Levitin limit in Eq.~\ref{eq:ML_Energy}, we obtain the maximum energy associated with the hydrodynamical bound: 
\begin{equation}
E_{\text {hydro}}(t)=\frac{\pi \hbar \ n_{\text {hydro }}^{\text {local }} P(t)}{l_{d}}
\end{equation}
For $P(t) \approx 100\,\text{cm}$ at $t = 24$ hours, a typical timescale of our \textit{Physarum} experiments, the maximum hydrodynamical energy evaluates to $E_{\text {hydro}} \approx 1.25 \times 10^{-32} \mathrm{~J} = 7.81 \times 10^{-14} \ \mathrm{eV}$, naturally in an extremely low-energy regime. The maximum number of hydrodynamical operations performed during a time interval from $0$ to
$t$ is given by
\begin{equation}\label{eq:Hyd_bound}
\mathcal{N}_{\text{hydro}}(t)=\frac{n_{\text{hydro}}^{\text{local}}}{l_d} \int_0^t dt^{\prime} P(t^{\prime}).
\end{equation}
\textcolor{black}{As discussed in Sec. 3, treating the system as driven and dissipative, with weak coupling between the system and environment, introduces a correction factor of $\pi/\sqrt{2}$ to the Margolus-Levitin bound for closed systems. The corresponding open-system hydrodynamical bound is therefore approximated as
\begin{equation}\label{eq:Hyd_bound_Open}
\mathcal{N}_{\mathrm{hydro}}^{\mathrm{open}}(t)
\approx
\frac{\pi}{\sqrt{2}}
\mathcal{N}_{\mathrm{hydro}}(t)=\frac{\pi \ n_{\text{hydro}}^{\text{local}}}{\sqrt{2}\ l_d} \int_0^t dt^{\prime} P(t^{\prime}).
\end{equation}}
This hydrodynamic bound is computed for all four experimental groups (young Japanese, old Japanese, Carolina, young Japanese-starved), and the resulting values are presented and compared in the Results.

\subsection{Chemical ATP bound}
\label{subsec:chemATP}

The chemical energy available to \textit{Physarum} can be experimentally linked to the organism's ATP concentration, which is highest along the advancing periphery ($\sim2$ mM) and decreases sigmoidally toward the center (in radially growing bodies) or toward the rear (in extended fronts). Interestingly, the degree of this gradient in maximum ATP concentration depends on the type of growth. During equiradial growth, the frontal ATP concentration decreases to about half at the center of the plasmodial biomass, as reported in \cite{ueda1987patterns}. When \textit{Physarum}'s growth becomes protuberant rather than radial, the ATP concentration at the center can fall to nearly an order of magnitude less than that at the front \cite{hirose1980changes}.

We can therefore write the energy $E(r,t) = \rho_E(r) V(t)$, where $V(t)$ represents the volume of a radially growing \textit{\textit{Physarum}} body at time $t$, $\rho_E(r) = \rho_0 (\tanh(1.472r/r_{\text{max}})+0.1)$ is a sigmoidally growing function of $r$ from the \textit{\textit{Physarum}} body center ($r=0$) to its perimeter ($r = r_{\text{max}}$) and has been determined from ATP assays \cite{hirose1980changes}. For $2$-mM ATP concentration at the advancing perimeter, $\rho_0 = 3.82 \times 10^{17}$ eV/cm$^3=61.2$ mJ/cm$^3$, using the fact that the conversion of ATP to adenosine diphosphate (ADP) releases $7.3$ kcal per mol ATP ($0.317$ eV per molecule). 
\textcolor{black}{\textit{Physarum} grows predominantly in the plane orthogonal to the surface normal vector, so we assume a constant thickness $\ell=100 \ \mu \mathrm{m}$ for the main calculations presented here. \textcolor{black}{This choice corresponds to the most frequently observed thickness range in our thickness analysis (see Methods and Section S14 of the Supplementary Material for details). However, sensitivity analysis of the bound over the biologically relevant range of $100$--$500~\mu\mathrm{m}$ is provided in the Supplementary Material.} 
Using this approximation across the body area $A$, we obtain}

\begin{equation}
\begin{split}
    E_{\text{chem}}(t) &= \frac{1}{r_{\text{max}}(t)} \int_0^{r_\text{max}} dr\, \rho_E(r)\, V(t) \\
         &= \frac{\rho_0 A(t) \ell}{r_{\text{max}}(t)} \int_0^{r_\text{max}} dr\, [\tanh(1.472r/r_{\text{max}})+0.1].
\end{split}
\end{equation}
Changing variables with the replacement $x = r/r_{\text{max}}$, we get the following value for the \textit{\textit{Physarum}} maximum accessible chemical energy: 
\begin{equation} \label{numtanh}
     E_{\text{chem}}(t) = \rho_0 A(t) \ell \int_0^1 dx [\tanh(1.472x)+0.1] = 0.664 \rho_0 A(t) \ell.
\end{equation} 

Modifying the integral in Eq.~\ref{numtanh} by accounting for distinct ATP distribution patterns with different body circularities \cite{ueda1987patterns, hirose1980changes} would increase the numerical value for the integral only by a few percent, as body circularities are mostly protuberant compared to equiradial over the course of the experiment (see Figure~\ref{Various_strains}a). Substituting this value of the maximum accessible ATP energy into Eq. ~\ref{eq:slimeops}, we obtain the maximum number of chemical operations performed in a given time interval:
\begin{equation}\label{eq:chem_ATP}
\mathcal{N}_{\text{chem}}(t)=\frac{0.664 \rho_0 \ell}{\pi \hbar} \int_0^t dt^{\prime} A(t^{\prime}).
\end{equation}
\textcolor{black}{
As discussed in the Methods and Supplementary Material, treating the system as driven and dissipative with weak coupling between the system and environment introduces a correction factor of $\pi/\sqrt{2}$ to the Margolus-Levitin bound for closed systems. The corresponding open-system chemical ATP bound is therefore estimated as
\begin{equation}\label{eq:chem_ATP_Open}
\mathcal{N}^{\mathrm{open}}_{\mathrm{chem}}(t)
\approx
\frac{\pi}{\sqrt{2}}
\mathcal{N}_{\mathrm{chem}}(t)=\frac{0.664 \rho_0 \ell}{\sqrt{2} \hbar} \int_0^t dt^{\prime} A(t^{\prime}).
\end{equation}
}\textcolor{black}{Eq.~\ref{eq:chem_ATP_Open} represents an idealized upper bound assuming maximal utilization of the available ATP-derived free energy for distinguishable physical-state transitions contributing to the computational bound. In a living biological system, however, several constraints restrict the fraction of ATP free energy that can be converted into meaningful work associated with such transitions. 
\textcolor{black}{Experimental studies on \textit{Physarum} suggest that the fraction of the ATP stock metabolically available for mechanical work, $\eta_{\mathrm{turnover}}$, ranges from $\sim 0.6\%$ to $10\%$ \cite{yoshimoto1981atp}; we thus adopt a representative conservative value of $\eta_{\mathrm{turnover}}=1\%$, near the lower end of the experimentally reported range. Furthermore, only a fraction of the ATP-derived metabolic free energy is converted into useful work, denoted by $\eta_{\mathrm{work}}$, while the remainder is dissipated through multiple channels, including viscous losses and other cellular processes.} In biological systems such as muscle tissue, mechanical work-conversion efficiencies are typically estimated to be in the range of $20\%-35\%$, with human muscle efficiency being of the order of $\sim20\%$ \cite{shephard1975efficiency}. \textcolor{black}{Since \textit{Physarum polycephalum} behaves as a viscoelastic sol-gel system with substantial viscous dissipation in addition to shuttle streaming, we expect its work-conversion efficiency to be lower than these reported values. We therefore adopt a representative value of $\eta_{\mathrm{work}}{\sim}10\%$. The corresponding sensitivity analysis is presented in Table S6 of the Supplementary Material.} 
The corresponding efficiency-corrected ATP-based computational bound may therefore be estimated as
\begin{equation}
\mathcal{N}_{\text {chem }}^{(\eta)}(t)=\eta_{\text {turnover }} \eta_{\text {work }} \mathcal{N}^{\mathrm{open}}_{\mathrm{chem}}(t)
\end{equation}
or equivalently,
\begin{equation}
\mathcal{N}_{\text {chem }}^{(\eta)}(t)=\frac{0.664 \ \eta_{\text {turnover}} \eta_{\text{work }} \rho_0 \ell}{\sqrt{2} \hbar} \int_0^t d t^{\prime} A\left(t^{\prime}\right).
\end{equation}
Taking $\eta_{\text {turnover }}=1 \%$ and $\eta_{\text {work }}=10 \%$ for \textit{Physarum}, the corresponding efficiency-corrected chemical ATP bound becomes $\mathcal{N}_{\text {chem }}^{(\eta)}(t)=10^{-3} \mathcal{N}^{\mathrm{open}}_{\mathrm{chem}}(t)$, corresponding to a reduction of approximately three orders of magnitude relative to the estimate based on the total ATP stock.} \textcolor{black}{The efficiency factors are applied only to the chemical bound because ATP represents the upstream energetic resource, whose available energy pool is constrained by ATP turnover and ATP-to-work conversion efficiencies. The other bounds—hydrodynamic, kinetic, and quantum-optical—instead characterize downstream processes driven by this available energy and are evaluated from their respective physical observables, which stem from the already constrained ATP pool.}

\subsection{Kinetic energy bound}

As \textit{Physarum} grows, its advancing front typically migrates at speeds ranging from a few millimeters per hour \cite{rolland2023behavioural} to four centimeters per hour \cite{boussard2021adaptive}, depending on environmental conditions. This advancing front is formed by an annular region of the \textit{Physarum} body located near the frontal boundary. The kinetic energy of the advancing front can be expressed as
\begin{equation}
\text{KE}(t)=\frac{1}{2} m_{\text{slime}}^{\text{front}}(t) \ v(t)^2
\end{equation}
where $m_{\text{slime}}^{\text{front}}(t)$ is the mass of the advancing portion of \textit{Physarum} at time $t$, and $v(t)$ is the speed of the front. \textcolor{black}{The effective front propagation speed is taken to be the rate of area expansion per unit advancing perimeter:
\begin{equation}\label{Eq:speed}
v(t)=\frac{1}{P(t)} \frac{d A}{d t}=\frac{\dot{A}(t)}{P(t)}
\end{equation}}

As an annulus at the perimeter is the primary mass that is advancing kinetically, we scale the mass accordingly based on its morphological area expansion.

Substituting this into the expression for the upper bound obtained from the macroscopic Margolus-Levitin limit, the upper bound on the number of operations that \textit{Physarum} can perform during its kinetic exploratory growth over a time interval from $0$ to $t$ is given by
\begin{equation}\label{Eq:General_KE_Bound}
\mathcal{N}_{\mathrm{KE}}(t)=\frac{1}{2\pi \hbar} \int_0^t d t^{\prime} m_{\text{slime}}^{\text{front}}(t^{\prime})v(t^{\prime})^2.
\end{equation}
We can express the mass of the \textit{Physarum} front as
\begin{equation}\label{eq:slime_mass}
m_{\text{slime}}^{\text{front}}(t)=\rho_m  f_{\mathrm{avg}} A(t) \ell,
\end{equation}
where $\rho_m = 1100~\text{kg}/\text{m}^3$ is the effective mass density \cite{rosina2025mathematical}, $A(t)$ is the area of the \textit{Physarum} body at time $t$, and $\ell=100 \ \mu$m is its effective thickness, \textcolor{black}{as discussed for the chemical ATP bound above.} \textcolor{black}{
Sensitivity analyses using representative thickness values of $100 \ \mu \mathrm{m}$ and $500 \ \mu \mathrm{m}$ are provided in Section S14 of the Supplementary Material.}
$f_{\mathrm{avg}} \in (0,1)$ is the time-averaged, kinetically weighted fraction of the macroplasmodial body that advances as the annular front over the analysis window. Substituting Eq.~\ref{eq:slime_mass} into the kinetic energy upper bound in Eq.~\ref{Eq:General_KE_Bound} gives

\textcolor{black}{
\begin{equation}\label{N_KEint}
\mathcal{N}_{\mathrm{KE}}(t)=\frac{\rho_m f_{\mathrm{avg}} \ell}{2 \pi \hbar} \int_0^t d t^{\prime} \frac{A\left(t^{\prime}\right) \dot{A}\left(t^{\prime}\right)^2}{P\left(t^{\prime}\right)^2}
\end{equation}
As discussed in Sec. 3, treating \textit{Physarum} as a weakly coupled driven-dissipative system introduces a correction factor of $\pi/\sqrt{2}$ relative to the closed-system Margolus-Levitin bound. The corresponding driven-dissipative kinetic-energy bound is therefore,
\begin{equation}\label{Eq:N_KEint_Open}
\mathcal{N}^{\mathrm{open}}_{\mathrm{KE}}(t)
\approx
\frac{\pi}{\sqrt{2}}
\mathcal{N}_{\mathrm{KE}}(t)=\frac{\rho_m f_{\mathrm{avg}} \ell}{2 \sqrt{2} \hbar} \int_0^t d t^{\prime} \frac{A\left(t^{\prime}\right) \dot{A}\left(t^{\prime}\right)^2}{P\left(t^{\prime}\right)^2}.
\end{equation}}
More specifically, $f_{\mathrm{avg}}$ is obtained by first computing the fraction $f(t) = \Delta A(t)/A(t)$ from the area time series data and then averaging over the 24-hour experimental window, weighted by the product of morphological indices in the integrand of Eq.~\ref{N_KEint}. Since $f(t)$ cannot be determined at the very first time point ($t_0 = 0$)—because it is defined through changes in area—we set the lower limit to $t_i = 0.5~\mathrm{hr}$. Accordingly, $f_{\mathrm{avg}}$ is computed over the interval $0.5~\mathrm{hr} \leq t \leq 24~\mathrm{hr}$: 

\textcolor{black}{\begin{equation}\label{favg}
f_{\mathrm{avg}}=\frac{\int_{0.5}^{24} f(t) A(t) \frac{\dot{A}(t)^2}{P(t)^2} d t}{\int_{0.5}^{24} A(t) \frac{\dot{A}(t)^2}{P(t)^2} d t}
\end{equation}}
Because $f(t)$ is a more erratic time-series than the morphological indices, which are more suitable to simple analytical fits, we can compare this time-averaged value for $f_{\mathrm{avg}}$ with direct numerical integration of the numerator in Eq.~\ref{favg}, replacing this value for \textcolor{black}{$f_{\text{avg}} \int_0^t d t^{\prime} A\left(t^{\prime}\right) \frac{\dot{A}\left(t^{\prime}\right)^2}{P\left(t^{\prime}\right)^2}$} in Eq.~\ref{N_KEint}.

\subsection{\textcolor{black}{Predicted and prospective quantum optical bound from simulations of actin fiber networks}}

\textcolor{black}{One author (PK) and coworkers have experimentally demonstrated that extremely large networks of tryptophan in microtubule architectures exhibit single-photon superradiance, as observed by their thermal quantum yields of fluorescence\cite{babcock2024ultraviolet}. When a single photon is coherently shared across many quantum two-level systems, here with tryptophan, superradiant states emerge with a collective radiative decay rate that is far more rapid---by a factor up to the number of two-level systems---than the spontaneous emission rate of each individual two-level system. Predictions for similar superradiant enhancements in actin fiber bundles were made shortly thereafter \cite{patwa2024quantum}.} \textcolor{black}{Such superradiant states, however, have not yet been empirically observed in \textit{Physarum} actin fiber networks; accordingly, we treat the quantum-optical bound developed below as predicted and prospective.}

These findings suggest that quantum degrees of freedom within such biomolecular mega-architectures of ultraviolet-photoexcited qubits could be utilized for logical operations and information processing, above the thermal noise floor. In these architectures, an upper bound on computation can be estimated based on the shortest lifetimes of the superradiant states. \textit{Physarum}'s vein network primarily consists of actin and myosin fibers, which are organized into bundles (see Figure~\ref{fig:Schematic}d).  The average number of actin fibers per unit area of each macroplasmodial body is given by $n_{\text{actin}}$. Thus, the total number of such fibers in a time-dependent area $A(t)$ is
\begin{equation}
N_{\text{fibers}}(t)=n_{\text{actin}}A(t).
\end{equation}

The shortest superradiant lifetime (for the state most strongly coupled to the electromagnetic field, in the single-photon limit) for a 19-filament actin bundle of a few microns is on the order of 10 picoseconds \cite{patwa2024quantum}. If $\tau$ denotes this shortest characteristic superradiant lifetime of an actin filament bundle, then $1/\tau$ is the maximum number of operations per second each filament bundle can perform, and the maximum number of ultraviolet quantum optical operations per second across the macroplasmodial body is thus given by
\begin{equation}\label{QO_bound_general}
N_{QO}(t) = \frac{N_{\text{fibers}}(t)}{\tau} 
= \frac{n_{\text{actin}}}{\tau} A(t).
\end{equation}
By comparing the upper bound on the rate of superradiant operations in Eq.~\ref{QO_bound_general} with the upper bound on the rate of operations in the macroscopic Margolus-Levitin limit in Eq.~\ref{eq:ML_Energy}, we obtain the maximum energy associated with the quantum optical bound: 
\begin{equation}
E_{\text {QO}}(t)=\frac{\pi \hbar \ n_{\text{actin}} A(t)}{\tau}
\end{equation}
For $A(t) \approx 20\,\text{cm}^2$ at $t = 24$ hours, a typical timescale of our \textit{Physarum} experiments, the maximum quantum optical energy evaluates to $E_{\text {QO}} \approx 1.33 \times 10^{-16} \mathrm{~J} = 0.827 \ \mathrm{keV}$, reflecting the ultraviolet excitations for each fiber scaled up across the organismal body. The maximum number of superradiant operations that can be performed over a time interval from $0$ to $t$ is then
\begin{equation} \label{maxQOops}
\mathcal{N}_{QO}(t) = \frac{n_{\text{actin}}}{\tau} \int_{0}^{t} dt' A(t').
\end{equation}
\textcolor{black}{As discussed in Sec. 3, treating \textit{Physarum} as a weakly coupled driven-dissipative system introduces a correction factor of $\pi/\sqrt{2}$ relative to the closed-system Margolus-Levitin bound. The corresponding driven-dissipative quantum optical bound is therefore
\begin{equation}\label{maxQOops_Open}
\mathcal{N}^{\mathrm{open}}_{\mathrm{QO}}(t)
\approx\frac{\pi}{\sqrt{2}}\mathcal{N}_{\mathrm{QO}}(t)= \frac{\pi n_{\text{actin}}}{\sqrt{2} \tau} \int_{0}^{t} dt' A(t').
\end{equation}}
This expression provides the superradiant upper bound on the number of ultraviolet-photoexcited quantum optical operations that can be performed by the actin fiber network in \textit{Physarum} over a given time interval.

\textcolor{black}{\subsection{Robustness of computational bounds to parameter variations}}
\textcolor{black}{The sensitivity analyses presented for the individual bounds examine their dependence on the representative parameter choices over the biologically relevant ranges considered (see Tables S4-S6 in the Supplementary Material). Collectively, these analyses show that variations in the corresponding parameters produce no more than approximately order-of-magnitude-level variations in the bounds, while preserving their overall order-of-magnitude hierarchy. Thus, although the precise numerical values depend on the representative parameter choices, the reported bounds remain stable over the parameter ranges considered.}

\section{Results}

We conducted experiments following the protocols described in Methods sections~8.1--8.3, to investigate the morphological indices in \textit{Physarum polycephalum}. These experiments utilized samples from various strains, feeding patterns, ages, and initial seeding biomasses of \textit{Physarum} (see Fig. S1 of the Supplementary Material). 
\textcolor{black}{We measured the area, perimeter, circularity, and fractal dimension of more than 500 macroplasmodial samples every 30 minutes, over 24–72 hours of growth.}
These parameters were then plotted against time to illustrate the growth morphology, and these morphological indices were then used to calculate computational capacities for different physical degrees of freedom in \textit{Physarum}, as described above. The results of these experiments are presented in the following subsections.

\subsection{Time-series data of morphological indices}

\subsubsection{Variations in strains and feeding conditions}
In this analysis, we considered three groups based on strain and feeding condition: young Japanese (age since revival from sclerotia $\leq 27$ days), Carolina (age since revival from sclerotia $\leq 29$ days) and starved young Japanese (age since revival from sclerotia $\leq 27$ days, not fed for three days). The aforementioned morphological indices were analyzed and compared across the groups. For each group, the indices were averaged across all samples, and these averages were plotted over time with their corresponding standard errors. As evident from Fig. S1 of the Supplementary Material, the average area and perimeter explored by the Japanese strain (see Fig. \ref{Various_strains}) are significantly larger than those of the Carolina strain. In addition, the starved Japanese strain explores the largest area and perimeter, likely due to its rapid growth in search of nutrients (see Fig. \ref{Various_strains}a-c). Despite these differences, the circularity dynamics between the Carolina and Japanese strains are very similar, with only minor variations (Fig. \ref{Various_strains}d). Adjustments to these dynamics can be exhibited by altering the starvation conditions. Notably, the averaged fractal dimension is highest for the starved young Japanese group, followed by the young Japanese and the Carolina, further confirming the growth distinctions between the Japanese and Carolina strains (see Fig. \ref{Various_strains}e).

While experiments for the young Japanese groups were limited to 24 hours, several of the Carolina strain experiments extended beyond this window. Out of all Carolina samples ($N=209$) with at least a 24-hour experimental window, $N=63$ were extended to 36 hours and $N=45$ were continued to 48 hours. For the Carolina strain, the extended time series of the morphological indices are shown in Fig. S2 of the Supplementary Material. 

\textcolor{black}{To assess the statistical robustness of the trends in the morphological quantities shown in Fig. ~\ref{Various_strains}, we performed bootstrap resampling analyses and computed $95 \%$ and $99.5 \%$ confidence intervals ,derived from 100,000 bootstrap resamples, for the corresponding group-aggregated mean quantities. Details of the bootstrap resampling procedure are discussed in the Methods section, while the corresponding confidence-interval plots are presented in Fig. S6 of the Supplementary Material. These confidence intervals estimate the variability of the inferred mean trajectories under replicate-level resampling and indicate that the principal temporal trends and relative distinctions between experimental groups shown in Fig. ~\ref{Various_strains} remain consistent across the experimental trials.}

\textcolor{black}{We additionally aggregated the data shown in Fig. 2 by scanner and analyzed the scanner-aggregated mean trajectories of the morphological quantities for each experimental group. Although some variability was observed among the mean trends corresponding to different scanners, no dominant systematic scanner-dependent trend was observed in the resulting trajectories. The corresponding scanner-aggregated plots are provided in Fig. S10 of the Supplementary Material.}

\begin{figure}
    \centering
    \hspace*{\fill}
    \begin{minipage}[b]{0.92\textwidth}
        \raggedleft
        \includegraphics[width=\textwidth]{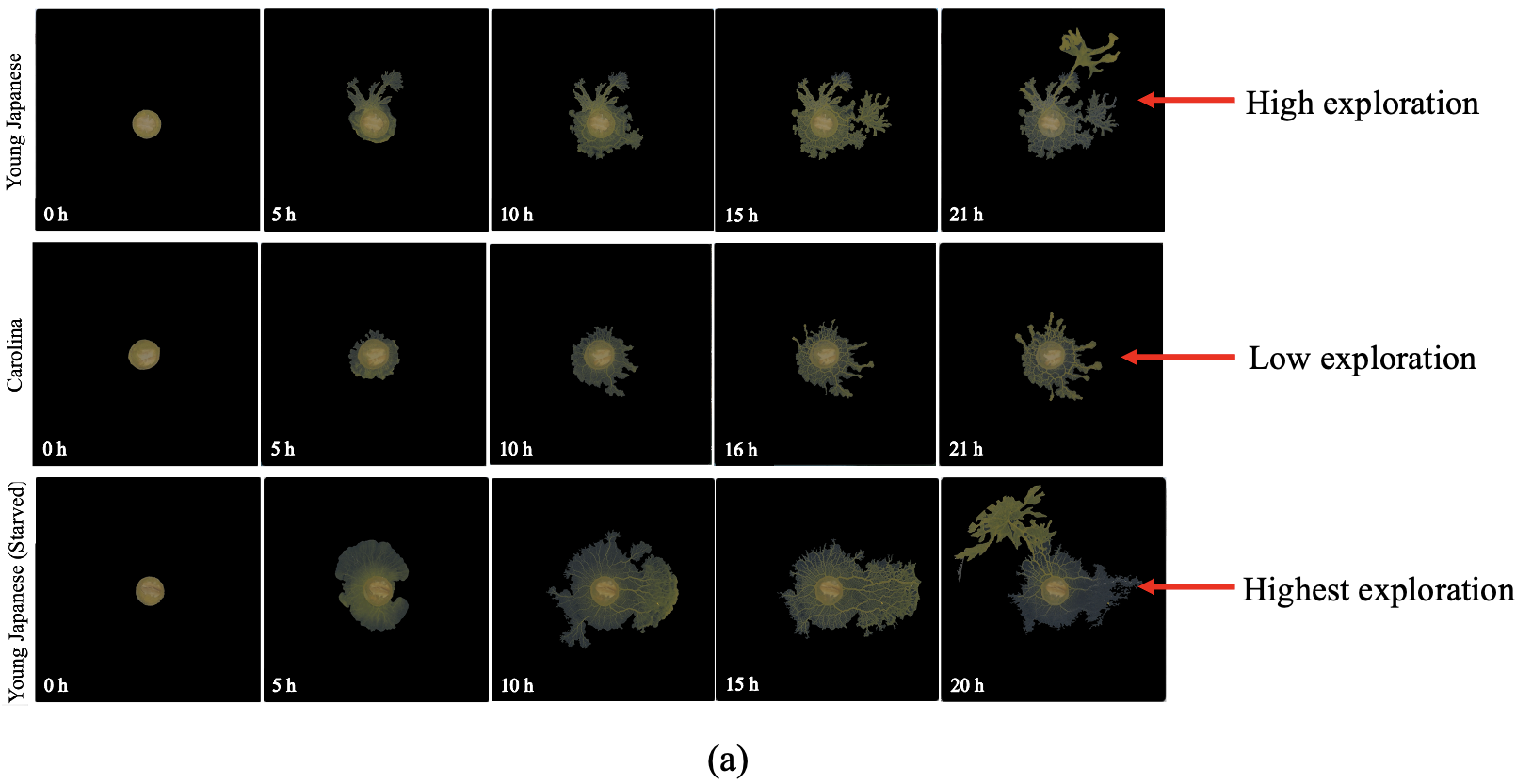}
    \end{minipage}
    \includegraphics[width=0.85\textwidth]{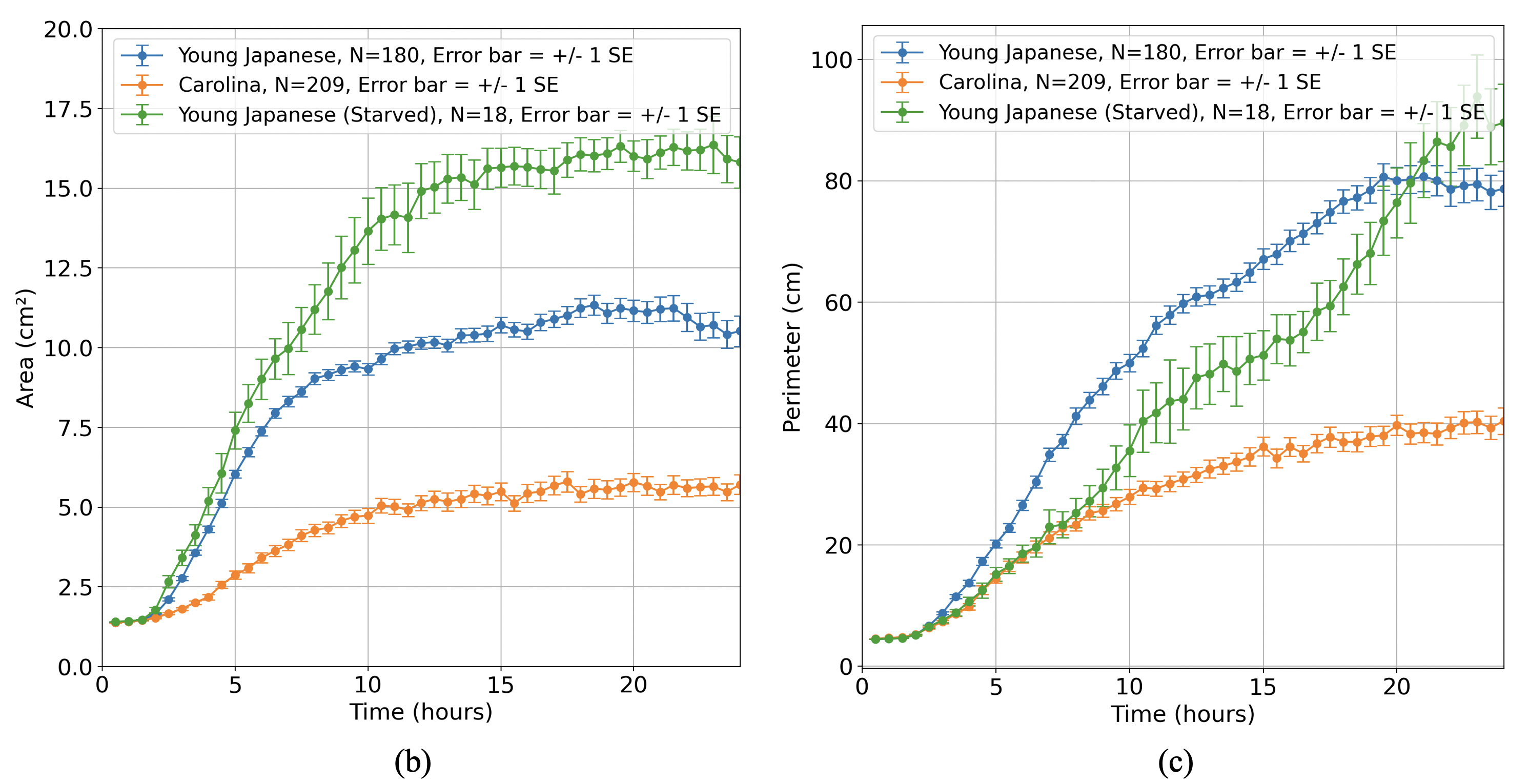} 
    \includegraphics[width=0.85\textwidth]{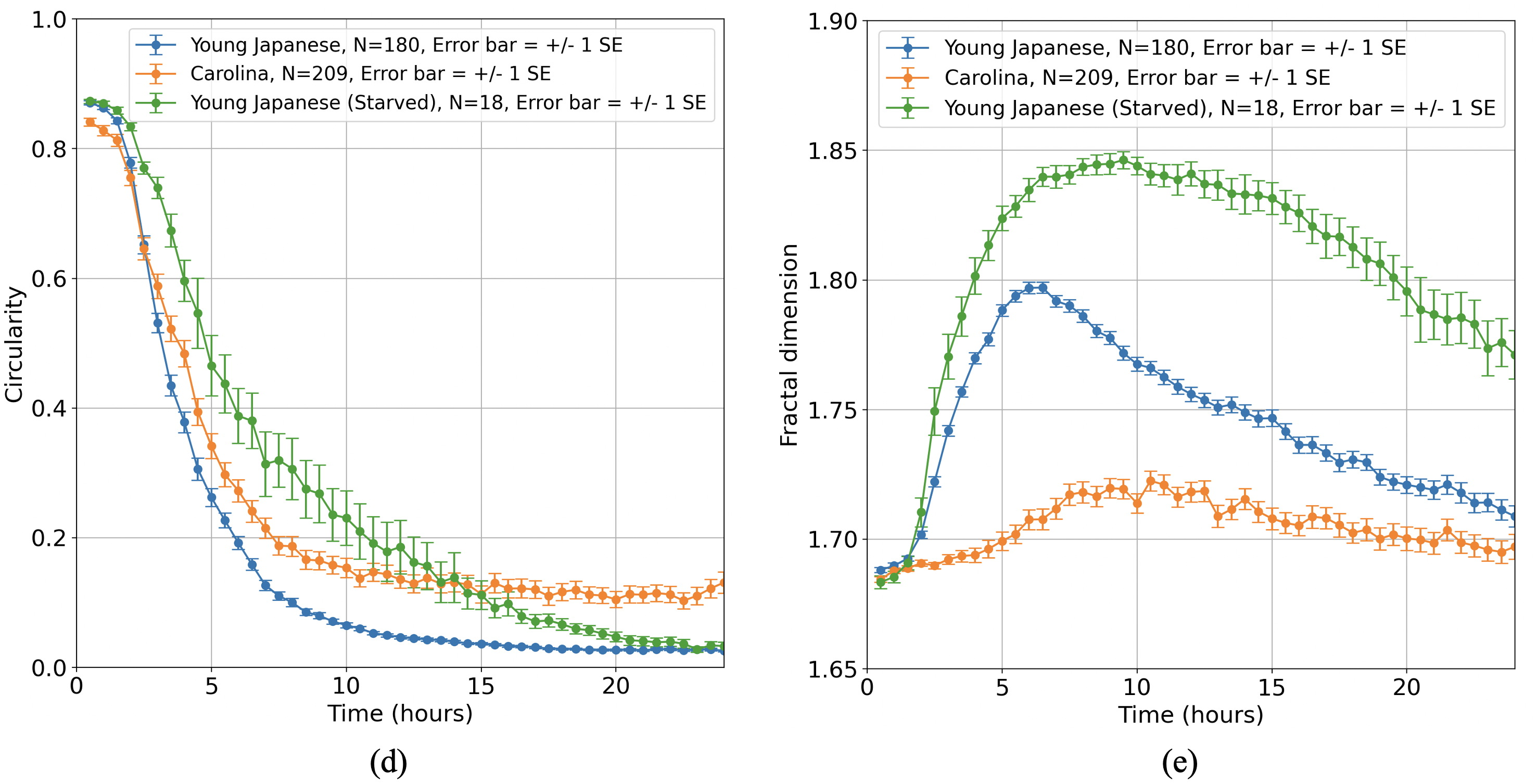}
\end{figure}

\begin{figure}
\centering
    \caption{\textbf{The morphological circularity of a \textit{\textit{Physarum}} body---proportional to the ratio of its area to its perimeter squared---exhibits a similar decay across the younger strains, with an earlier peak in their fractal dimensions while exploring a 2D agar surface.} (a) Time-lapse snapshots of \textit{\textit{Physarum}} growth shown at $\sim$5-hours intervals over the first $\sim$20 hours for three young strains (age since revival from sclerotia$\leq$ 29 days) : Japanese (top row), Carolina (middle row), and starved Japanese (bottom row). (b) area, (c) perimeter, (d) circularity, and (e) fractal dimension as a function of time averaged across replicates in each group.} 
    \label{Various_strains}
\end{figure}

\subsubsection{Variations in the age since revival from sclerotia}

 For the Japanese strain, we compared morphological quantities for two groups categorized based on age since revival from sclerotia. The first group consists of younger \textit{Physarum} batches, aged $\leq 27$ days since the revival from sclerotia, while the second group comprises older batches, aged $\geq 49$ days since the revival from sclerotia. The younger group consisted of both starved and typically fed Japanese samples. The morphological indices were averaged across samples for both young and old \textit{Physarum} and are presented together in Fig. \ref{Age_groups}.

\begin{figure}
    \centering
    \hspace*{\fill}
    \begin{minipage}[b]{0.9\textwidth}
        \raggedleft
        \includegraphics[width=\textwidth]{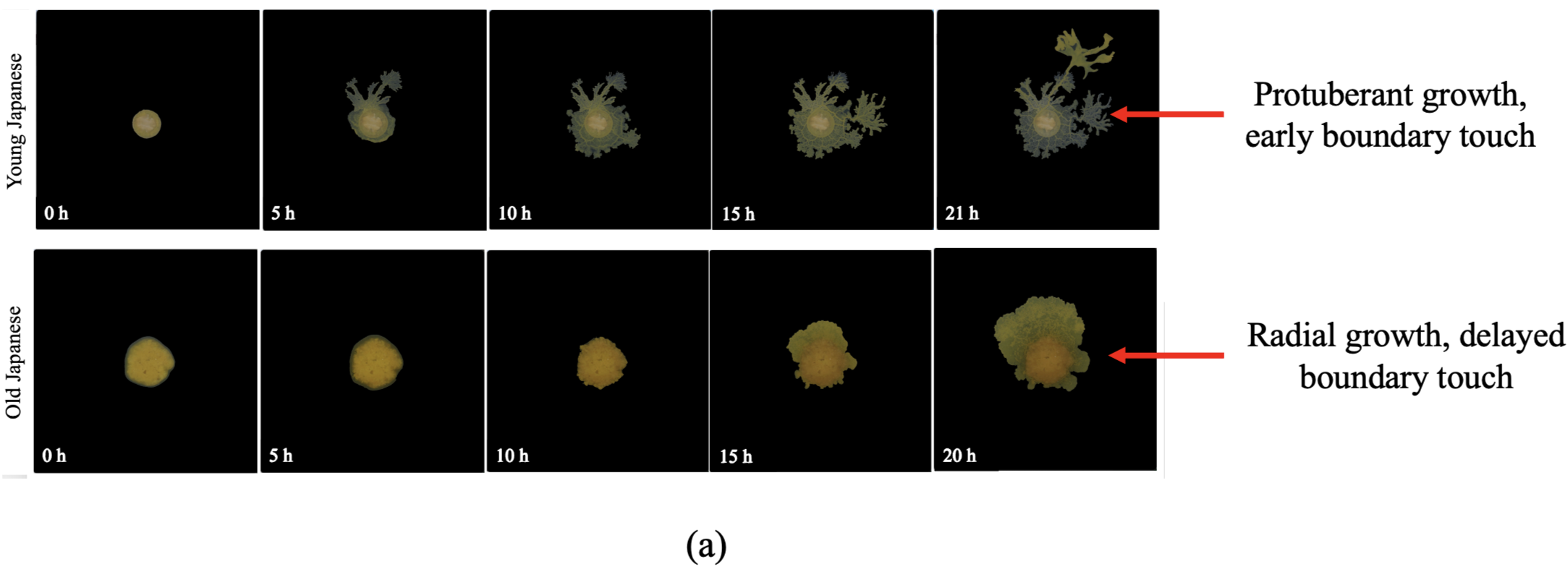}
    \end{minipage}
    \includegraphics[width=0.85\textwidth]{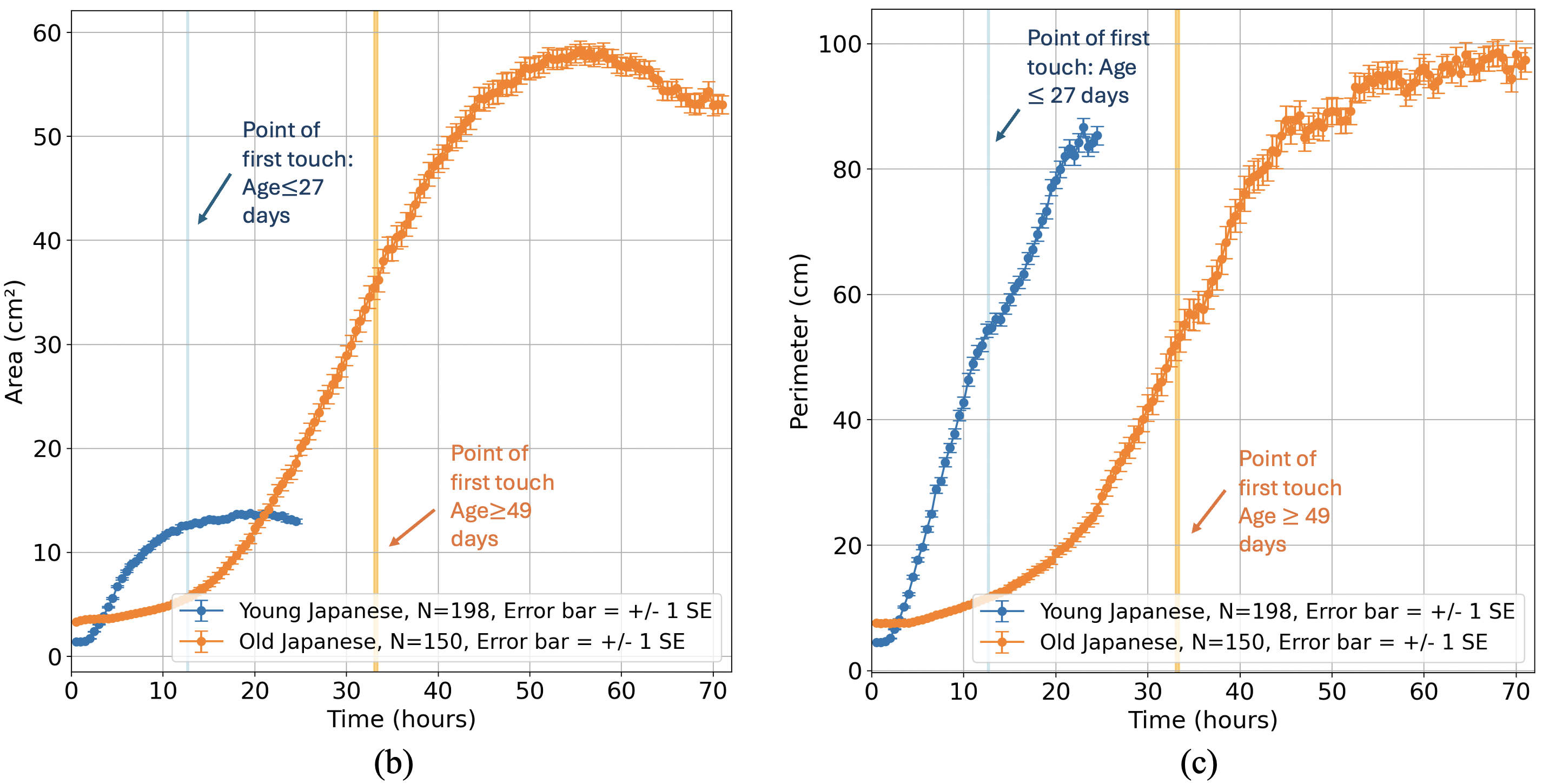}
    \includegraphics[width=0.85\textwidth]{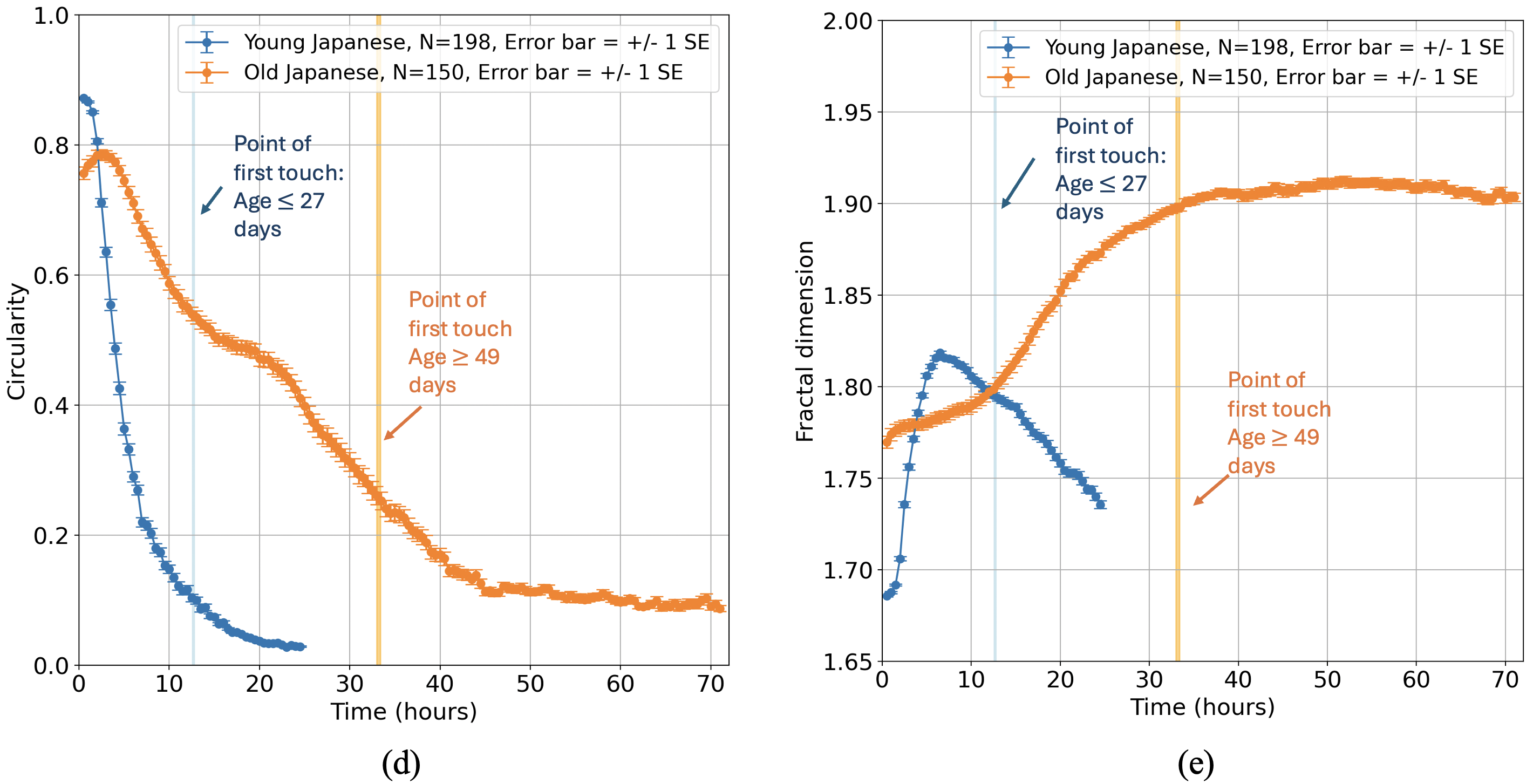}
\end{figure}
\begin{figure}
    \centering
    \caption{\textbf{The young Japanese group exhibits
rapid, protuberant initial growth with early peaks in its fractal dimension, whereas the old Japanese group grows more slowly and equiradially at the outset but eventually stabilizes at a higher fractal dimension.} (a) Time-lapse snapshots shown at $\sim$5-hour intervals over the first $\sim$20 hours for young (top row, age since revival from sclerotia $\leq$27 days) and old (bottom row, age since revival from sclerotia $\geq$49 days) Japanese strains.  Averaged (b) area, (c) perimeter, (d) circularity, and (e) fractal dimension for the young (age since revival from sclerotia $\leq$ 27 days) and old (age since revival from sclerotia $\geq$ 49 days) Japanese samples. The blue and orange bars indicate the earliest point at which the organism touches the boundary of the dish for young and old Japanese groups, respectively, with the width of each bar representing the standard error across samples in the corresponding group. \textcolor{black}{The growth snapshots of the representative young Japanese strain sample from Fig.~\ref{Various_strains} are reproduced as the upper row of panel (a) to show the differences in growth between representative young and old Japanese samples.}}
    \label{Age_groups}
\end{figure}

It can be observed that the younger group initially explores a larger area and has a greater perimeter compared to the older group. However, as time progresses, the exploration of the older group accelerates (see Fig. \ref{Age_groups}a-c). Additionally, the point of first touch to the agar plate boundary for the younger group occurs significantly earlier ($\sim10$ hours), nearly one-third the time required by the older group ($\sim30$ hours). The point of first touch achieved by both young and old \textit{Physarum} is reflected in their circularity dynamics. The circularity of the younger \textit{Physarum} decreases rapidly as it begins to grow in a more protuberant manner, whereas the older \textit{Physarum} exhibits more equiradial growth, resulting in a slower decrease in circularity (see Fig. \ref{Age_groups}d). In addition to circularity, significant distinctions between young and old \textit{Physarum} can be observed in the fractal dimension plot. The fractal dimension curve for younger \textit{Physarum} shows a peak between 5 and 10 hours, while the curve for older \textit{Physarum} lacks this feature (see Fig. \ref{Age_groups}e), stabilizing at more than 35 hours but at a higher peak value than for the younger strain.

\subsubsection{Variations in biomass and vein network structure}

To examine the effect of the biomass and vein network on morphological indices, we classified our experiments for old Japanese (age since revival from sclerotia$\geq49$ days) into two groups: (1) the vein network disrupted and (2) the vein network connected (see Fig. \ref{Vein_structure}a). These groups were further divided into subgroups based on their initial biomasses. The vein network disrupted group was first divided into ten biomass ranges and morphological indices were averaged over the respective biomass ranges. The average circularity and fractal dimension for these biomass ranges are shown in Fig. \ref{Vein_structure}b-c. From the circularity plot (Fig. \ref{Vein_structure}b), a notable revival is observed around the 30-40 hours interval. It is also evident that the highest initial biomass range exhibits the greatest circularity in this interval, highlighting the impact of initial biomass variation on morphology. Similarly, in the fractal dimension plot (Fig. \ref{Vein_structure}c), the highest initial biomass curve shows distinct behavior, exhibiting a peak within the first 10 hours of growth. This feature is absent in the other initial biomass ranges, further demonstrating the unique influence of higher initial biomass on fractal morphology.

Panels \ref{Vein_structure}d–h present the averaged morphological indices for groups with disrupted and connected vein networks, with standard error bars. This comparison was performed to assess how morphological indices differ between these two initial network configurations. As shown in Fig. \ref{Vein_structure}d, both groups initially explore similar areas, but the vein network-connected group surpasses the vein network-disrupted group around 15–20 hours and maintains a larger area thereafter. Similarly, in Fig. \ref{Vein_structure}e, both groups begin with comparable perimeter values, but the vein network-connected group exhibits a higher perimeter after approximately 20 hours. In the circularity plot (Fig. \ref{Vein_structure}f), we observe that the vein network-connected group shows a revival between 15 and 25 hours, while the vein network-disrupted group does not show a revival for the averaged circularity values. Regarding fractal dimension, initially, the vein network-disrupted group has a higher fractal dimension value than the vein network-connected group (Fig. \ref{Vein_structure}g). However, later in the experiment, the vein network-connected group's fractal dimension surpasses that of the vein network-disrupted group after $\sim$20 hours. These observations demonstrate that morphological growth trajectories are sensitive to the initial structure of the vein network.
 
\begin{figure}
    \centering
    \hspace*{\fill}
    \begin{minipage}[b]{0.9\textwidth}
        \raggedleft
        \includegraphics[width=\textwidth]{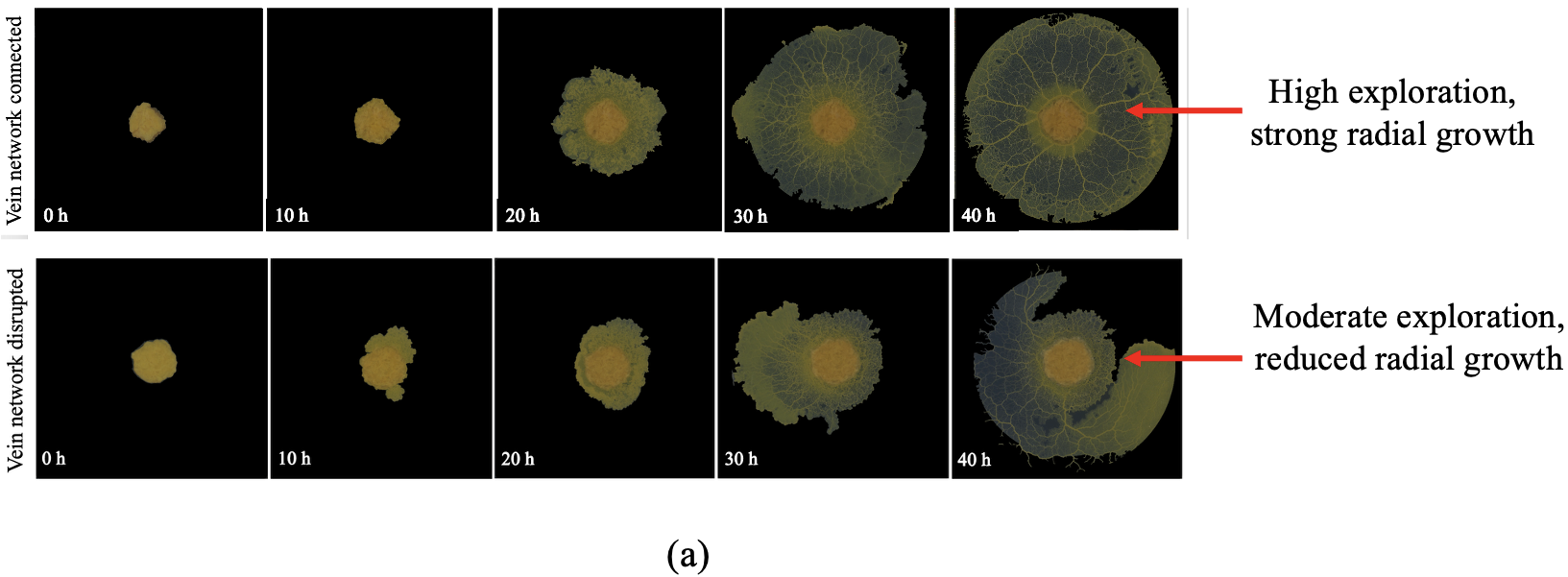}
    \end{minipage}
    \includegraphics[width=0.9\textwidth]{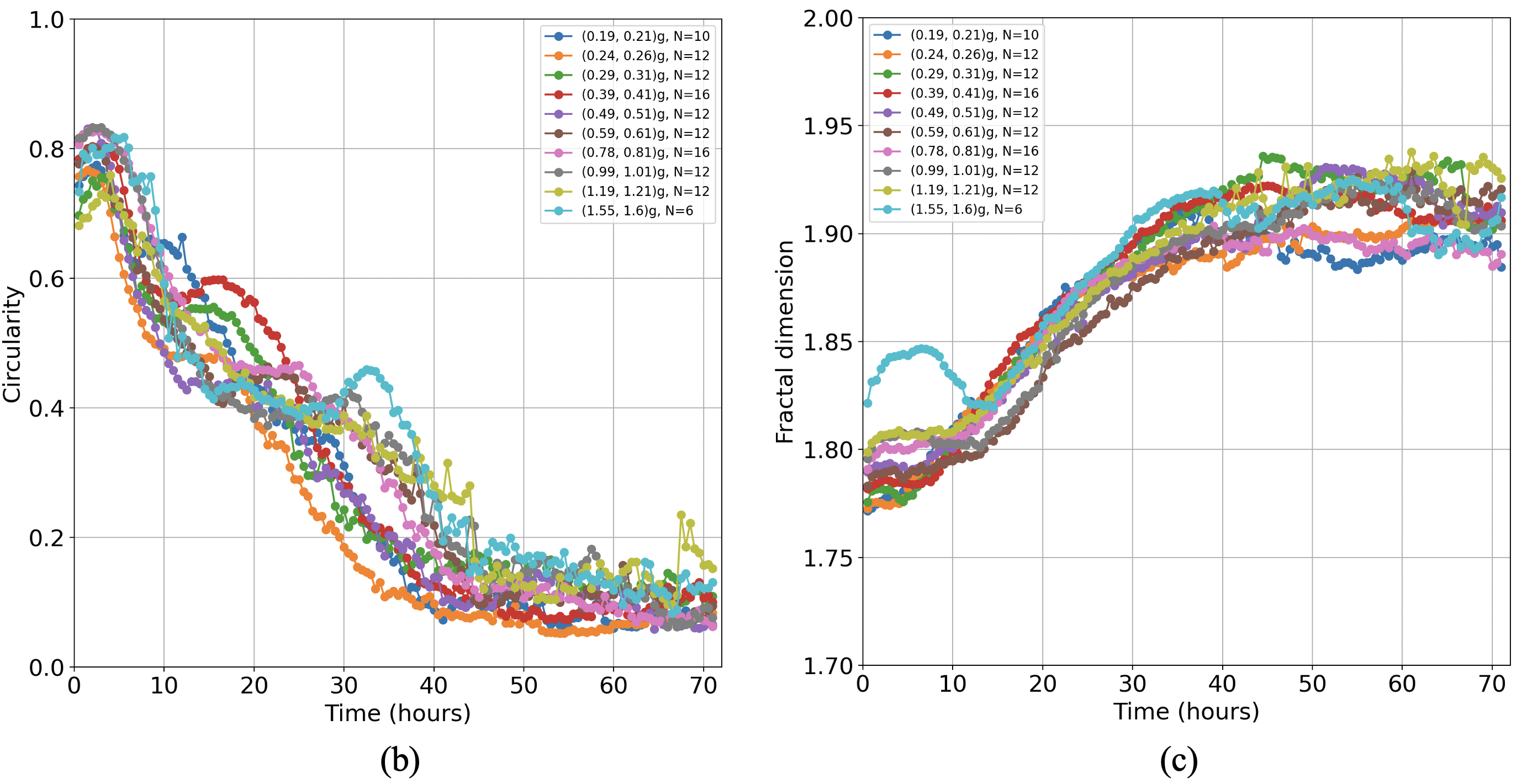}
    \includegraphics[width=0.9\textwidth]{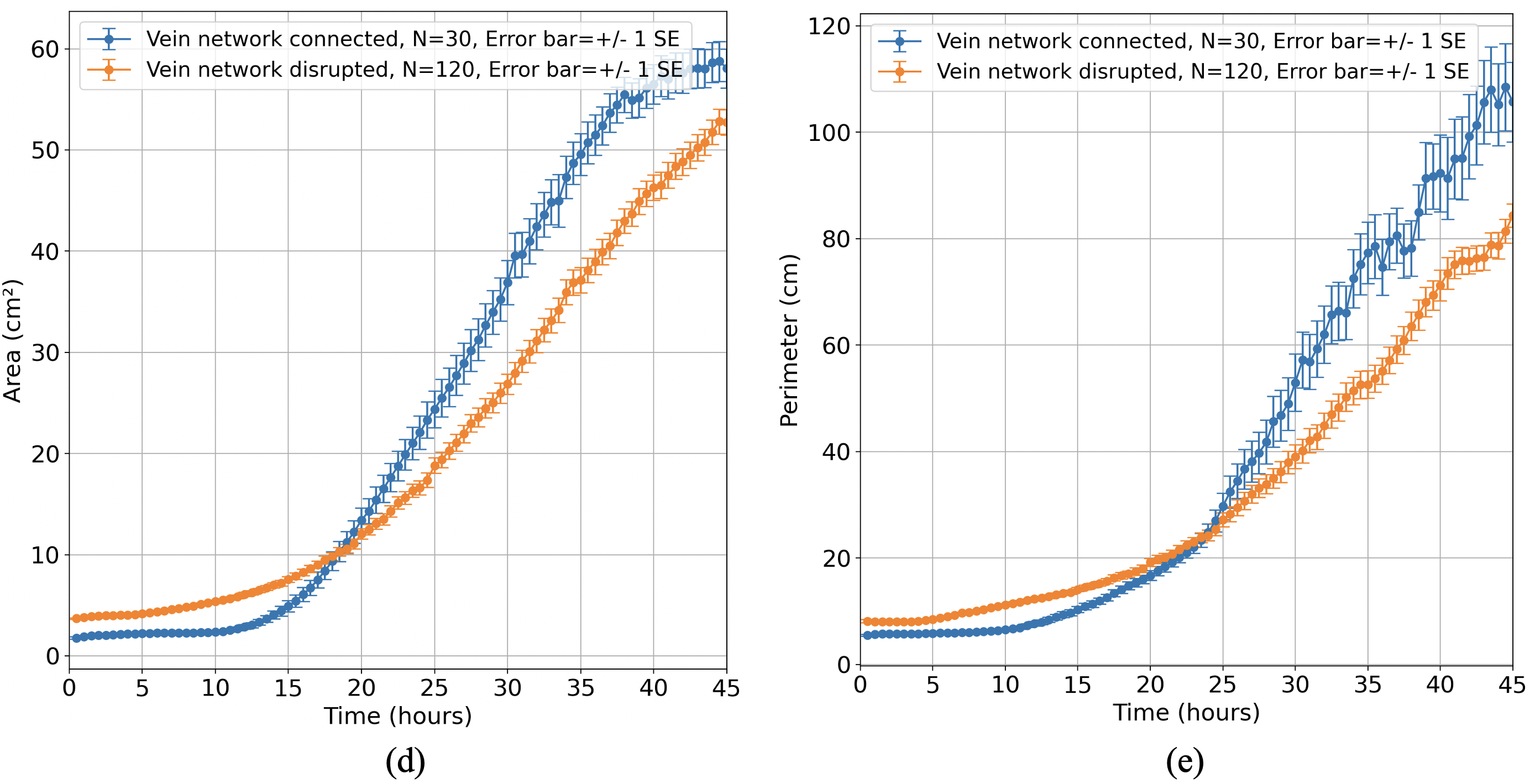}
\end{figure}
\begin{figure}[h!]
    \centering
    \includegraphics[width=0.9\textwidth]{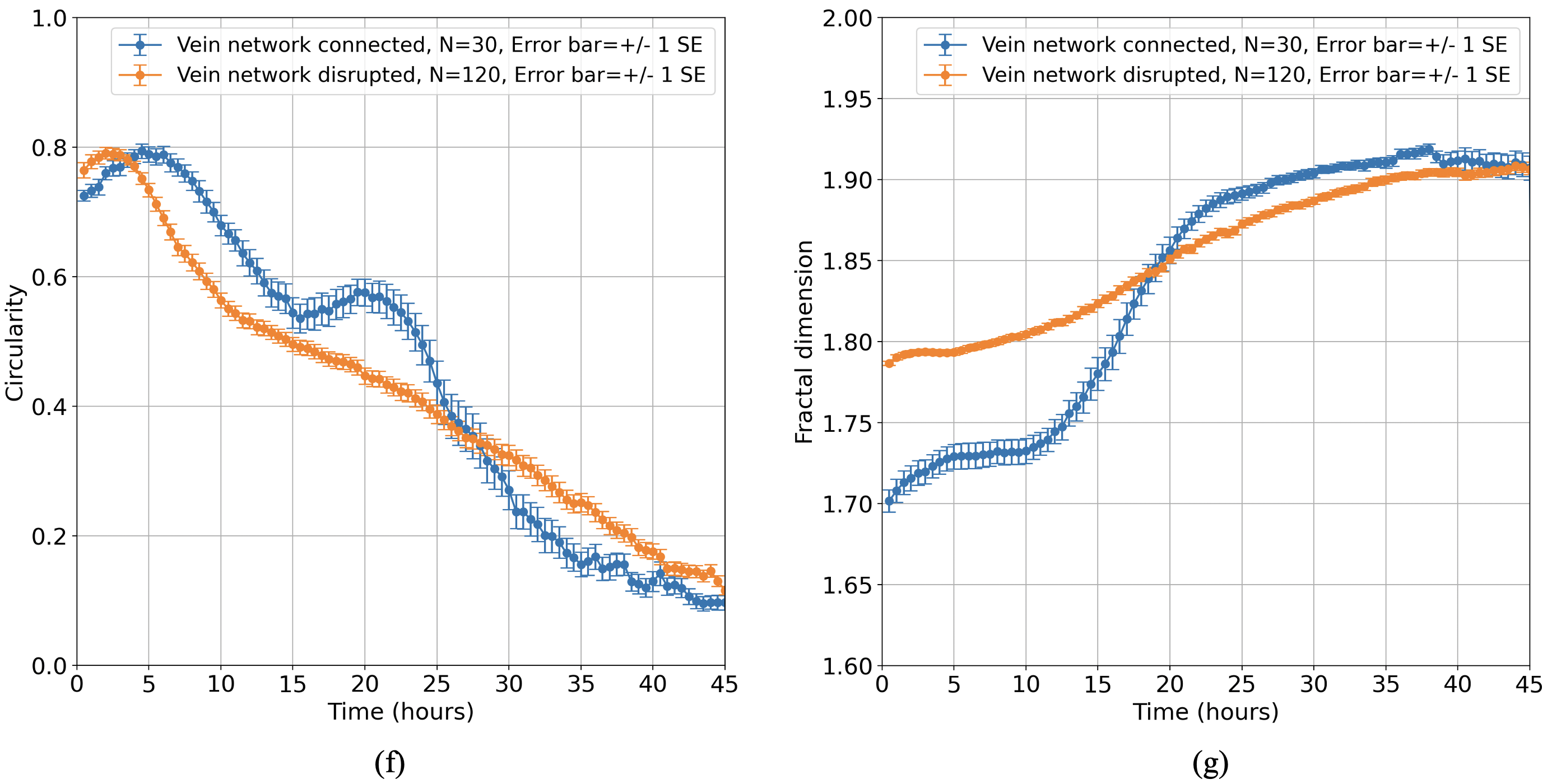}
    \caption{\textbf{The vein network-connected Japanese \textit{\textit{Physarum}} group initially exhibits higher circularity and lower fractal dimension values compared to the vein network-disrupted group, before gradually transitioning to lower circularity and slightly higher fractal dimension as they explore larger area and perimeter values over time.} (a) Time-lapse snapshots shown at $\sim$10-hours intervals over the first $\sim$40 hours for two old Japanese samples (age since revival from sclerotia $\geq$49 days): vein network connected (top row) and vein network disrupted (bottom row).
   Averaged (b) circularity  and (c) fractal dimension  for ten biomass ranges for vein network-disrupted group. Averaged (d) area, (e) perimeter, (f) circularity, and (g) fractal dimension across all biomasses for both vein network-connected and -disrupted groups.}
    \label{Vein_structure}
\end{figure}

\subsection{\label{app:subsec1}Scaling laws for area, perimeter, and circularity of \textit{Physarum polycephalum} macroplasmodial body}

For the Japanese strain groups, namely the young Japanese, old Japanese, and young Japanese-starved, the area and perimeter increase monotonically over time and eventually saturate, marking the transition of the organism into a non-equilibrium steady state (NESS). This behavior is well captured by a single sigmoid function fit to the time series of both the area and the perimeter. Specifically, we model these morphological indices as
\begin{equation}\label{Area_fit_Japanese}
A_{\text{fit, Japanese}}(t) = \frac{\alpha}{1 + e^{-\beta(t - \gamma)}},
\end{equation}
\begin{equation}\label{Perimeter_fit_Japanese}
P_{\text{fit, Japanese}}(t) = \frac{\delta}{1 + e^{-\eta(t - \theta)}}.
\end{equation}
Here, \( \alpha \) and \( \delta \) represent the asymptotic values of area and perimeter, respectively. The parameters \( \beta \) and \( \eta \) control the growth rates, while \( \gamma \) and \( \theta \) determine the corresponding inflection time points. 
\( \alpha \) has dimensions of area \([L^2]\), \( \delta \) has dimensions of length \([L]\), \( \beta \) and \( \eta \) have dimensions of inverse time \([T^{-1}]\), and \( \gamma \) and \( \theta \) have dimensions of time \([T] \).

For the Carolina strain, out of a total of $N = 209$ samples, $N = 195$ exhibit sigmoid behavior. A smaller subgroup $N = 14$ extending up to 48~hours displays a distinct bi-sigmoid pattern with two successive growth phases. The subgroup-aggregate morphological indices for these samples ($N = 14$) are shown in Fig.~S3 of the Supplementary Material. As a result, the group-aggregate area and perimeter time series, averaged across all examined samples at each time point, display bi-sigmoid trends for area and perimeter (see Fig.~S9a,b in the Supplementary Material), though the second growth phase is largely averaged out by the supermajority of sigmoid samples. To consistently account for this variability, we employ a more general bi-sigmoid model to describe the area and perimeter growth in the Carolina strain:
\begin{equation}\label{Area_fit_Carolina}
A_{\text{fit, Carolina}}(t) = \frac{\alpha_1}{1 + e^{-\beta_1(t - \gamma_1)}} + \frac{\alpha_2}{1 + e^{-\beta_2(t - \gamma_2)}},
\end{equation}
\begin{equation}\label{Perimeter_fit_Carolina}
P_{\text{fit, Carolina}}(t) = \frac{\delta_1}{1 + e^{-\eta_1(t - \theta_1)}} + \frac{\delta_2}{1 + e^{-\eta_2(t - \theta_2)}}.
\end{equation}
In this case, \( \alpha_1, \alpha_2 \) and \( \delta_1, \delta_2 \) represent the area and perimeter contributions from the two growth phases; \( \beta_1, \beta_2, \eta_1, \eta_2 \) are the respective growth rates; and \( \gamma_1, \gamma_2, \theta_1, \theta_2 \) specify the inflection time points. The dimensions are analogous to the single-sigmoid case: each \( \alpha_i \) has dimensions of area \([L^2]\), each \( \delta_i \) has dimensions of length \([L]\), each \( \beta_i, \eta_i \) has dimensions of inverse time \([T^{-1}]\), and each \( \gamma_i, \theta_i \) has dimensions of time \([T] \).

Fig.~S9a in the Supplementary Material shows the area fits for the mean area for each of the four groups. The models described by Equations \ref{Area_fit_Japanese} and \ref{Area_fit_Carolina} effectively capture the observed growth dynamics ($R^{2}>0.97$), demonstrating a close match with the discrete time-series data. The fitted parameter values for the various strains and conditions are summarized in Table S1 of the Supplementary Material. 

A useful metric to describe the growth of \textit{\textit{Physarum}} is its circularity,
\begin{equation} \label{circularity1}
    C(t) = 4\pi \frac{A(t)}{P(t)^2},
\end{equation}
which in the case of a perfect circle is exactly unity, because $P^2 = (2\pi r)^2 = 4 \pi A$. While it may seem natural to fit circularity using the fits for area and perimeter, it is important to consider the critical caveat that the errors in area and perimeter propagate into circularity. Therefore, we fit the circularity separately. For this purpose, we utilized a similar (but reverse, decaying) sigmoid model to fit the circularity for the Japanese strains:
\begin{equation}
C_{\text{fit, Japanese}}(t) = 1 - \frac{\phi}{1 + e^{-\kappa (t - \xi)}}.
\label{Circ_simplified}
\end{equation}
Since the area and perimeter of the Carolina strain are described by a bi-sigmoid model, its circularity—which depends on the ratio of area to perimeter squared—can be effectively captured by a decaying bi-sigmoid model:
\begin{equation}
C_{\text{fit, Carolina}}(t)=1-\left(\frac{\phi_1}{1+e^{-\kappa_1\left(t-\xi_1\right)}}+\frac{\phi_2}{1+e^{-\kappa_2\left(t-\xi_2\right)}}\right)
\label{Circ_simplified_2}
\end{equation}
Fitting the experimental circularity values with the models in Eqs.~\ref{Circ_simplified}-\ref{Circ_simplified_2} resulted in excellent agreement, with $R^2$ values of 0.96 or higher across all groups. The area, perimeter, and circularity time-series data plots with corresponding analytical model fits are presented in Fig.~S9 in the Supplementary Material. The fit parameters for the evolution of area, perimeter, and circularity are summarized in Table S1 of the Supplementary Material. 

\subsection{Determination of transition time to NESS}

In response to the boundary conditions imposed by the agar plate, \textit{Physarum}’s growth exhibits a non-equilibrium steady state (NESS), during which the explored area plateaus and changes only minimally over time. We have shown that the growth of the Japanese strain exhibits a single NESS, which can be accurately modeled using a single sigmoid function. In contrast, the Carolina strain displays two distinct NESS phases, necessitating a bi-sigmoid model. To identify the transition time(s) to NESS, we computed the time derivative of the area curve, $\frac{dA}{dt}$, obtained from a sigmoid (in the case of Japanese) or bi-sigmoid fit (in the case of Carolina). We define the transition time to NESS,  $t_{\text{NESS}}$, as the earliest time such that:
\[
\frac{dA}{dt} \leq \epsilon, \quad \text{where } \epsilon = 0.15 \times  \left(\frac{dA}{dt}\right)_{\text{max}},
\]
and for all $t\geq t_{\text{NESS}}$, the first derivative of the area curve will fall below the critical threshold set by $\epsilon$.

To investigate dynamics of the transition to the NESS, we examined the second time derivative of area across all strains. The inflection point—defined by $\frac{d^2A}{dt^2} = 0$—marks the onset of deceleration in area growth. We then evaluated the second derivative at the empirically determined 15\% growth-rate cutoff time for each sample.

Across strains, $\frac{d^2 A}{d t^2}$ decreased substantially beyond the inflection point, reaching values between $-0.1$ (young Japanese and young Japanese-starved) and $-0.03$ (Carolina) at the 15\% cutoff, compared to the corresponding maxima of $+0.28$ (young Japanese and young Japanese-starved) and $+0.08$ (Carolina). This decline quantifies the marked slowing of growth dynamics relative to the inflection point. At this 15\% cutoff, \textit{Physarum} had already covered $\sim$95\% of the area it would eventually reach at NESS. These observations support our choice of a 15\% threshold as a consistent and operationally meaningful marker for identifying the transition to NESS. Beyond this point, $\left|\frac{d^2A}{dt^2}\right|$ continued to decline toward zero in all cases. The area fits for all groups, along with their first and second derivatives and the 15\% threshold, are shown in Fig.~S13 of the Supplementary Material.

Using this criterion, the NESS transition times ($t_{\text{NESS}}$) were calculated for the four experimental groups: For the young Japanese, \textit{Physarum} reached a NESS at approximately $11.17$ hours, while for the young Japanese-starved, the transition occurred at $13.19$ hours. \textcolor{black}{For the Carolina, the first transition occurred at $12.63$ hours, more than an hour after the young Japanese group but about half an hour before the young Japanese-starved group, with a second NESS transition expected around $51.48$ hours---predicted by extrapolating the fit beyond the available imaging data.} By contrast, the old Japanese transitioned to the NESS much later, at approximately $51.43$ hours. The computed transition times are summarized in Table S2 of the Supplementary Material.

\textcolor{black}{We additionally evaluated the robustness of the $15\%$ derivative threshold used to define the transition to the NESS. The derivative threshold was varied from $10\%$ to $20\%$, and for all experimental groups the resulting NESS transition times remained within $10\%$ of those obtained using the $15\%$ derivative threshold. The corresponding transition times for different derivative-threshold values are provided in the Supplementary Material.}

\subsection{Behavior of sigmoid integrals affecting computational capacity bounds}

For a general sigmoid function of the form  
\[
f(t) = \frac{\alpha}{1 + e^{-\beta (t-\gamma)}},
\]
the indefinite integral is given by
\begin{equation} \label{sigmoidint}
\int f(t)\, dt = \frac{\alpha}{\beta}\,\ln\!\left(1+e^{\beta (t-\gamma)} \right) + \text{const}.
\end{equation}
For early times \((t \ll \gamma)\), we can set \(x = e^{\beta (t-\gamma)}\) with \(x \ll 1\).  
Using the approximation \(\ln(1+x) \approx x\), we obtain  
\[
\int f(t)\, dt \approx \frac{\alpha}{\beta} e^{\beta (t-\gamma)} + \text{const},
\]
which shows that the integral initially follows an exponential growth law.

For late times \((t \gg \gamma)\), the exponential dominates, so $\ln\!\left(1 + e^{\beta (t-\gamma)}\right) \approx \beta (t-\gamma)$ and hence,  
\[
\int f(t)\, dt \approx \alpha t + \text{const}.
\]
Thus, any bound that depends on the time integral of a sigmoid area or perimeter fit grows linearly in time once the organism has reached its non-equilibrium steady state (NESS).

The hydrodynamic bound, the chemical ATP bound, and the quantum optical bound can each be expressed as integrals of the area or perimeter. Since integrals of sigmoid functions exhibit an exponential rise at early times and a linear tail at late times, these three bounds inherit the linear growth regime beyond the NESS. In the next section, this late-time behavior is demonstrated explicitly by fitting the later time points with a linear function for each bound (except the kinetic energy bound), confirming the expected linear tail beyond the NESS. In contrast, the kinetic energy bound integral, according to Eq.~\ref{N_KEint}, involves both the area and the squared time derivative of the perimeter. Since the speed decays to zero once perimeter growth saturates, this bound does not show a linear tail at late times but instead saturates.

\subsection{Analytical upper bounds on morphological operations of \textit{\textit{Physarum}}}

\subsubsection{Hydrodynamic cytosol bound}
\label{subsec:Hyd_bound_values}

Using the general integral for the hydrodynamic cytosol upper bound expressed by Eq. \ref{eq:Hyd_bound_Open}, with the perimeter fit for the Japanese strain groups in Eq.~\ref{Perimeter_fit_Japanese}, we obtain

\textcolor{black}{\begin{equation}\label{Hyd_bound_with_fit_Japanese}
\mathcal{N_{\text{hydro, Japanese}}^{\text{open}}}(t) = \frac{\pi n_{\text {hydro }}^{\text {local }}}{\sqrt{2} \ l_d} \frac{\delta}{\eta} \left[ \ln \left(1 + e^{\eta (t-\theta)} \right)  \right],
\end{equation}
up to a constant term defined by $\pi n_{\text{hydro}}^{\text{local}} \, \delta / (\sqrt{2} \ l_d \, \eta)\ \ln(1 + e^{-\eta \theta})$ from the lower bound of the integral ($t=0$)}. For all values of \( \eta \) and \( \theta \) used in our fits, the product \( \eta\theta \) is greater than or equal to 2.48 (see Table S1 of the Supplementary Material). As a result, the term \( \ln(1 + e^{-\eta\theta}) \) remains close to zero, resulting in only negligible changes to the bounds in Eq.~\ref{Hyd_bound_with_fit_Japanese}. Including this constant term does not affect the order of magnitude or the relative order of the estimated bounds across experimental groups.

Similarly, using the bi-sigmoid perimeter fit from Eq. \ref{Perimeter_fit_Carolina} for the Carolina strain, we obtain
\textcolor{black}{
\begin{equation}\label{Hyd_bound_with_fit_Carolina}
\mathcal{N}_{\text{hydro, Carolina}}^{\text{open}}(t) = \frac{\pi n_{\text {hydro}}^{\text {local}}}{\sqrt{2} \ l_d} \left[
\frac{\delta_1}{\eta_1} \ln \left(1 + e^{\eta_1(t - \theta_1)} \right)
+ \frac{\delta_2}{\eta_2} \ln \left(1 + e^{\eta_2(t - \theta_2)} \right)
\right].
\end{equation}}
Here, $n_{\text {hydro }}^{\text {local }}$ and $l_{d}$ represent the number of operations per second and the typical width for a single oscillator, as described in Section 5. Eqs. \ref{Hyd_bound_with_fit_Japanese} and \ref{Hyd_bound_with_fit_Carolina} provide estimates of the upper bounds on the number of hydrodynamic cytosol operations for the four groups. For each sample of \textit{Physarum}, the perimeter data were modelled analytically using the corresponding sigmoid or bi-sigmoid functions, and the resulting fit parameters were used to calculate the bounds for that individual macroplasmodial body. These values were then averaged within each experimental group to obtain a group-aggregate bound. The computed bounds over a 24-hour interval are summarized in Table S2 of the Supplementary Material.

\begin{figure}[h!]
    \centering
    \includegraphics[width=0.65\textwidth]{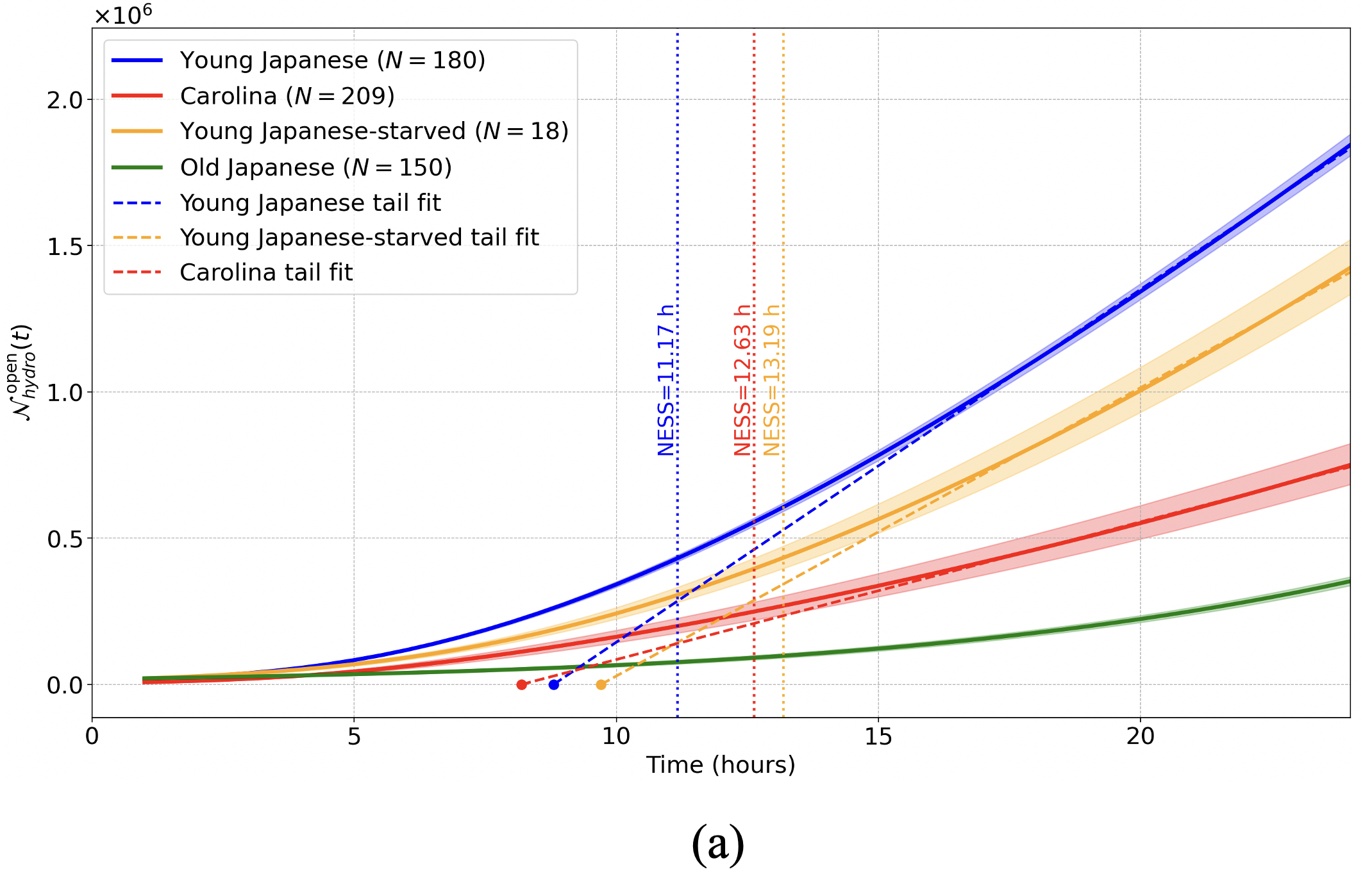}
    \includegraphics[width=0.65\textwidth]{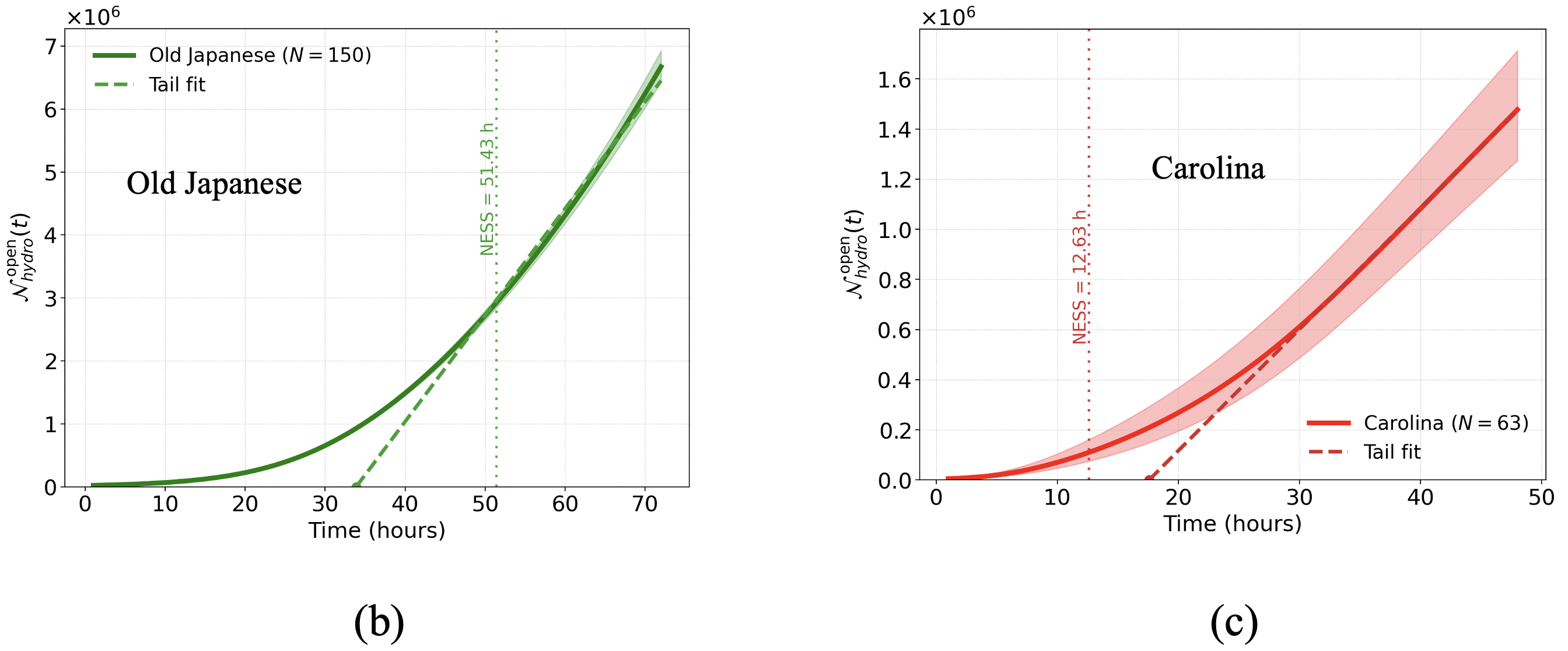}
    \caption{\textbf{The hydrodynamic bound obtained over a 24-hour interval is higher in the more active young Japanese groups compared to the old Japanese and Carolina groups, closely mirroring their perimeter growth trends shown in Fig. \ref{Various_strains}c.} The figure shows (a) the hydrodynamic bound for each group, averaged across all group samples, over a 24-hour interval, with the tail fitted using a linear function to capture long-time behavior. The vertical dotted line indicates the NESS, defined from area stabilization, which precedes the onset of the long-time linear regime of the hydrodynamic bound defined from perimeter stabilization. The $x$–axis intercepts of the late-time fits mark the inflection points of the perimeter sigmoids in Fig.~S9b of the Supplementary Material, yielding values close to the $\theta$ estimates in Table~S1, except for the Carolina subgroup ($N = 63$) extending to 48~h, where the intercept lies at a weighted average of  the two inflection points of the perimeter bi-sigmoid. The plot is shown extended to 72 hours in panel (b) for the old Japanese group and to 48 hours in panel (c) for the Carolina strain subgroup ($N = 63$). The transition to the NESS in the old Japanese group occurs well beyond 24 hours.}
    \label{Hydrodynamic_bound}
\end{figure}

The hydrodynamic bound $\mathcal{N_{\text{hydro}}}(t)$ over time, up to 24 hours, is shown in Fig. \ref{Hydrodynamic_bound}a for the four groups. \textcolor{black}{Among all groups, the young Japanese consistently exhibits the highest hydrodynamic bound ($1.84 \times 10^6$), followed by the young Japanese-starved ($1.42 \times 10^6$). This stands in contrast to the lower values observed in the old Japanese ($3.53 \times 10^{5}$) and the Carolina ($7.51 \times 10^{5}$).} This trend closely follows the perimeter growth observed in each group, which is expected given that the hydrodynamic bound scales directly with the time-integral of the perimeter of the organism. These findings indicate that the hydrodynamic bound increases as the number of independent pseudopod-like oscillators in the advancing periphery increases, that is, as the perimeter increases.

Since the hydrodynamic bound depends on the integral of the perimeter, it shows the typical sigmoid integral behavior: an exponential rise at early times and a linear tail at later times. As shown in Fig.~\ref{Hydrodynamic_bound}a, the late-time portion of the hydrodynamic bound for each group was fitted with a linear function, yielding $R^2 > 0.95$. The vertical dotted line in each plot marks the transition to NESS, determined from the stabilization of the area. This transition occurs slightly earlier than the point where the hydrodynamic number of operations begins to follow the long-time linear trend. The difference arises because the NESS time is defined from the stabilization of the area, whereas the linear regime in the hydrodynamic bound reflects the stabilization of the perimeter. 

The $x$-axis intercepts of the fitted linear tails for the different groups generally correspond to the time points representing the inflections of their respective perimeter sigmoids in Fig.~S4b, yielding values that closely match the $\theta$ values reported in Table~S1. For the Carolina strain, which in general exhibits bi-sigmoid growth (Eq.~35), this holds for the first linear regime defined by $\theta_{1} \ll t < \theta_{2}$, where the bound values are dominated by the first growth-phase term. Consequently, this intercept lies close to $t_{x_1}=\theta_{1}$, similar to the uniphasic Japanese cases. The tail fits shown for Carolina in Fig.~\ref{Hydrodynamic_bound}a are extrapolated from $t = 24~\mathrm{h}$, which lies between the two inflection points of the mean perimeter bi-sigmoid (see Fig.~S9 of the Supplementary Material); hence, the fitted intercept coincides with $t_{x_{1}} = \theta_{1}$. In contrast, in the region $t \gg \theta_{2}$, both growth-phase terms contribute, and the intercept shifts to the weighted average of the two inflection times, rather than occurring at $\theta_{2}$. Thus, the $\theta_{2}$ values do not coincide with the intercepts of the second linear fits (see Sec.~S17 of the Supplementary Material for proof and further details).

The tail fit of the bound values for the Carolina subgroup ($N = 63$) that extends up to 48~h is shown in Fig. ~\ref{Hydrodynamic_bound}c. This fit, corresponding to the regime $t > \theta_{2}$, also intercepts the x-axis not at $\theta_{1}$ or $\theta_{2}$ but at their weighted average, $t_{x_2} = (\delta_{1} \theta_{1} + \delta_{2} \theta_{2})/(\delta_{1} + \delta_{2})$. Bound values for the 72~h experimental duration of the old Japanese group are shown in Fig.~\ref{Hydrodynamic_bound}b, where the transition to the NESS occurs well beyond 24~h. The bound values for the extended Carolina subgroup ($N = 63$) are shown up to 48~h in Fig.~\ref{Hydrodynamic_bound}c.

\subsubsection{Chemical ATP bound}

Eq.~\ref{eq:chem_ATP_Open} provides the upper bound on the number of logical operations due to the exhaustive conversion of chemical ATP to ADP in a given experimental interval from $0$ to $t$. Substituting the sigmoid area fits from Eq.~\ref{Area_fit_Japanese} for the Japanese strain groups into Eq.~\ref{eq:chem_ATP_Open}, we obtain:
\textcolor{black}{
\begin{equation}\label{General_chem_ATP_2}
\mathcal{N}_\text{chem, Japanese}^{\text{open}}(t)=\frac{0.664 \ \eta_{\text {turnover}} \eta_{\text{work }}  \rho_0 \ell}{\sqrt{2} \hbar}\int_0^t \frac{\alpha}{1+e^{-\beta(t-\gamma)}} d t
\end{equation}
By solving this integral, as shown in Eq.~\ref{sigmoidint}, we arrive at the following expression for the maximum number of chemical logical operations for Japanese strain groups in a given time interval:
\begin{equation}\label{Chemical_ATP_Japanese}
\mathcal{N}_\text{chem, Japanese}^{\text{open}}(t)=\frac{0.664 \ \eta_{\text {turnover}} \eta_{\text{work}} \rho_0 \ell \alpha}{\sqrt{2} \beta  \hbar}\left[\ln \left(1+e^{\beta (t-\gamma)}\right)\right],
\end{equation}
up to a constant term defined by $0.664  \ \eta_{\text {turnover}} \eta_{\text{work }} \rho_{0} \ell \, \alpha / (\sqrt{2}\beta \, \hbar)\ \ln(1 + e^{-\beta \gamma})$ from the lower bound of the integral ($t=0$)}.
Since the product \( \beta\gamma \) is greater than or equal to 1.4 for all values of $\beta$ and $\gamma$ (see Table S1 of the Supplementary Material), the associated constant term \( \ln(1 + e^{-\beta\gamma}) \) remains close to zero. This ensures that the bounds in Eq.~\ref{Chemical_ATP_Japanese} are robust to the inclusion or exclusion of this term, with no impact on the order of magnitude and ranking of the bounds across the experimental groups. Similarly, we obtain the expression for the maximum number of chemical logical operations for the Carolina strain, but using the bi-sigmoid area fit given in Eq. \ref{Area_fit_Carolina}: 
\textcolor{black}{
\begin{equation}\label{Chemical_ATP_Carolina}
\mathcal{N}_{\text{chem, Carolina}}^{\text{open}}(t) = \frac{0.664 \ \eta_{\text {turnover}} \eta_{\text{work}} \rho_{0}\, \ell }{\sqrt{2} \, \hbar} \left[
\frac{\alpha_1}{\beta_1} \ln \left(1 + e^{\beta_1 (t - \gamma_1)} \right)
+ \frac{\alpha_2}{\beta_2} \ln \left(1 + e^{\beta_2 (t - \gamma_2)} \right)
\right].
\end{equation}
}
Eqs.~\ref{Chemical_ATP_Japanese} and ~\ref{Chemical_ATP_Carolina} provide estimates of the upper bound on the number of chemical ATP operations that \textit{Physarum} can perform over an experimental time interval from $0$ to $t$. For each \textit{Physarum} macroplasmodial sample, the individual area time-series data were analytically modelled using a sigmoid function, and the resulting fits were substituted into the integrals above to compute the chemical ATP upper bound for each sample. These sample-specific bounds were then averaged to obtain a mean group-aggregate bound. The values of $\mathcal{N}_\text{chem}^{\text{open}}(t)$ for a 24-hour interval are summarized in Table S2 of the Supplementary Material, which reflects how the fitting parameters for area growth vary across strains, ages, and feeding conditions.

\begin{figure}
    \centering
    \includegraphics[width=0.65\textwidth]{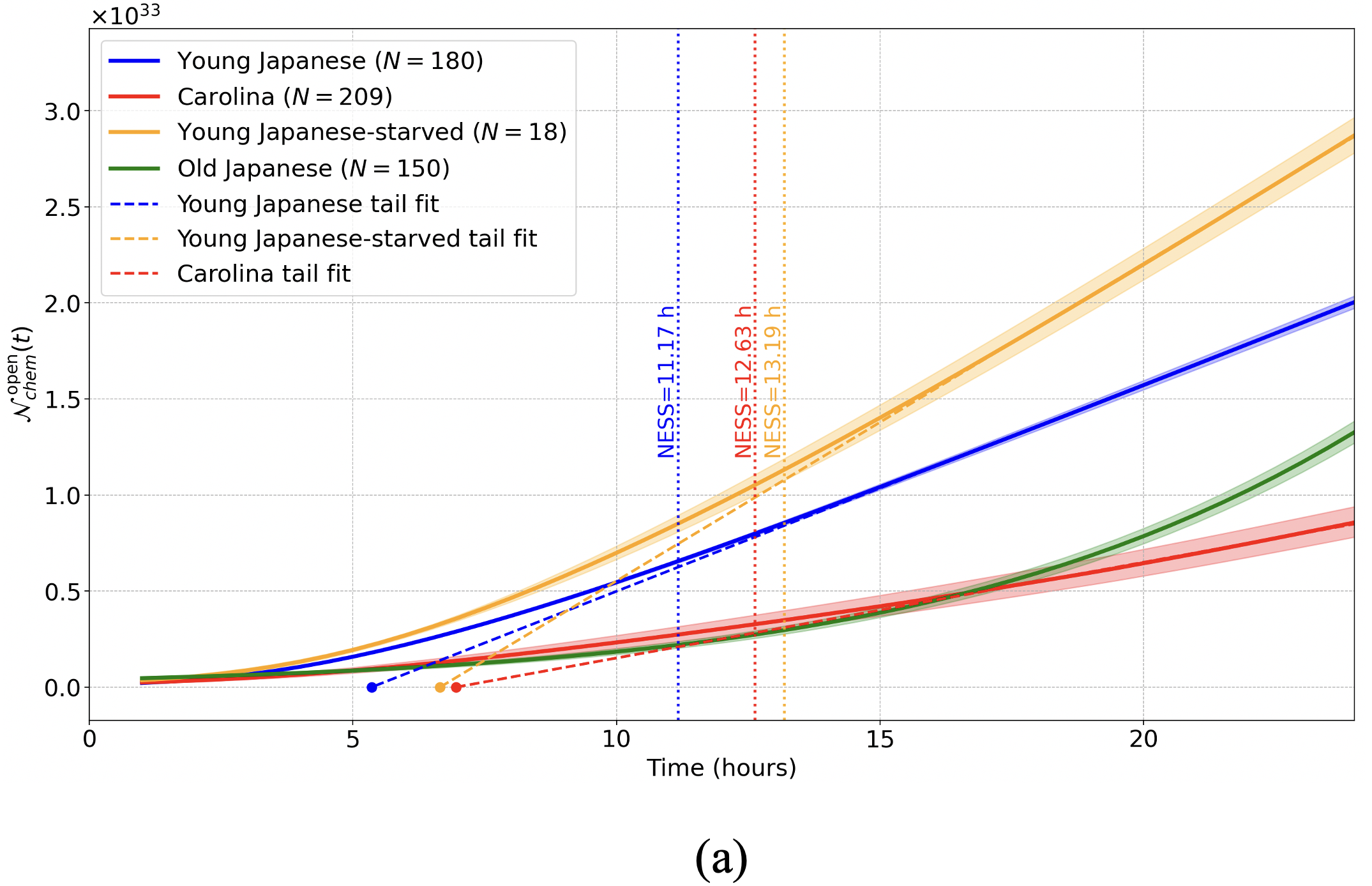}
    \includegraphics[width=0.65\textwidth]{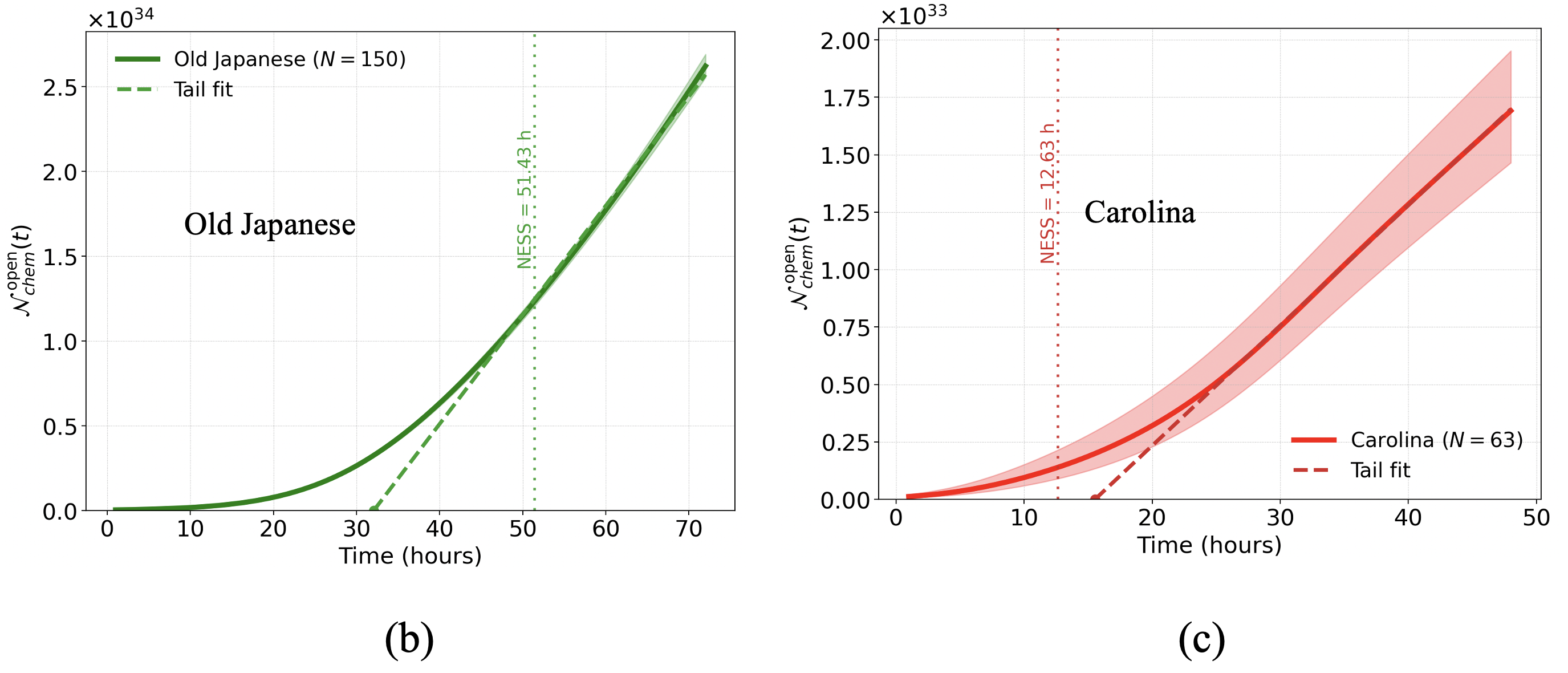}
    \caption{\textbf{The computational capacity upper bounds for maximum conversion of ambient chemical ATP to ADP, obtained over a 24-hour interval, are greater in the rapidly growing young Japanese groups compared to the old Japanese and Carolina groups, and can be inferred from the area growth patterns shown in Fig. \ref{Various_strains}b.} The figure shows the number of operations obtained through chemical ATP conversion for the four groups over a 24-hour interval. The long-term behavior is captured by fitting the tail with a linear function. The time point at which the bound begins to follow this long-time linear trend is marked as the onset of the NESS. Vertical dotted lines indicate the slightly earlier time, where the system is transitioning to the NESS. The $x$–axis intercepts of the late-time fits mark the inflection points of the area sigmoids in Fig.~S9a of the Supplementary Material, yielding values close to the $\gamma$ estimates in Table~S1, except for the Carolina subgroup ($N = 63$) extending to 48~h, where the intercept lies at a weighted average of  the two inflection points of the area bi-sigmoid. The plot is shown extended to 72 hours in panel (b) for the old Japanese group and to 48 hours in panel (c) for the Carolina strain subgroup ($N = 63$). The transition to the NESS in the old Japanese group occurs well beyond 24 hours.}
    \label{Chemical_ATP_Plot}
\end{figure}

\textcolor{black}{In Fig.~\ref{Chemical_ATP_Plot}a, we plot the chemical ATP upper bound, $\mathcal{N}_{\text{chem}}(t)$, over various time intervals up to 24 hours to examine its temporal evolution, with error bars indicating $\pm 1$ standard error. Among the four groups, the young Japanese-starved group exhibited the highest number of logical operations over 24 hours, with a total of $2.87 \times 10^{33}$ operations. In contrast, the Carolina group showed the lowest value, reaching $8.57 \times 10^{32}$ operations. The remaining groups—young Japanese and old Japanese—showed intermediate values of $2 \times 10^{33}$ and $1.33 \times 10^{33}$, respectively.}

Similarly, the young Japanese-starved group shows the highest number of operations due to its rapid exploratory growth in search of nutrients (see Fig. \ref{Chemical_ATP_Plot}). In contrast, the Carolina strain, which exhibits limited area expansion in the morphological plots, performs fewer operations compared to the Japanese strain groups. This highlights how the morphological growth dynamics directly influence the organism's computational capacities using diverse physical degrees of freedom. That the chemical ATP operations bound is vastly larger than the others, by more than ten orders of magnitude, reflects the Margolus-Levitin theorem in Eq.~\ref{Nmax} and thus the potentiality of accessible energy flow from the primary chemical conversion of ATP to ADP, toward other biophysical behaviors in the amoeboid cell: cytosolic flow patterns, kinetic motions of the macroplasmodial body, and photoexcitatory processes. It should be remembered that a portion of this energy is not converted to useful computation, but rather lost as heat to uncontrolled mechanical degrees of freedom. Still, some of this heat may be recovered through nonlinear couplings in the bath focusing energy back into harnessible functional vibrations \cite{azizi2023examining,bajpai2025tracking}, so keeping this limit in mind as an absolute chemical upper bound is imperative.

The chemical ATP bound is governed by the integral of the area, and therefore follows an exponential rise before crossing over into a linear regime. As shown in Fig.~\ref{Chemical_ATP_Plot}, a linear fit to the late-time portion of each group’s curve gave excellent agreement, with $R^2 > 0.95$. The vertical dotted lines in the figure mark the NESS transition obtained from the area dynamics. Since the time to transition to NESS is defined as the point just before the onset of steady-state behavior, this transition occurs slightly earlier than the stage where the chemical ATP operations align with the long-time linear growth. This consistency reinforces both the reliability of our NESS estimate and the validity of the late-time scaling. The $x$-axis intercepts of the fitted linear tails for the different groups generally correspond to the inflection points of their respective area sigmoids in Fig. S9a of the Supplementary Material, yielding values that closely match the $\gamma$ values reported in Table~1. This applies to all cases shown in Fig.~\ref{Chemical_ATP_Plot}a (shown up to 24~h) except for the linear tail of the Carolina strain ($t \gg \gamma_{2}$) observed in the subgroup of Carolina samples extending up to 48~h (Fig.~\ref{Chemical_ATP_Plot}c). In this case, the intercept does not lie near $\gamma_{2}$ but instead shifts to $t_{x_2} = (\alpha_{1} \gamma_{1} + \alpha_{2} \gamma_{2})/(\alpha_{1} + \alpha_{2})$, due to contributions from both growth phases. The bound values for the old Japanese group, shown for the 72~h experimental duration, are presented in Fig.~\ref{Chemical_ATP_Plot}b, while those for the extended Carolina subgroup ($N = 63$) are shown up to 48~h in Fig.~\ref{Chemical_ATP_Plot}c.

\subsubsection{Kinetic energy bound} \label{sect:KEbound}
\textcolor{black}{
The kinetic energy upper bound quantifies the computational capacity arising from the motion of \textit{Physarum}'s active boundary. This bound, denoted by \( \mathcal{N}_{\text{KE}} \), is given by the integral expression in Eq.~\ref{Eq:N_KEint_Open}.
We use the sigmoid area fit for the Japanese strain from Eq. \ref{Area_fit_Japanese}, and obtain the speed of the advancing front from the rate of area expansion per unit advancing perimeter, where $A(t)$ and $P(t)$ are given by their analytical fits in Eqs.~\ref{Area_fit_Japanese} and \ref{Perimeter_fit_Japanese}, respectively, as follows:
\begin{equation}
v(t)=\frac{\dot{A}(t)}{P(t)}
=\frac{\alpha \beta}{\delta}
\frac{\left(1+e^{-\eta(t-\theta)}\right)}
{\left(1+e^{-\beta(t-\gamma)}\right)^2}
\end{equation}
Substituting the area and velocity functions into Eq.~\ref{Eq:N_KEint_Open}, we get
\begin{equation}
\mathcal{N}_{\mathrm{KE,Japanese}}^{\text{open}}(t)=\frac{\rho_m f_{\mathrm{avg}}^{\mathrm{J}}\ell\,\alpha^3\beta^2}{2\sqrt{2}\hbar\,\delta^2}
\int_0^t dt'\, \frac{\left(1+e^{-\eta(t'-\theta)}\right)^2 e^{-2\beta(t'-\gamma)}}{\left(1+e^{-\beta(t'-\gamma)}\right)^5}.
\end{equation}
Changing variables with the substitution $z^{\prime}=e^{-\beta\left(t^{\prime}-\gamma\right)}$, we arrive at
\begin{equation}\label{z-sub}
\mathcal{N}_{\mathrm{KE, Japanese}}^{\text{open}}(t)
=-\frac{\rho_m f_{\mathrm{avg}}^{\mathrm{J}}\ell \alpha^3 \beta}
{2 \sqrt{2}\hbar \delta^2}\int_{z_0}^{z} dz'\,
\frac{\left(1+b z'^{\,a}\right)^2 z'} {\left(1+z'\right)^5},
\end{equation}
where we have set $\frac{\eta}{\beta}=a$ and $e^{\eta(\theta-\gamma)}=b$.
Swapping the limits in Eq.~\ref{z-sub}, we obtain
\begin{equation}\label{z-sub1}
\mathcal{N}_{\mathrm{KE, Japanese}}^{\text{open}}(t)
=\frac{\rho_m f_{\mathrm{avg}}^{\mathrm{J}}\ell \alpha^3 \beta}
{2 \sqrt{2}\hbar \delta^2}\int_{z}^{z_0} dz'\,
\frac{\left(1+b z'^{\,a}\right)^2 z'} {\left(1+z'\right)^5},
\end{equation}
The integral in Eq.~\ref{z-sub1} admits a closed-form analytical representation in terms of Gaussian hypergeometric functions. Defining the antiderivative \(\mathcal{F}(z')\) as
\begin{equation}
\mathcal{F}(z')=-\frac{4z'+1}{12(1+z')^4}+\frac{2b\,z'^{\,a+2}}{a+2}
\,{}_2F_1(5,a+2;a+3;-z')+\frac{b^2 z'^{\,2a+2}}{2a+2}\,{}_2F_1(5,2a+2;2a+3;-z'),
\end{equation}
the kinetic-energy bound can be written compactly as
\begin{equation}\label{eq:NKE_Japanese}
\mathcal{N}_{\mathrm{KE, Japanese}}^{\text{open}}(t)=\frac{\rho_m f_{\mathrm{avg}}^{\mathrm{J}}\ell \alpha^3 \beta}
{2\sqrt{2}\hbar \delta^2}\left[\mathcal{F}(z_0)-\mathcal{F}(z)\right].
\end{equation}
which gives the upper bound on the number of kinetic logical operations performed by Japanese strain groups of \textit{Physarum}, due to the motional energy of their advancing perimeters.
}
\textcolor{black}{
Similarly, for Carolina, we can express the kinetic energy bound as follows, by substituting the bi-sigmoid area and perimeter fits into Eq.~\ref{N_KEint}:
\begin{equation}\label{eq:NKE_Carolina}
\mathcal{N}_{\mathrm{KE, Carolina}}^{\text{open}}(t)=
\frac{\rho_m f_{\mathrm{avg}}^{\mathrm{C}}\ell}{2\sqrt{2}\hbar}\int_{0}^{t}
dt'\,\frac{\left[\dfrac{\alpha_1}{1+e^{-\beta_1(t'-\gamma_1)}}
+\dfrac{\alpha_2}{1+e^{-\beta_2(t'-\gamma_2)}}\right]
\left[\dfrac{\alpha_1\beta_1 e^{-\beta_1(t'-\gamma_1)}}
{\left(1+e^{-\beta_1(t'-\gamma_1)}\right)^2}+
\dfrac{\alpha_2\beta_2 e^{-\beta_2(t'-\gamma_2)}}
{\left(1+e^{-\beta_2(t'-\gamma_2)}\right)^2}
\right]^2}{\left[\dfrac{\delta_1}{1+e^{-\eta_1(t'-\theta_1)}}+
\dfrac{\delta_2}{1+e^{-\eta_2(t'-\theta_2)}}\right]^2
}.
\end{equation}
}
\textcolor{black}{Unlike the analytical kinetic energy bound for Japanese strains obtained in Eq.~\ref{eq:NKE_Japanese}, the kinetic energy bound for the Carolina strain in Eq.~\ref{eq:NKE_Carolina} does not admit a similarly compact analytical form, and we therefore numerically evaluated the expression in Eq.~\ref{eq:NKE_Carolina}. Using Eqs.~\ref{eq:NKE_Japanese} and ~\ref{eq:NKE_Carolina}, we evaluated the kinetic energy bound for each \textit{Physarum} macroplasmodial sample and then averaged the results within each of the four groups to obtain estimates of the group-aggregate means.}
We evaluated this computational capacity rate over a 24-hour period to estimate the maximum number of kinetic logical operations performed by each group due to the motional energy of their advancing perimeters. The average values of \( \mathcal{N}_{\text{KE}}\) for the four groups are summarized in Table S2 of the Supplementary Material.

\begin{figure}
    \centering
    \includegraphics[width=0.5\textwidth]{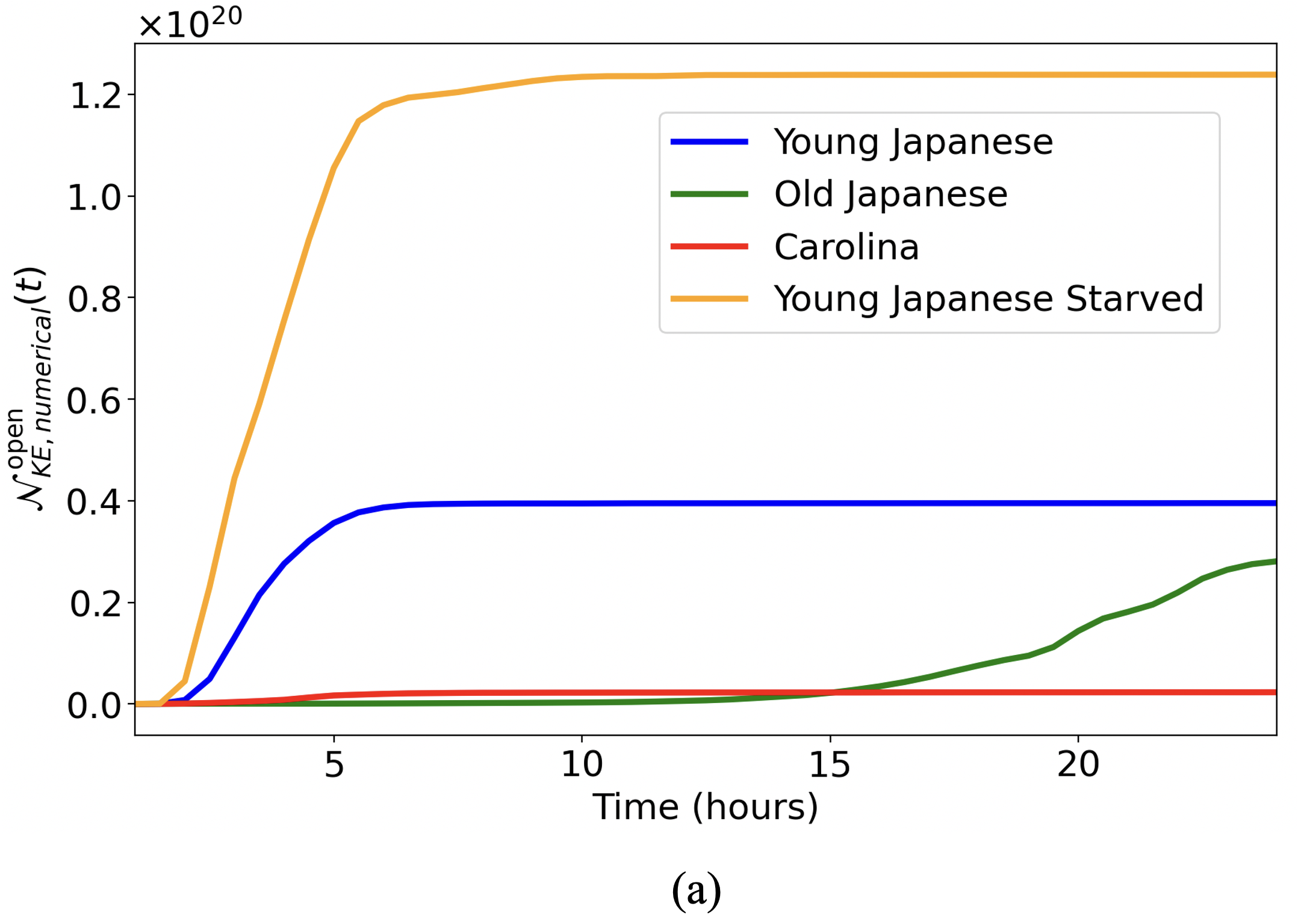}
    \includegraphics[width=0.5\textwidth]{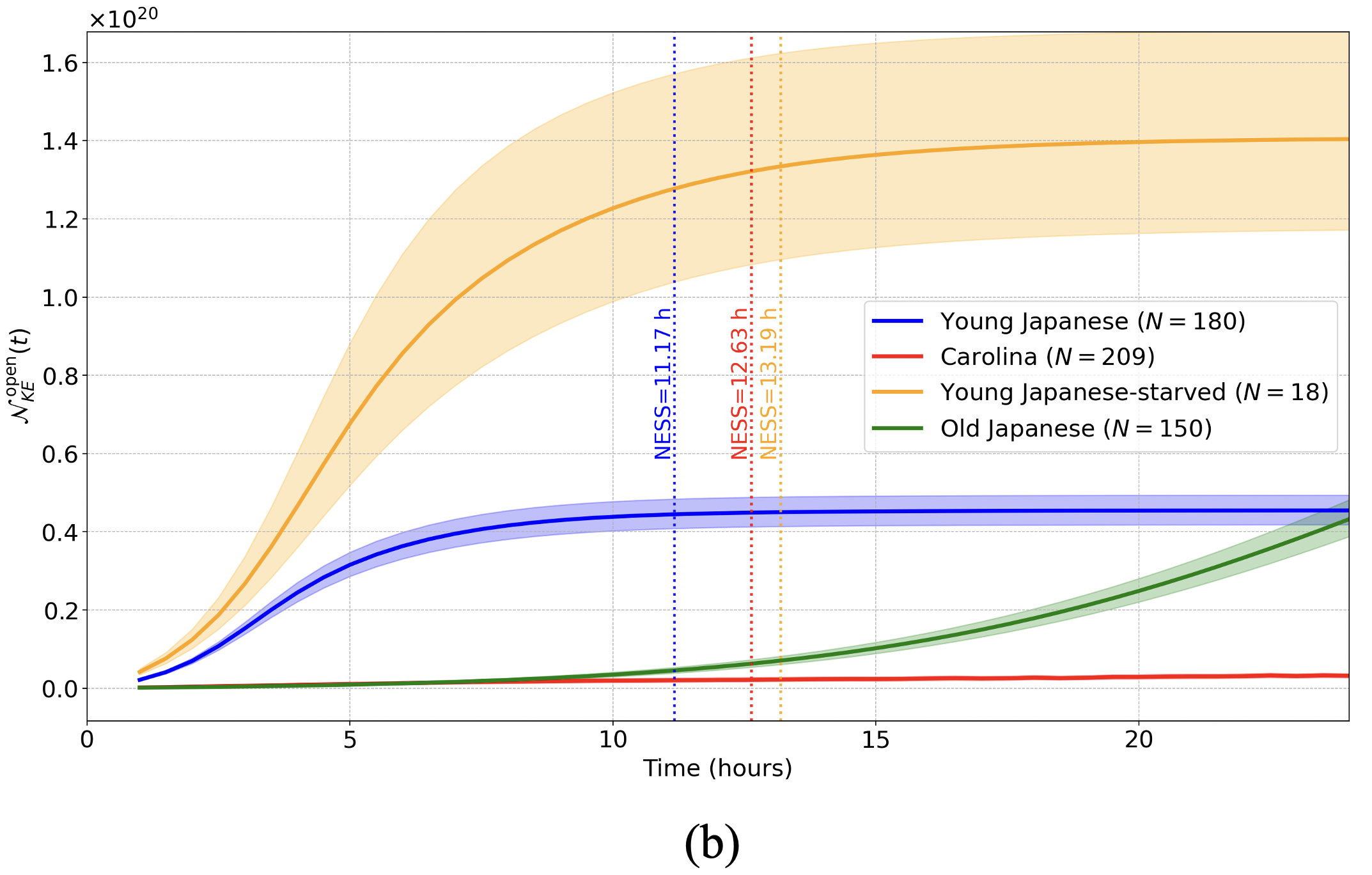}
    \includegraphics[width=0.5\textwidth]{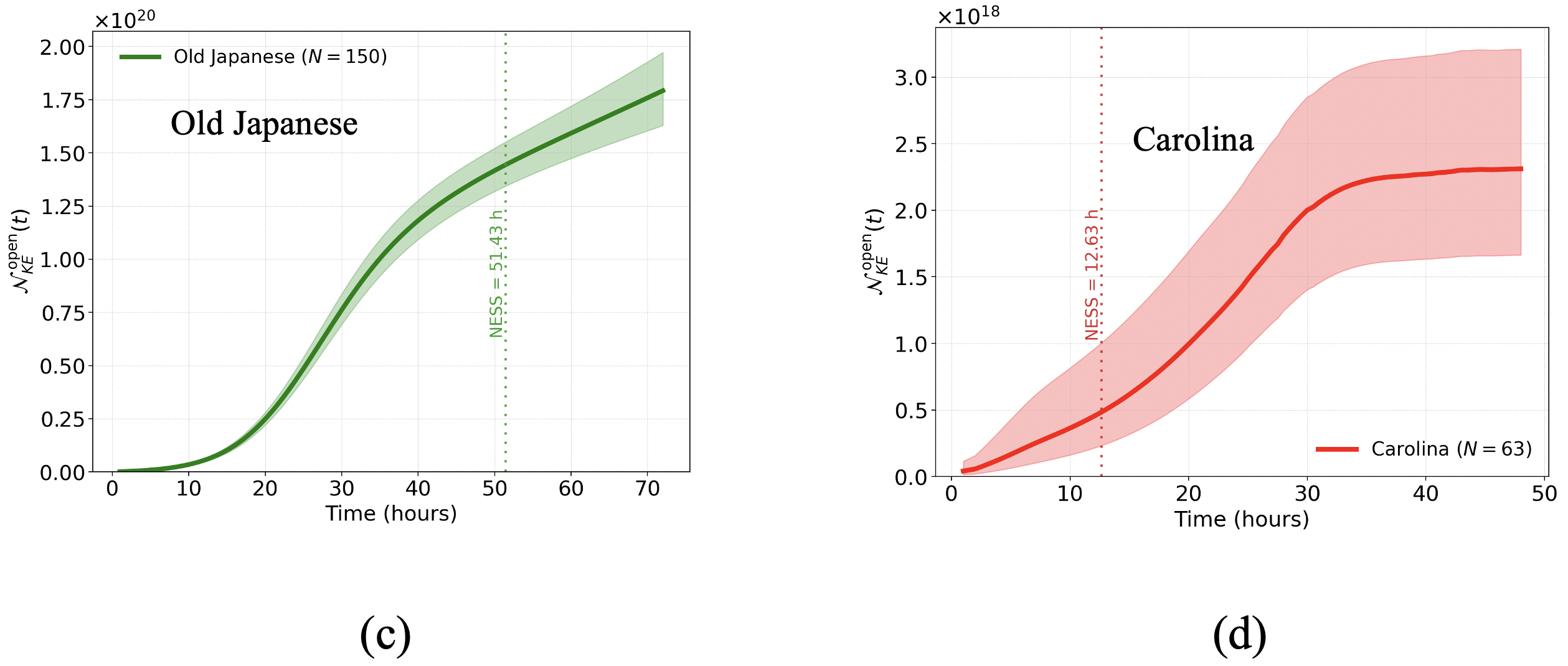}
    \caption{\textcolor{black}{\textbf{The kinetic energy bounds, while reflecting the same rapid growth trends in the more active, young Japanese groups, exhibit distinct motional computational capacities due to advancing \textit{Physarum} fronts.}
    The figure shows the maximum number of logical operations performed from the motion of the outer annulus of \textit{Physarum}'s body over a 24-hour interval, computed from (a) numerically integrating the expression in Eq.~\ref{eq:NKE_num} assuming a time-dependent ratio of the area of the advancing annular region to the total area, $f(t)$, and (b) the analytical expressions in Eqs.~\ref{eq:NKE_Japanese} and \ref{eq:NKE_Carolina} assuming a time-averaged ratio of the area of the advancing annular region to the total area $f_{\mathrm{avg}}$. The trends obtained from the analytical expressions are qualitatively similar to those obtained from the numerical integration. Vertical dashed lines mark the time points at which \textit{Physarum} transitions to the NESS. The plot is shown extended to 72 hours in panel (c) for the old Japanese group and to 48 hours in panel (d) for the Carolina strain subgroup ($N=63$)}}
    \label{KE_bound}
\end{figure}

\textcolor{black}{
The time-dependent kinetic energy bound, $\mathcal{N}_{\text{KE}}(t)$, is plotted for the four groups in Figure~\ref{KE_bound}b up to 24 hours, with shaded regions indicating $\pm1$ standard error. Of course, this bound is higher for the more active, younger Japanese groups. The young Japanese-starved group shows the highest kinetic energy bound, reaching approximately $1.40 \times 10^{20}$ operations, followed by the young Japanese and old Japanese groups at $4.54 \times 10^{19}$ and $4.32 \times 10^{19}$ operations, respectively. In contrast, the Carolina group exhibits a significantly lower kinetic energy bound of approximately $3.18 \times 10^{18}$ operations. This distinction highlights the substantial differences in motility and morphological activity across strains, and how computational capacities are distributed into kinetic or motional degrees of freedom in each organism.}

\textcolor{black}{For comparison, we also found the time-dependent kinetic energy bound by numerically computing the following integral:
\begin{equation}
\mathcal{N}_{\text{KE,num}}^{\text{open}}(t) =\frac{\rho_m \ \ell}{2 \sqrt{2} \hbar}\int_{0}^{t} f(t')\,A(t')\,
\frac{\dot{A}(t')^{2}}{P(t')^{2}}\,dt',\qquad 
\text{where} \quad 
f(t') = \frac{\Delta A(t')}{A(t')}.
\label{eq:NKE_num}
\end{equation}}

Here, $\Delta A(t_{i+1}) = A(t_{i+1}) - A(t_i)$ represents the change in area between successive time points. The integration was performed using the trapezoidal method over the interval $0.5 \le t \le 24$~hours, with a uniform time step of $\Delta t =t_{i+1}-t_i= 0.5$~hours. \textcolor{black}{These numerically computed, time-dependent kinetic energy bounds for all experimental groups, evaluated up to $t = 24$~hours, are shown in Fig.~\ref{KE_bound}a}. While the overall trends resemble those from the analytical expressions in Eqs.~\ref{eq:NKE_Japanese} and \ref{eq:NKE_Carolina}, some differences emerge. 

\textcolor{black}{At 24 hours, the young Japanese-starved group exhibits the highest kinetic energy bound in both the numerically and analytically obtained bounds, followed by the young Japanese group. The kinetic-energy bound for the young Japanese group is approximately one-third of that of the young Japanese-starved group, consistent across both the numerical and analytical approaches. The Carolina group exhibits the lowest kinetic energy bound in both cases. One difference between the two approaches, however, is observed in the relative separation between the young Japanese and old Japanese kinetic energy bounds at 24 hours. In the numerical bound, the old Japanese group exhibits a lower kinetic energy bound than the young Japanese group. In contrast, in the analytically obtained bound, the corresponding values for these two groups remain very close to each other. This discrepancy mainly arises because $f(t)$ is treated as time-dependent in the numerical case, but replaced with its time-averaged value $f_{\text{avg}}$ in the analytical case.}

\textcolor{black}{Per Eqs. \ref{Eq:speed} and \ref{Eq:N_KEint_Open}, it should be noted that the kinetic energy bound depends on the integral of the product of the area and the squared ratio ($v^2$) of the rate of area expansion to the perimeter.} Unlike the other three bounds, whose integrals reduce to a linear scaling at long times, this form produces a more intricate dependence that does not necessarily exhibit a simple linear regime at long times (see Fig.~\ref{KE_bound}). Consequently, late-time linear fits are not shown for the kinetic energy bound. \textcolor{black}{Instead, because this bound involves the rate of area expansion, it approaches saturation once the area stabilizes in the NESS. This saturation is clearly observed in the young Japanese and Japanese-starved groups, where the area stabilizes within the experimental window (see Fig.~\ref{Various_strains}b). A subgroup of Carolina samples extending beyond 24~hours ($N=63$) also exhibits this saturation after approximately 36~hours (see Fig.~\ref{KE_bound}d), even though a majority of Carolina samples ($N=146$) showcases a similar saturation before 24 hours (see Fig. \ref{Various_strains}b). In contrast, for the old Japanese group, the area changes appreciably even near the end of the experimental window, preventing the onset of as pronounced a saturation effect (see Fig.~\ref{KE_bound}c).}

\subsubsection{\textcolor{black}{Predicted and prospective quantum optical bound from simulations of actin fiber networks}}

The quantum optical bound is derived by considering the number of actin fibers within a given area of \textit{Physarum}'s body. On average, there are two cytoplasmic actin fibrils per 1000~\textmu m$^2$ of \textit{Physarum} area, as determined by rhodamine-phalloidin staining and fluorescence microscopy in axenic cultures of \textit{Physarum} \cite{ohl1991studies}. In our own histological staining of actomyosin fibrils (see Fig.~\ref{fig:Schematic}), we observed 2–3 actin fibrils per 1,000 $\mu$m$^2$ (in Fig.~\ref{fig:Schematic}d), confirming this earlier estimate. Therefore, the number of fibers in an area $A(t)$ for an arbitrary \textit{Physarum} sample can be roughly estimated as:
\[
N_{\text{fibers}} = n_{\text{actin}} A(t) = \frac{2 \text{ fibers}}{1000\, \mu\text{m}^2} A(t).
\]
The characteristic superradiant lifetime for a 19-filament actin bundle ($\tau$) is approximately $10^{-11}$ s, or 10 ps \cite{patwa2024quantum}. Hence, the quantum optical operations per second for a \textit{Physarum} body of area $A(t)$ at time $t$ can be written from Eq. \ref{QO_bound_general}, which can be integrated over the interval from $0$ to $t$ per Eq.~\ref{maxQOops} to obtain the maximum number of superradiant operations achievable by the macroplasmodial body. The prefactor before the integral in Eq.~\ref{maxQOops} is given by $\frac{n_{\text{actin}}}{\tau} = 2 \times 10^{20}$ m$^{-2}$ s$^{-1}$. 

\begin{figure}
    \centering
    \includegraphics[width=0.65\textwidth]{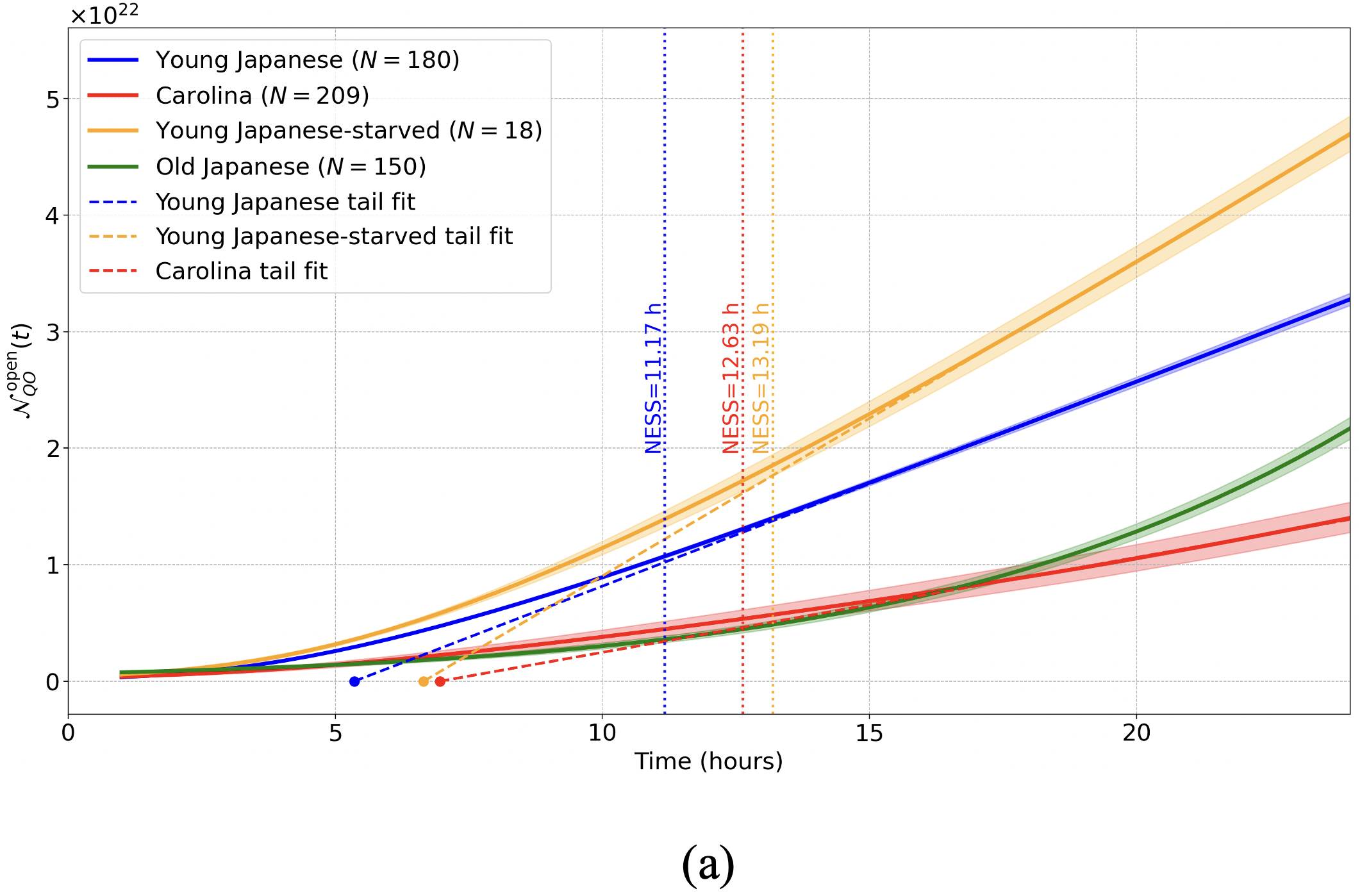}
    \includegraphics[width=0.65\textwidth]{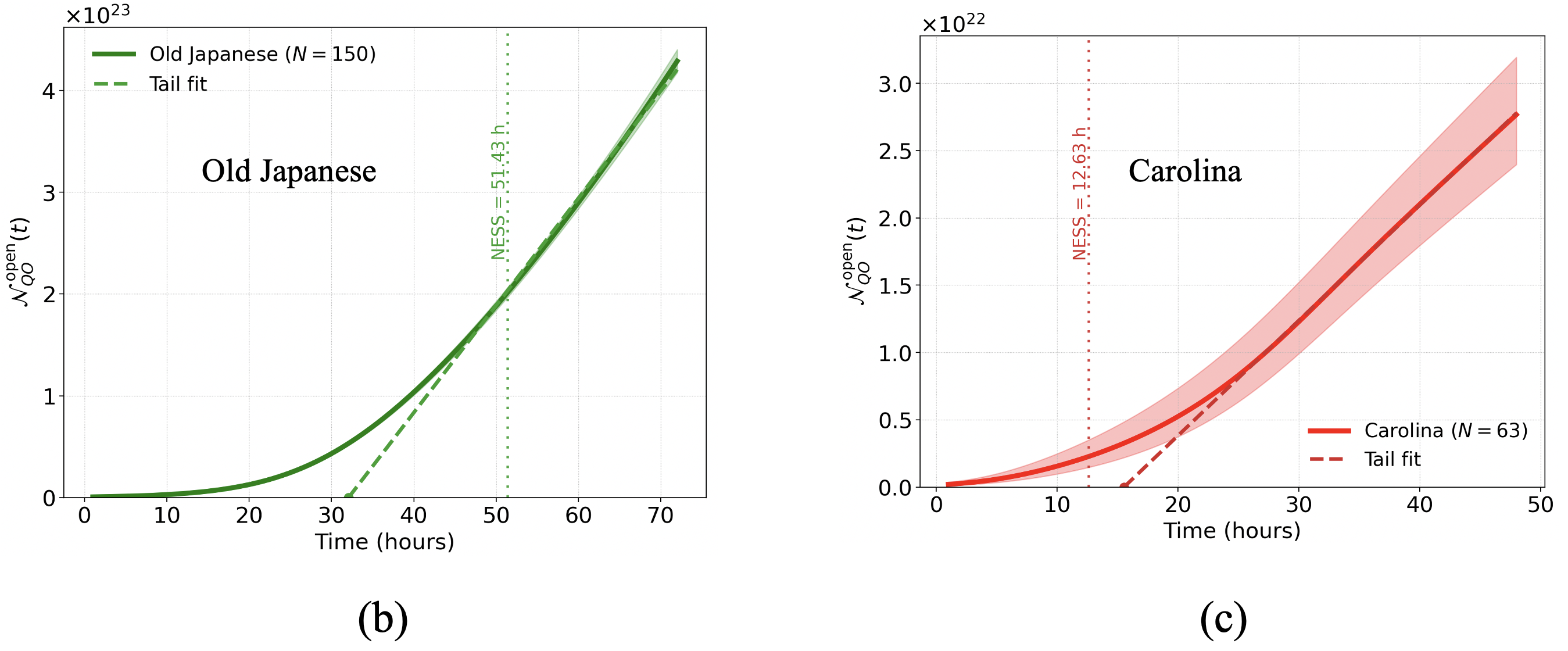}
    \caption{\textbf{The quantum optical bound, similar to the chemical ATP bound, shows higher values for young Japanese groups, reflecting the growth of their morphological areas and the commensurate number of superradiant protein fibers theoretically predicted in each sample.} The figure shows (a) the maximum number of operations predicted from superradiant states in the actomyosin fiber network in \textit{Physarum} bodies over the time course of our experiments. To capture the long-term dynamics, the late-time portion of the curve is fitted with a linear function. The point where the bound begins to align with this linear trend is taken as the onset of the NESS, while vertical dotted lines denote the time of transition to the NESS. The $x$–axis intercepts of the late-time fits mark the inflection points of the area sigmoids in Fig.~S9a of the Supplementary Material, yielding values close to the $\gamma$ estimates in Table~S1, except for the Carolina subgroup ($N = 63$) extending to 48~h, where the intercept lies at a weighted average of  the two inflection points of the area bi-sigmoid. The plot is shown extended to 72 hours in panel (b) for the old Japanese group and to 48 hours in panel (c) for the Carolina strain subgroup ($N = 63$). The transition to the NESS in the old Japanese group occurs well beyond 24 hours.}
    \label{QO_Bound}
\end{figure}

We substituted the sigmoid area fit for the Japanese strain groups (Eq.~\ref{Area_fit_Japanese}) and the bi-sigmoid area fit for the Carolina strain group (Eq.~\ref{Area_fit_Carolina}) to obtain closed-form expressions for the quantum optical bounds:
\textcolor{black}{
\begin{equation}\label{QO_Bound_Japanese}
\mathcal{N}_{\mathrm{QO, Japanese}}^{\text{open}}(t)=\frac{\pi n_{\text{actin}}}{\sqrt{2}\tau}\frac{\alpha}{\beta}\left[\ln \left(1+e^{\beta (t-\gamma)}\right)\right]
\end{equation}
\begin{equation}\label{QO_Bound_Carolina}
\mathcal{N}_{\text{QO, Carolina}}^{\text{open}}(t) = \frac{\pi n_{\text{actin}}}{\sqrt{2}\tau} \left[
\frac{\alpha_1}{\beta_1} \ln \left(1 + e^{\beta_1 (t - \gamma_1)} \right)
+ \frac{\alpha_2}{\beta_2} \ln \left(1 + e^{\beta_2 (t - \gamma_2)} \right)
\right],
\end{equation}}
up to constant terms proportional to $\ln(1 + e^{-\beta \gamma})$, $\ln(1 + e^{-\beta_{1} \gamma_{1}})$, and $\ln(1 + e^{-\beta_{2} \gamma_{2}})$, respectively, derived from the lower integral limits ($t=0$). These constant terms contribute negligibly to the magnitude of the bounds for the fitted values of the parameters (see Table S1 of the Supplementary Material) and do not affect the order of magnitude or the hierarchy of bound values across experimental groups.

Using Eqs. \ref{QO_Bound_Japanese} and \ref{QO_Bound_Carolina}, we computed the quantum optical upper bounds numerically for each \textit{Physarum} sample across all experimental groups. For each group, the values were averaged to produce a mean group-aggregate quantum optical bound. These results are reported at the 24-hour mark in Table S2 of the Supplementary Material. These quantum optical bounds, together with $\pm 1$ standard error, are shown as a function of time in Fig.~\ref{QO_Bound}a for the four experimental groups. \textcolor{black}{The young Japanese-starved group exhibited the highest quantum optical bound, approximately $4.7 \times 10^{22}$ operations in 24 hours. In contrast, the Carolina group showed the lowest bound of approximately $1.4 \times 10^{22}$ operations. The young Japanese and old Japanese groups showed intermediate values at about $3.28 \times 10^{22}$ and $2.17 \times 10^{22}$ operations, respectively.}

Similar to the chemical ATP bound, the quantum optical bound is also governed by the integral of the area, and thus exhibits an exponential increase at early times followed by a transition to linear growth at later times. A linear fit to the late-time portion of each group’s curve yielded $R^2 > 0.95$. The vertical dotted lines (Fig.~\ref{QO_Bound}) indicate the NESS transitions as determined from the area dynamics. Since the time to transition to NESS is defined as the point just before the onset of steady behavior, this transition is observed slightly earlier than the stage where the quantum optical operations align with the long-time linear trend. This agreement supports both the robustness of our NESS estimate and the validity of the late-time scaling. The $x$-axis intercepts of the fitted linear tails for the different groups generally correspond to the inflection points of their respective area sigmoids in Fig. S9a of the Supplementary Material, yielding values that closely match the $\gamma$ values reported in Table~1. This applies to all cases shown in Fig.~\ref{QO_Bound}a (shown up to 24~h) except for the linear tail of the Carolina strain ($t \gg \gamma_{2}$) observed in the subgroup of Carolina samples extending up to 48~h (Fig.~\ref{QO_Bound}c). In this case, the intercept does not lie near $\gamma_{2}$ but instead shifts to $t_{x_2} = (\alpha_{1} \gamma_{1} + \alpha_{2} \gamma_{2})/(\alpha_{1} + \alpha_{2})$, due to contributions from both growth phases. The bound values for the old Japanese group, shown for the 72~h experimental duration, are presented in Fig..~\ref{QO_Bound}b, while those for the extended Carolina subgroup ($N = 63$) are shown up to 48~h in Fig.~\ref{QO_Bound}c.

\subsection{Allometric scaling of the chemical ATP bound}
\begin{figure}[h!]
    \centering
    \includegraphics[width=0.9\linewidth]{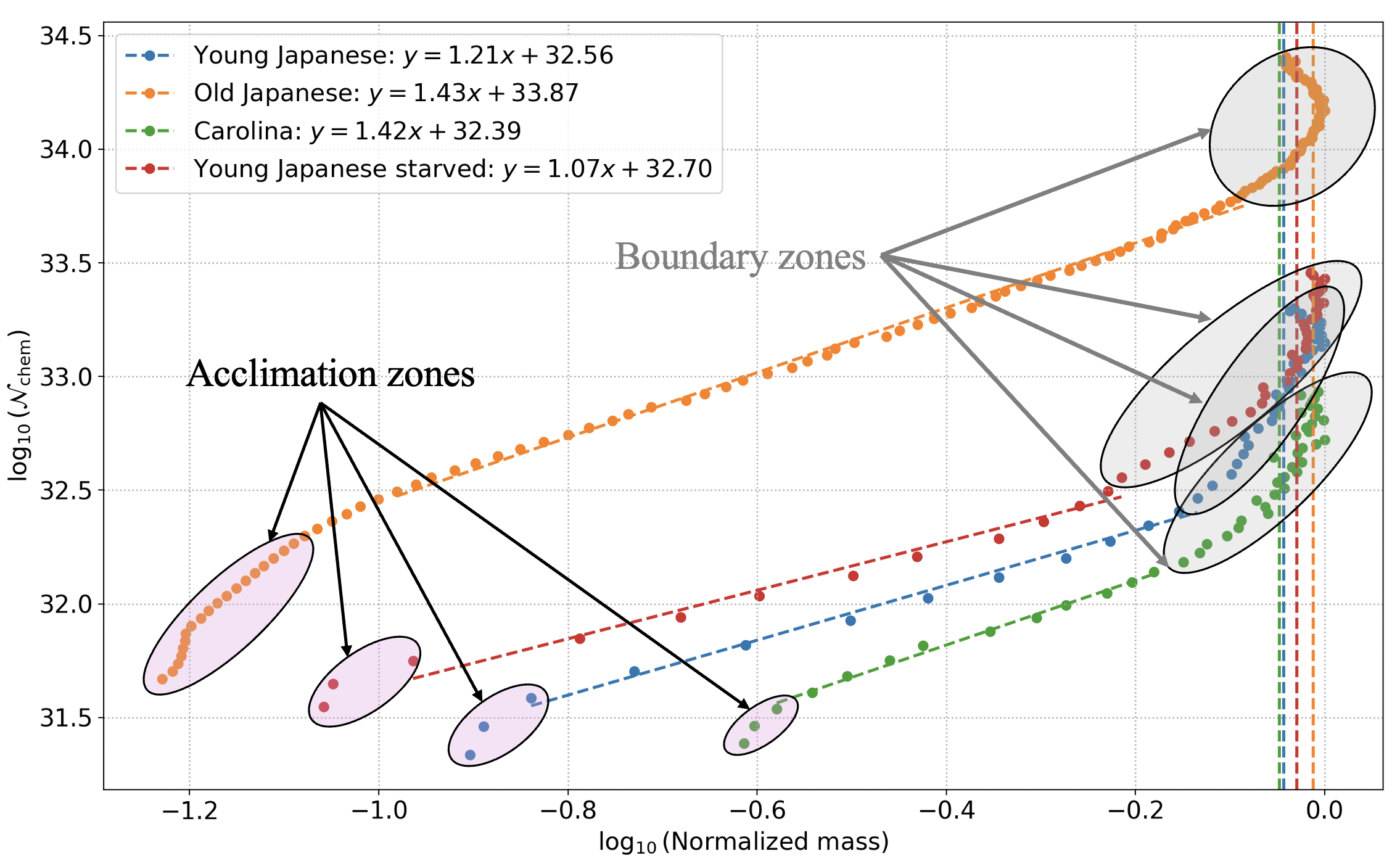}
   \caption{\textbf{The chemical ATP bound scales with the normalized organismal mass in the intermediate regime by an allometric exponent that varies with strain type, age, and the fractal network of the \textit{Physarum} samples.} The figure shows the log--log plot of the chemical ATP bound versus the normalized organismal mass $M/M_{0}$, where $M_{0}$ denotes the maximum mass reached during the experimental window for each group. The plots are linear, indicating self-similar or fractal behavior, except at the extremes: the early-time acclimation zone reflects the adaptation of \textit{Physarum} to the agar-filled petri dish, while the boundary zone corresponds to the regime where edge effects dominate. The vertical dashed lines indicate each group's respective transition to the NESS, which occurs within the boundary zones.
}
    \label{fig:Chem_ATP_Scaling_R}
\end{figure}

Allometric relations between metabolic rate and body mass are well established \cite{kleiber1947body, west2002allometric}, motivating us to investigate how the chemical ATP bound---an upper limit on the number of logical operations performed by \textit{Physarum} through the conversion of ambient ATP to ADP during a given interval---scales with organismal mass across the experimental groups. To quantify this relationship, we calculated the chemical ATP bound from the analytical expressions in Eqs.~\ref{Chemical_ATP_Japanese}-~\ref{Chemical_ATP_Carolina} over a 24-hour interval. The average mass $M(t)$ for each group at time point $t$ was then obtained from the group-aggregate mean area $A(t)$, using $M(t) = \rho_m A(t) l$, where $\rho_m = 1100~\text{kg}/\text{m}^3$ is the average mass density and $\ell=100  \ \mu$m is the average thickness of the \textit{Physarum} body. To find the normalized mass for each group, the mass $M(t)$ was divided by the maximum mass (acquired in the NESS) specific to the group, $M_0$. The resulting relationship between $\log_{10}(\mathcal{N}_{\text{chem}})$ and $\log_{10}(M/M_0)$ is shown in Fig.~\ref{fig:Chem_ATP_Scaling_R}. Since this is a log--log plot, the slope corresponds to the allometric exponent, indicating how the chemical operations scale with the mass of the \textit{Physarum} body. For completeness, the plot of the chemical ATP bound $\mathcal{N}_{\text{chem}}$ as a function of organismal mass $M$ is shown in Fig. S16 of the Supplementary Material.

For all groups, we observe a dominant linear region, with deviations appearing only at the extremes: at the beginning (acclimation zone, shaded in pink) and towards the end of the experiment (boundary zone, shaded in gray). The acclimation zone corresponds to the initial growth phase, where the organism adapts to the arena (the agar-filled petri dish) and exhibits an early, rapid rise in the number of operations. In contrast, the intermediate linear regime shows a steady scaling of the number of operations with increasing mass. Finally, in the boundary zone, the organism detects the dish boundary and boundary effects begin to dominate, culminating in a steep rise in the values of the chemical ATP bound. We fitted the intermediate linear regime with straight lines, obtaining slope values (allometric exponents) for the different groups.

\textcolor{black}{We observe that the faster-growing young Japanese and young Japanese-starved groups (age since revival from sclerotia $\leq 29$ days) exhibit smaller allometric exponents (1.21 and 1.07, respectively). By contrast, the more slowly growing Carolina strain and old Japanese group (age since revival from sclerotia $\geq 49$ days) show larger allometric exponents, approximately 1.4. This hierarchy is nontrivial, as the slower-growing groups exhibit larger allometric exponents despite their reduced growth rates. \textcolor{black}{The allometric exponent can be written as
\begin{equation}
\sigma
=\frac{d\log N_{\mathrm{chem}}}{d\log\widetilde{M}}
=\frac{\frac{1}{N_{\mathrm{chem}}}\frac{dN_{\mathrm{chem}}}{dt}}
{\frac{1}{\widetilde{M}}\frac{d\widetilde{M}}{dt}}=\frac{\widetilde{M}}{N_\text{chem}}\frac{dN_\text{chem}}{d\widetilde{M}},
\end{equation}
where $\widetilde{M}=M/M_0$. This shows that $\sigma$ is proportional to the ratio between the time rates of change in the number of chemical ATP operations and in normalized mass. Consequently, larger values of $\sigma$ correspond to regimes in which the rate of increase in the number of chemical ATP operations with respect to the normalized mass grows, as observed for the Carolina and old Japanese strains, whose relative rate of mass increase (growth rate) is comparatively smaller. By contrast, the young Japanese and young Japanese-starved strains exhibit faster growth and therefore larger relative rates of mass increase, causing the corresponding allometric exponents to remain comparatively smaller.}
This ordering also appears correlated with the temporal behavior of morphological complexity measures, including circularity and fractal dimension (see Fig. \ref{fig:circ_frac}). In the faster-growing groups with lower allometric exponents, the peak changes in circularity and fractal dimension occur approximately simultaneously (Figs. \ref{fig:circ_frac}a,b). In contrast, for the slower-growing Carolina and old Japanese groups, which exhibit larger allometric exponents, the largest changes in circularity and fractal dimension occur at distinct times and do not coincide (Figs. \ref{fig:circ_frac}c,d). These observations suggest a potential relationship between allometric scaling behavior, growth dynamics, and fractality in \textit{Physarum}. Although metabolic scaling relationships are often associated with fractal organization in biological systems \cite{ronellenfitsch2019phenotypes,katifori2010damage, bhattacharyya2022memory, bhattacharyya2023memory, fessel2012physarum}, further investigation will be required to establish a quantitative connection between the observed allometric exponents and the fractal dimension in \textit{Physarum}}.

The allometric scalings obtained here are higher than the celebrated $3/4$ exponent from Kleiber’s law, which has been verified to hold for animals spanning over 27 orders of magnitude in size. However, there are important distinctions in our case. First, \textit{Physarum} is a unicellular organism, and previous work \cite{niklas2006phyletic} has shown that for various unicellular organisms, the allometric exponent can range between $3/4$ and $1$, consistent with our observations. Second, \textit{Physarum} macroplasmodia undergo nuclear division rather than cell division, as in animals. These features contribute to a higher allometric exponent in \textit{Physarum}, with the precise value depending on strain conditions and the fractal structure of the organism.

\subsection{Relationships between circularity, fractal dimension, and computational capacity limits}

\begin{figure}[h]
    \centering
    \includegraphics[width=1\linewidth]{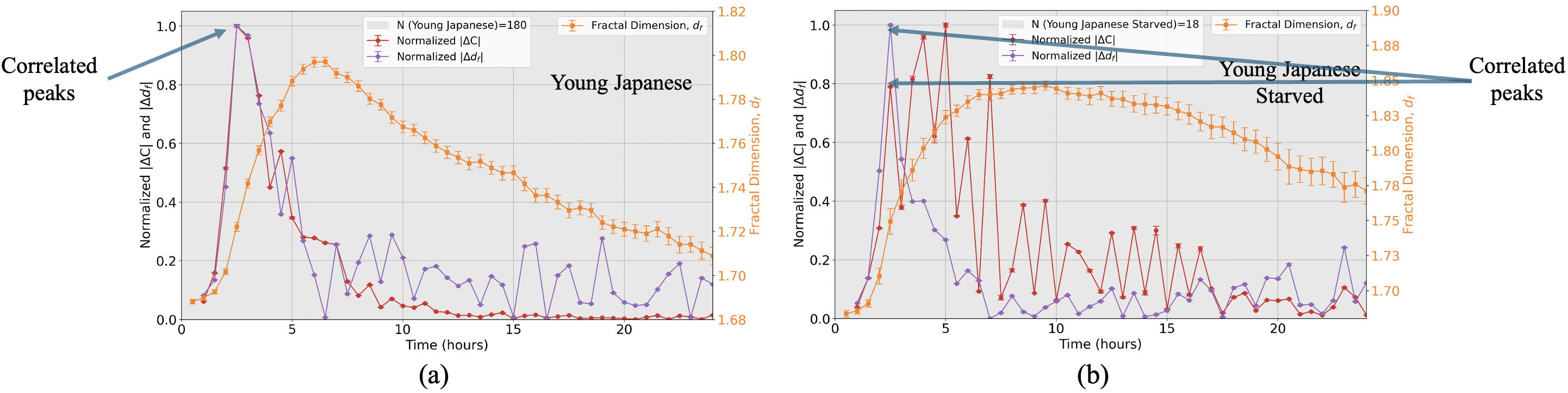}
    \includegraphics[width=1\linewidth]{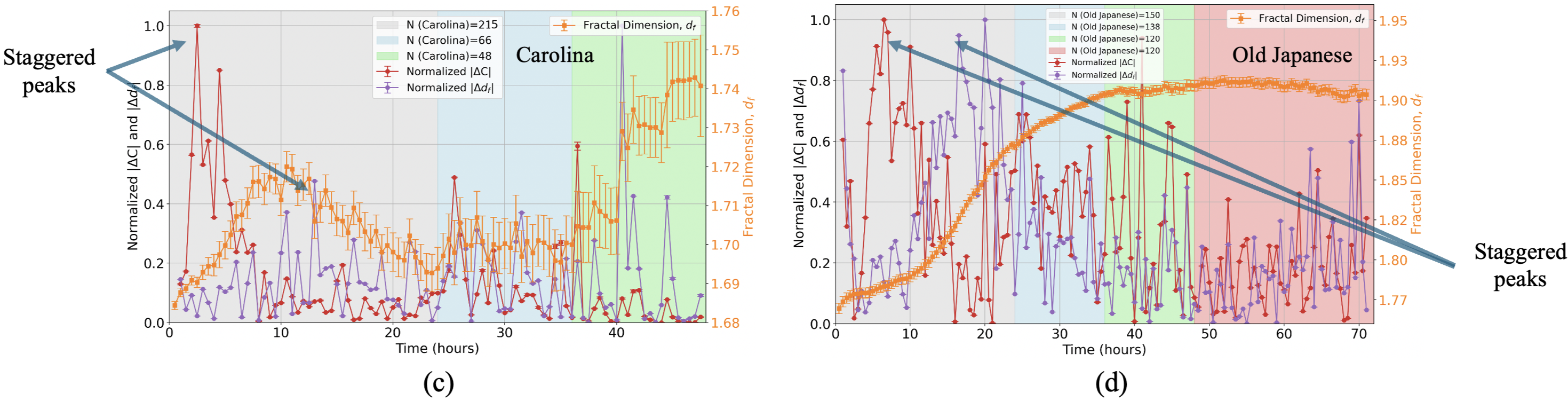}
   
\caption{\textbf{Stronger correlation between normalized changes in circularity and fractal dimension is observed in the young Japanese and young Japanese-starved groups, compared to the Carolina and old Japanese groups, and is reflected in an early peak in absolute fractal dimension between five and 10 hours for the young Japanese groups.} The figure shows the evolution of the normalized changes in mean group-aggregate circularity ($|\Delta C|$) and fractal dimension ($|\Delta d_f|$), as well as the absolute fractal dimension ($d_f$), over time for the four experimental groups: (a) young Japanese, (b) young Japanese-starved, (c) Carolina, and (d) old Japanese.
\label{fig:circ_frac}
}

    \label{Direct_assoc}
\end{figure}

It is important to consider whether a direct relationship exists between the morphological indices of circularity and fractal dimension, and the derived upper bounds on computational capacities. To investigate this, we analyze how the normalized changes in mean group-aggregate circularity ($|\Delta C|$) and fractal dimension ($|\Delta d_f|$) between successive time points evolve across the four experimental groups. These dynamics, presented in Fig. \ref{Direct_assoc}, illustrate how morphological complexity and body shape regularity evolve under varying conditions imposed by the strain, age, and feeding condition of \textit{Physarum polycephalum}.

In the young Japanese (Fig.~\ref{Direct_assoc}a) and young Japanese-starved (Fig.~\ref{Direct_assoc}b) groups, the trajectories of $|\Delta C|$ and $|\Delta d_f|$ closely follow each other throughout the experiment, indicating a strong correlation. Particularly during the early phase of the experiment, both morphological indices attain peak values at or near the same time, suggesting a strong coupling between morphological shape change and perimeter jaggedness during the initial growth window. 

In contrast, for the Carolina (Fig.~\ref{Direct_assoc}c) and old Japanese (Fig.~\ref{Direct_assoc}d) groups, although $|\Delta C|$ is initially very high—indicating early, marked changes in shape regularity—$|\Delta d_f|$ remains low, suggesting that these changes are not accompanied by increases in structural complexity and that the correlation between $\Delta C$ and $|\Delta d_f|$ is weaker. Interestingly, for these two groups, peaks in $|\Delta d_f|$ appear offset and staggered from peaks in $\Delta C$, on timescales varying from about five to more than $10$ hours. While the peak values for absolute fractal dimension ($d_f$) in young Japanese and young Japanese-starved groups occur between five and $10$ hours, the peaks for $d_f$ in Carolina and old Japanese groups are attained only in their late-time exploration, after 48 hours, reflecting these two groups' slower initial changes in $|\Delta d_f|$. 

This distinction is particularly striking when compared alongside the derived upper bounds on computational capacities (see Table S2 of the Supplementary Material): young Japanese and young Japanese-starved groups exhibit the highest computational bounds, while Carolina and old Japanese show consistently lower values, sometimes by as much as one to three orders of magnitude. Thus, the stronger early correlation between $|\Delta C|$ and $|\Delta d_f|$ appears to serve as a morphological signature of higher computational activity, which is not surprising considering the relationship of these quantities to important biophysical characteristics including changes in overall body shape and metabolic energy consumption at the periphery, respectively (see Section \ref{subsec:chemATP}). Notably, the young Japanese and young Japanese-starved strains---which show strong correlations between $|\Delta C|$ and $|\Delta d_f|$---also expand their area and perimeter more rapidly than the other two groups (see Fig. \ref{Various_strains}b-c). In contrast, the two groups (Carolina and old Japanese) in which peak values of these quantities are staggered, particularly during their early growth stages, tend to exhibit slower exploratory growth and reduced computational capacities. This indicates a direct association between morphological growth, structural complexity, and computational potential. These findings highlight the importance of the temporal coordination and tradeoff between morphological uniformity and complexity in shaping the amoeboid organism’s ability to perform information-processing tasks as it dynamically evolves in response to its local environment.

\newpage

\section{Discussion}

We observed distinct morphological growth patterns of \textit{Physarum} across different strains, ages, initial biomasses, vein networks, and feeding conditions. The Carolina strain exhibited slower expansion in both area and perimeter compared to the Japanese strain. In contrast, the starved Japanese samples expanded more rapidly across the plate in search of nutrients, resulting in greater area coverage than both the normally fed Japanese and Carolina batches. These divergent growth patterns are strain-specific, likely reflecting the distinct environmental origins of each strain. Environmental factors are known to strongly influence \textit{Physarum} morphology and contraction patterns—for instance, differences in temperature or light (UV) irradiation can induce veinless or net-like morphologies, respectively, and alter streaming dynamics~\cite{nakagaki2000interaction}. Thus, the more rapid growth of the Japanese strain and the slower growth of the Carolina strain suggests different responses to environmental stimuli, and further ecological studies would be required to establish direct links to their native environments.

Similar differences in growth between the Japanese and Carolina strains were also reported in ~\cite{vogel2015phenotypic}. However, two key distinctions exist between the trends reported there, and ours for the morphological indices. First, in \cite{vogel2015phenotypic}, the area of both Japanese and Carolina strains does not stabilize within the reported experimental duration, since values beyond 10 hours were not included. In contrast, our data show that both groups reach the nonequilibrium steady state (NESS) only after about 11 hours. Second, the circularity values in \cite{vogel2015phenotypic} are distinct from ours, for the following reason: We define circularity as $4 \pi \times \mathrm{Area}/\mathrm{Perimeter}^2$, whereas the authors of \cite{vogel2015phenotypic} define it as $\mathrm{Area}/\pi d^2$, with $d$ being the maximum distance of the \textit{Physarum} body periphery from the center of the plate.

We categorized the Japanese batches into two age groups: young, whose age since revival from sclerotia is 27 days or less, and old, whose age since revival from sclerotia is 49 days or more. The younger group initially exhibits faster growth and first touches the boundary of the dish much earlier ($\sim10$ hours) than the older group and achieves a NESS around 20 hours. In contrast, the older group starts with slower, equiradial expansion, but eventually explores a much larger area than the younger group, reaching NESS after a significantly longer period ($\sim40$ hours). The time to reach a NESS was also estimated from the stabilization of the area explored by \textit{Physarum}, providing an analytical estimate consistent with the observed transitions. The difference in NESS time between the different age groups may be attributed to the higher metabolic rate in the younger samples, which promotes more vigorous, protuberant growth in the early phase (within 10 hours). Meanwhile, the older group, operating with a lower metabolic rate, appears to assess the arena and shows no protuberant behavior during this early stage. These findings highlight how sample age influences morphological dynamics. Our results are consistent with previous observations \cite{rolland2023behavioural}, where aging from zero to 30 weeks markedly reduced migration speed, with further aging to 100 weeks characterized by more gentle declines. Decision-making performance was evaluated by offering plasmodia (6–99 weeks old) a choice between agar gel bridges differing in food quality (high-quality, 5\% w/v vs. low-quality, 2.5\% w/v oat gels) or aversiveness (high-aversive, 0.6$\%$ w/v vs. low-aversive, 0.4$\%$ w/v NaCl or NaNO$_3$) and computing the fraction of distance travelled on the preferred bridge. This performance remained largely unaffected under non-aversive conditions but increased with age under aversive conditions.

The state of the vein network and the choice of seeding biomass in \textit{Physarum} significantly influence its growth dynamics. The old Japanese batches were subdivided into vein network–connected and vein network–disrupted groups. Within the disrupted group, further division into ten biomass ranges showed that the samples with the largest biomass exhibited a revival in circularity and an earlier peak in fractal dimension values. Initially, the vein network–connected group exhibited higher circularity and lower fractal dimension values compared to the vein-disrupted group. At later time points, however, this trend reversed: the connected group showed lower circularity and higher fractal dimension, ultimately exploring a larger area with greater perimeter values owing to the connectivity of the vein network. These observations suggest that both initial biomass and the structural integrity of the vein network play important roles in regulating the morphological exploration of the local environment by \textit{Physarum}.

Upper bounds on the maximum number of operations performed in the universe have been estimated by Lloyd \cite{lloyd2002computational}, and for carbon-based life on Earth by the senior author of this work (PK) \cite{kurian2025computational}—both derived from fundamental physical constraints on quantum matter operating within relativistic event horizons and under thermodynamic law. \textcolor{black}{Extending the closed-system Margolus–Levitin (ML) limit to dissipative-driven biological systems such as \textit{Physarum} through the introduction of a dissipation-driven correction factor, we derived estimates for four distinct upper bounds on the slime mold’s computational capacities, each corresponding to a different operational regime defined by the exploitation of distinct physical degrees of freedom.} Specifically, we obtained: (i) a hydrodynamic cytosol bound, based on \textit{Physarum}’s well-documented peristaltic oscillations; (ii) a chemical ATP bound, derived from the distribution of ATP across \textit{Physarum}’s body as described in \cite{hirose1980changes, ueda1987patterns}; (iii) a kinetic energy bound, resulting from the mechanical energy of the advancing annular region at the periphery; and (iv) a quantum optical bound, arising from superradiant states predicted in the actin fiber network.

The hydrodynamic bound depends on the perimeter of \textit{Physarum}'s body and the width of the individual pseudopod-like oscillators at the advancing front. \textcolor{black}{We find that the mean value of this bound is on the order of $10^{6}$ operations over 24 hours for the young Japanese groups, and on the order of $10^{5}$ operations for the old Japanese and Carolina groups.} These differences mirror trends in perimeter growth: young Japanese groups, being more active explorers, exhibit faster perimeter expansion, which is reflected in their higher hydrodynamic bounds compared to slow-growing old Japanese and Carolina groups.

The expression for the chemical ATP bound depends on both the distribution of ATP concentration across the body of \textit{Physarum} and the area it covers. \textcolor{black}{The mean chemical ATP bound is on the order of $10^{33}$ operations per 24 hours for the young Japanese, young Japanese-starved, and old Japanese groups, and on the order of $10^{32}$ for the Carolina group}. The trends indicate higher accessible energies from ATP hydrolysis in the more active young Japanese groups as compared to old Japanese and Carolina. 
\textcolor{black}{The kinetic energy bound is determined by the area explored by \textit{Physarum} and the ratio of the rate of area expansion to the perimeter. For the young Japanese group, this bound is on the order of $10^{20}$ operations per 24 hours, whereas for the young Japanese-starved and old Japanese groups it is on the order of $10^{19}$ operations, followed by the Carolina group on the order of $10^{18}$ operations, consistent with the slower growth characteristics observed in Carolina. The kinetic energy bound is higher for the young Japanese-starved group than for the young Japanese group, reflecting a higher rate of area growth in the former.}

\textcolor{black}{The predicted quantum optical bound is determined by the area explored by \textit{Physarum}, the number of actin fibrils within that area, and the superradiant lifetime of each actin filament bundle. This bound is on the order of $10^{22}$ operations per 24 hours across all experimental groups, with the young Japanese groups exhibiting comparatively larger values than the old Japanese and Carolina groups, although all remain within the same order of magnitude.} These values mirror the total area attained by each group and are consistent with their relative levels of activity. Though the quantum optical bounds are many orders of magnitude lower than the chemical ATP bounds, they are 2-3 orders of magnitude larger than the kinetic energy bounds and many orders of magnitude greater than the hydrodynamic bounds. The inclusion of intranuclear and, if present, cytoplasmic microtubules in this estimate would further increase the maximum number of superradiant operations that \textit{Physarum} can perform in a given time interval, using its protein fiber networks of tryptophan quantum emitters. These biomolecular qubits drastically increase the functional computational capacity of protein fiber networks in \textit{Physarum}, by at least 15 orders of magnitude from the expected hydrodynamic bounds.

Potentiality of the flow of chemical energy from ATP hydrolysis to other degrees of freedom in the organism is to be expected, as a significant portion of this energy is transferred to uncontrolled thermalized motions as heat, and not directed toward useful morphological computation. \textcolor{black}{As a single \textit{Physarum} macroplasmodium can achieve an absolute maximum number of chemical operations of $\sim10^{33}$ per day, over the approximately one billion-year history of \textit{Physarum}'s ancestral supergroup Amoebozoa on Earth such a continuously growing primordial cell can have performed no more than $\sim10^{44}$ (chemical) ops}. For comparison, this value is of the same order as the maximum number of logical operations---in one second---that could be performed if one gram of \textit{Physarum} or any other rest mass were converted completely into its relativistic energy equivalent (according to $E=mc^2$), in a controlled and harnessible fashion for computation. Such a feat is, of course, not even remotely accessible because it would require controllable thermonuclear explosions to unlock the organism's rest mass-energy.

Since the bounds depend on the area, perimeter, and/or rate of area expansion, the more active strains that explore space more rapidly generally exhibit higher values. The young Japanese-starved group, which explores the largest area and perimeter, consistently showed the highest bounds, though detailed measurements of ATP turnover rates in these samples would be warranted to assess how tight those bounds can be made in the case of chemical operations. By contrast, the Carolina strain, with the smallest area and perimeter, exhibited the lowest values. The young Japanese group followed closely behind the starved group, whereas the old Japanese group was closer to Carolina, reflecting reduced metabolic capacity with age. Overall, these trends highlight how both strain and age shape growth dynamics and thereby the associated computational capacities.

As noted in the Results, both area and perimeter were analytically modelled with either sigmoid (or bi-sigmoid) models. In section 7.4, we show that the integral of a sigmoid function at long times approaches a linear function. Consequently, all bounds that depend solely on area or perimeter—except for the kinetic energy bound—exhibit a linear tail at long times. We analytically modelled the late-time evolution of these bounds with a linear function and obtained an excellent fit, confirming the expected linear dependence at later times. The onset of this linear regime coincides with the transition to the NESS, indicating that \textit{Physarum} reaches the NESS once the number of operations begins to scale linearly with time. This linear scaling is consistent with the hypothesis that \textit{Physarum} scales its computational capacity at most with the area of its body: Though the fractal dimension $d_f$ of its boundary can reach values close to two, at long times the perimeter remains effectively one-dimensional in the bounded arena, yielding a time-scaling exponent close to unity.

Although our experiments involved a significant number of samples (more than 100 per group, except for the starved one), we acknowledge that the behavior of \textit{Physarum} is highly sensitive to environmental conditions, which can substantially influence its growth dynamics---and consequently, the area and perimeter values used in estimating the computational upper bounds. Nevertheless, under standardized conditions---specifically, temperatures between 25--30\,\textdegree{}C and relative humidity $\geq$\,75\%---one  may observe reproducible trends that yielded stable estimates of the derived computational bounds. While the precise magnitudes of these bounds may vary depending on specific environmental factors, we expect their orders of magnitude to remain robust across comparable experimental setups.

\textcolor{black}{This study presents the first known framework for bounding the computational capacities of an aneural living organism based on its morphology. The underlying physical bounds are derived from our derivation of the open, driven-dissipative, macroscopic limit of the Margolus–Levitin theorem, which is applicable to any set of distinguishable logical states and can thus be broadly applied to any physical system, including biological systems. Extending this morphology-based approach to other organisms, however, will require organism-specific characterization of growth indices, dynamical evolution, and physical constraints to derive similar limits and represents an important direction for future investigation. More broadly, this framework may facilitate the derivation of analogous physical bounds for a broader class of systems--both animate and inanimate--that compute using their physical substrates, as in reservoir computing.} In such systems, the highly nonlinear body is dynamically perturbed with a linear stimulus and probed, imaged, or read out to update the stimulus, thereby performing computational tasks \cite{nakajima2020physical}. Living examples include the brain \cite{cai2023brain} and multicellular collectives \cite{nikolic2023computational}, while non-living implementations span water \cite{fernando2003pattern}, mechanical oscillators \cite{coulombe2017computing}, and brain-inspired networks \cite{cucchi2021reservoir}. A particularly striking subtype is the liquid state machine (LSM), which demonstrates how inanimate fluids can encode and process information through transient dynamic states. One compelling example of a LSM is a liquid computer in which a bucket of water was used to perform an XOR logic operation \cite{fernando2003pattern}. These embodied computing systems are subject to geometric and allometric constraints, suggesting that their computational capacities—like that of \textit{Physarum}—can be linked to their morphological structure and dynamics.

\section{Materials and Methods} 

We tracked the growth of the macroplasmodia of the slime mold \textit{Physarum polycephalum}. Two strains were studied — Japanese (Sonobe) and Carolina — to investigate strain-specific morphological patterns. The Japanese strain samples were further divided into three subgroups based on age and feeding conditions: young Japanese (age since revival from sclerotia $\leq$27 days), old Japanese (age since revival from sclerotia $\geq$49 days), and young Japanese-starved (age since revival from sclerotia $\leq$27 days, but deprived of feeding for one additional day compared to the other groups).

\subsection{Sample preparation}

\subsubsection{Young Japanese}

\textit{Physarum} was grown from sclerotia on 10 cm-deep plain 1\% agar plates with oats individually scattered across the surface. The oats were replaced every other day. Once the \textit{Physarum} grew to cover the plate in two days, they were moved to 15-cm plates by cutting a section of \textit{Physarum}-covered agar and placing on the new plate’s agar surface with the same single-layer oat arrangement. After growing to cover the larger plate in two days, a section of \textit{Physarum}-covered agar was placed on the edge of the surface of a 25-cm rectangular plate with a 3-4 cm strip on the opposite end of the dish covered in oats. In front of these oats were individually placed oats positioned about 1 cm apart from each other to form a less-dense feeding section. The oats were then sprayed with tap water to hydrate them. To acquire a biopsy punch of \textit{Physarum} for each timepoint, from a two-day old plate, a 1.5-mL Eppendorf tube base was pressed into the agar surrounding a singular oat which was covered in \textit{Physarum}. This agar-oat-\textit{Physarum} section was then placed in the center of a freshly poured, 10 cm-deep well petri dish. 

\subsubsection{Old Japanese}

\textit{Physarum} samples were taken from two-day old, 25-cm rectangular plates with a dampened 3-4 cm strip of oats at one end and old oats with \textit{Physarum} on the other end. For the vein network-connected condition, the lawn section in the middle of the plate was divided into sections of varying surface areas (6 $\times$ 60*35 mm, 6 $\times$ 30*35 mm, 12 $\times$ 30*17.5 mm sections). Each section was gently removed and placed in the middle of a 1\% plain agar plate within the bounds of a 16-mm diameter circle in the center. For the scraped (vein network-disconnected) condition, the center lawn was scraped from its surface and cut repeatedly using a cell lifter. The total mass of the semi-homogenized section was recorded and six sections of 1/12, 1/24, and 1/48 of the total mass were divided and placed in the center of a 10 cm-deep well plate within a 16-mm diameter circular boundary. 

\subsubsection{Carolina}

\textit{Physarum} obtained from Carolina Biologicals was grown from sclerotia on 10 cm-deep plain 1\% agar plates with oats individually scattered on the plate surface. These were changed every other day. For the oat condition, samples were taken from two-day old plates following the technique above for young Japanese. For the semi-defined media (SDM) condition, a section of lawn-growth \textit{Physarum} was scraped and placed on a 1.7\% SDM agar plate, prepared according to the recipe in ~\cite{mccullough1976defined}. Plates were passaged every several days, or until \textit{Physarum} covered the surface of the plate. The biopsy punches consisted of SDM agar and lawn-growth  \textit{Physarum}. For the SDM-to-oat conditions, sections of SDM agar with \textit{Physarum} growth were placed on plain agar plates with oats scattered in a single layer on the surface of the plain agar. After wandering off the SDM agar, the biopsy punch was removed so the only nutrient source was oats. Samples from this condition were taken from two-day old plates of varying passage exposure time to the oat-only condition.

\subsubsection{Young Japanese (starved)}

The sample preparation method was similar to that used for the young Japanese group described above, except that the samples were taken from a three day-old plate instead of a two day-old plate, resulting in a starved condition.

\subsection{Image acquisition}

The imaging platform was developed based on the framework presented in \cite{murugan2021mechanosensation} to enable high-throughput imaging over extended durations (see Fig. S15 of the Supplementary Material). Thus, we captured high-resolution, 1600-dpi snapshots of \textit{Physarum}’s growth every 30 minutes over a period of 24 to 72 hours. The setup supported parallel trials on up to 18 dishes simultaneously, using three scanners with six dishes per scanner, thereby significantly reducing the time required to generate biological replicates. All captured images were automatically written to the host computer for processing (see Fig. S1 of the Supplementary Material).

\subsection{Analysis}
Custom-written \href{https://github.com/sanghita211/pypetana}{codes} 
were modified to create an updated implementation (\href{https://github.com/suyashb4/PyPETANA2.0.git}{PyPETANA2.0}, available on GitHub), which was employed for processing and analyzing the scanner-acquired images. The pipeline first applies lossless compression to reduce the large image files. Since each scanner image contains six experimental plates, the images are automatically segmented into image files with an individual plate. Each plate image was converted to grayscale, subjected to noise reduction, and masked to remove spurious edges, after which thresholding was applied to segment and track the \textit{Physarum} body. We modified these codes to implement adaptive thresholding, in which different threshold values are applied to regions near the center versus the boundary of the plate, thereby reducing imaging artifacts that occur at the edges. Specifically, a smaller threshold was used for the central region (from the plate center to 85\% of its radius), and a higher threshold was used for the outer 15\% near the boundary, ensuring that \textit{Physarum} was accurately segmented while excluding spurious particles or liquid drops at the edges. From the processed images, we extracted key morphological indices, including explored area, perimeter, circularity, and fractal dimension.
\textcolor{black}{
\subsection{Robustness analysis of morphological quantities}
\subsubsection{Trial exclusion criteria} Experimental plates exhibiting severe segmentation artifacts arising from glare, water droplets, or failed contour detection were excluded from the analysis, resulting in the exclusion of 6 out of more than 500 plates. In a small subset of trials ($\sim 10$ plates), a few isolated time points exhibited imaging artifacts. These artifact-affected time points were excluded from the fitting procedure to ensure accurate extraction of the underlying dynamical trends in the data. After correcting for these artifacts, all sigmoidal fits showed strong agreement with the data ($R^2 > 0.9$).
\subsubsection{Boundary threshold}
The results presented in this work are based on an outer boundary threshold value of 65, which yielded clean and reliable segmentation for more than $95\%$ of the analyzed plates across all experimental groups. To assess robustness with respect to boundary masking, we recomputed the morphological quantities using multiple boundary-threshold values ranging from 45 to 85. This analysis was performed on representative subsets ($N=25$ each) for all experimental groups except the smaller young Japanese-starved group, for which all samples ($N=18$) were considered. We found that a threshold value of 65 provides an optimal balance between avoiding spurious classification of petri-dish boundary regions as 	\textit{Physarum} (observed for a boundary threshold of 45) and avoiding the omission of fine \textit{Physarum} structures near the boundary (observed for a boundary threshold of 85). However, the group-aggregate mean morphological trends remained robust over the reasonable boundary-threshold range of 55-85. The corresponding robustness analyses are provided in Section S5 of the Supplementary Material.
\subsubsection{Circularity and fractal dimension }The circularity is defined as $C=4 \pi A / P^2$. Because both area $A$ and perimeter $P$ are extracted from the same segmentation and scale consistently with box size, any scale-dependent variations affect numerator and denominator in a compensatory manner, rendering the circularity index robust to reasonable changes in box-counting scale. For each time point, the fractal dimension of the 	\textit{Physarum} body was computed using 21 box sizes ranging from 5 pixels (0.19 mm) to 200 pixels (7.84 mm), with the fractal dimension extracted as the negative slope of the corresponding box-counting relation. \textcolor{black}{Representative videos for each experimental group, showing a linear trend in the log–log box-counting plots over the entire time series and demonstrating the stability of the extracted fractal dimension across scales, are provided as Supplementary Movies S1--S4.}
\subsubsection{Derivative threshold}
We additionally evaluated the robustness of the $15\%$ derivative threshold used to define the transition to the nonequilibrium steady state (NESS). The derivative threshold was varied from $10\%$ to $20\%$, and for all experimental groups the resulting NESS transition times remained within $10\%$ of those obtained using the nominal 15\% derivative threshold. The corresponding transition times for different derivative-threshold values are provided in Table S3 of the Supplementary Material.}

\textcolor{black}{
\subsection{Thickness estimation and sensitivity analysis}
Assuming that the total volume (and hence total mass) of the \textit{Physarum} body remains approximately conserved during its morphological evolution on the agar plate, we estimated the thickness time series for $N=150$ samples from the old Japanese group, for which we had measurements of initial biomasses, using
\begin{equation}
l(t)=\frac{M_0}{\rho_m A(t)}
\end{equation}
where \(M_0\) is the total plasmodium mass, \(\rho_m\) is the mass density, and \(A(t)\) is the area covered by the \textit{Physarum} body at time \(t\). For each sample, the time-averaged thickness was computed as
\begin{equation}
\bar{l}=\frac{1}{T}\int_0^T l(t)\,dt
\end{equation}
where \(T\) is the total experimental duration. The resulting thickness time series were then averaged across samples to obtain the mean thickness evolution, yielding a time-averaged mean thickness of approximately \(500~\mu\mathrm{m}\). To further characterize the distribution of thickness values, we generated histograms showing the percentage of experimental time spent within different thickness ranges spanning \(0\)--\(3000~\mu\mathrm{m}\). We found that the \textit{Physarum} body spends the largest fraction of time near relatively small thickness values, with nearly $30\%$ of the experimental duration occurring near $100~\mu\mathrm{m}$ thickness (within the range $50\text{--}150~\mu\mathrm{m}$), while only about $5\%$ of the experimental duration lies within the range of thickness values $450\text{-}550~\mu\mathrm{m}$. Motivated by these observations, we take \(100~\mu\mathrm{m}\) and \(500~\mu\mathrm{m}\) as representative, biologically relevant thickness values (mode and time-averaged mean, respectively) and evaluate the mass-dependent bounds for these two representative cases. Although the absolute numerical values of our computational capacity bounds depend on an assumed thickness, the qualitative trends and principal conclusions remain robust for both representative thickness values considered here. The histogram of time spent versus thickness, along with the mean thickness time series for the old Japanese group, is shown in Figure S12 of the Supplementary Material.
}

\textcolor{black}{
\subsection{Bootstrap resampling}
Bootstrap resampling is a statistical method used to estimate uncertainty and confidence intervals by repeatedly generating synthetic datasets through random sampling with replacement from the original dataset. In this procedure, each selected sample is returned to the dataset before the next sampling step, allowing the same sample to be selected multiple times across different resampled datasets. For each experimental group and morphological quantity, the group-aggregate means were recomputed for every resampled dataset at each time point, enabling empirical estimation of confidence intervals and statistical robustness. A total of 100,000 bootstrap resampling iterations were performed for each morphological observable and experimental group. Confidence intervals corresponding to both the $95 \%$ and $99.5 \%$ levels were computed. The resulting bootstrap trajectories and confidence intervals for the morphological quantities are shown in Figure S6 of the Supplementary Material.}

\subsection{Coupled harmonic oscillators in the macroscopic Margolus-Levitin limit}

As detailed in Section \ref{sec:macMarg-LevinPhys}, for a single quantum harmonic oscillator with a non-degenerate spectrum, the mean energy approaches half of the maximum accessible energy: $\left\langle \mathcal{E} \right\rangle = \frac{1}{2} E_{\text{max}}$. However, this equation is altered when the system exhibits degeneracy. Degenerate energy levels tend to shift the mean energy closer to the maximum energy. 

It is well established that \textit{Physarum} behaves as a multi-oscillator system. More precisely, it functions as a collection of coupled, three-dimensional oscillators. To examine the mean energy of such systems in the macroscopic limit, we first consider the uncoupled case and then extend the analysis to progressively more complex, coupled quantum harmonic oscillator systems.

\subsubsection{Systems of Coupled $3$-D Isotropic Harmonic Oscillators}

We now turn to the case of coupled harmonic oscillator systems, which more accurately reflect the behavior of \textit{Physarum} as a spatially extended and interacting system, though this is an approximation as \textit{Physarum} is a highly nonlinear system. Let the position and momentum of the $i^{\text{th}}$ oscillator be denoted by $r_i$ and $p_i$, respectively. With periodic boundary conditions and only nearest-neighbor couplings between oscillators, the Hamiltonian of the system can then be written as: 
\begin{equation}
\hat{H}=\sum_{i=1}^\eta \left(\frac{p_i^2}{2 m}+\frac{1}{2} m \omega^2 r_i^2\right)+\frac{1}{2} k \sum_{i=1}^\eta\left|r_i-r_{i+1}\right|^2, \  \ r_{\eta+1}\equiv r_{1} 
\end{equation}
Here, $
r_i^2=x_i^2+y_i^2+z_i^2, \ p_{i}^2=p_{x_i}^2+p_{y_i}^2+p_{z_i}^2, \text{and} \ \eta$ is the number of oscillators.

Writing the above Hamiltonian only in one of the Cartesian axes (dimensions),
\begin{equation}\nonumber
\begin{aligned}
\hat{H}_x=\frac{1}{2 m}\left(p_ {x_1}^2+p_{x_2}^2+...+p_{x_{\eta}}^2\right)+\frac{1}{2} m \omega^2\left(x_1^2+x_2^2+...+x_\eta^2\right) \\
+ \ k\left[x_1^2+x_2^2+...x_{\eta}^2-\left(x_1 x_2+x_2 x_3+...+x_{\eta-1} x_{\eta}+x_{\eta} x_1\right)\right], \ \ x_{\eta+1}\equiv x_{1}
\end{aligned}
\end{equation}
we can express this as
\begin{equation}
\hat{H}_x=\frac{1}{2 m} p_x^{\top} p_x+\frac{1}{2} x^{\top} \kappa_{x} \ x.
\end{equation}
Next, we can find the eigenvalues of the matrix $\kappa_{x}$ and obtain the frequencies of the modes, which are identical for the $\kappa_y$ and $\kappa_z$ matrices in the case of $\eta$ isotropic oscillators.

For the case of three coupled 3D harmonic oscillators, the ratio of the mean energy to the maximum energy cutoff is given by (see Section S7 of the Supplementary Material for proof)
\begin{equation}
\frac{\langle\mathcal{E}\rangle}{E_{\max }}=\frac{\left(\frac{6}{7}+\frac{3 \tilde{\omega}}{4 \beta}\right)}{\left(1+\frac{\tilde{\omega}}{\beta}\right)},
\end{equation}
where $\beta = \tfrac{N_{1}}{N_{2}}$ denotes the ratio of the two distinct normal-mode cutoff states, and $\tilde{\omega} = \tfrac{\Omega^{\prime}}{\Omega}$ denotes the ratio of the two distinct normal-mode frequencies of the system (one degenerate). For various choices of $\beta$ and $\tilde{\omega}$, the ratio $\langle\mathcal{E}\rangle / E_{\max }$ varies between $0.75$ and $0.86$, which can be verified in the limits as $\tilde{\omega}\rightarrow \infty$ and $\tilde{\omega}\rightarrow 0$, respectively. 
If we make approximations for the uncoupled limit of our coupled-oscillator system with equal mode cutoffs ($\tilde{\omega}=\beta=1$), the ratio is approximately $0.80$, which is smaller than the value of $0.90$ obtained for the system of nine uncoupled 1D oscillators.

Similar expressions are obtained for the four- and five-coupled oscillator systems (see Supplementary Material), where the ratio $\langle\mathcal{E}\rangle / E_{\max }$ varies within the same interval (between 0.75 and 0.86). This range is determined by the degeneracy structure of the system. The maximum degeneracy (per dimension) $G$ is defined as the largest number of normal modes per dimension sharing the same frequency. For $\eta$ = 3, 4, or 5 coupled 3D oscillators, no frequency occurs more than twice, so the maximum degeneracy per dimension is $G = 2$. Thus, for the three- to five-oscillator cases in 3D, $G=2$ and $d=3$, giving an upper bound of $\frac{6}{7} \approx 0.86$, as noted earlier (see Supplementary Material for further details).

In general, the ratio $\langle\mathcal{E}\rangle / E_{\max }$  varies between a lower value $d /(d+1)$---when all modes are non-degenerate ($G=1$)---and an upper value $Gd/( Gd+1)$, where $d$ is the dimensionality of the system and $G$ is the maximum degeneracy per dimension. Thus, the general expression for the upper bound on the ratio is
\begin{equation}\label{Ratio_3D}
\frac{\langle \mathcal{E} \rangle_{Gd}}{E_{\max}} = \frac{G d}{G d + 1}.
\end{equation}
As the number of oscillators $\eta$ increases, the maximum degeneracy per dimension cannot exceed $\eta$, yielding the absolute upper bound on the ratio as
\begin{equation}
\sup \left(\frac{\langle \mathcal{E} \rangle_{Gd}}{E_{\max}}\right) = \frac{\eta d}{\eta d + 1},
\end{equation}
which tends to unity as $\eta \to \infty$. In this regime, the mean energy approaches the maximum energy of the system, showing that larger oscillator numbers (or higher dimensionality) drive $\langle \mathcal{E} \rangle/E_{\max}$ arbitrarily close to one.

\textcolor{black}{
\subsection{Derivation of correction to Margolus-Levitin closed-system bound from driven-dissipative channels in open coupled-oscillator systems}
The upper bound on the number of logical operations given by Eq.~\ref{eq:slimeops} is obtained from the Margolus-Levitin quantum speed limit (QSL), originally derived for closed systems undergoing unitary evolution. However, \textit{Physarum} is inherently an open, driven, and dissipative biological system that may be modeled at a coarse-grained level as a system of coupled oscillators \cite{saigusa2008amoebae, kuramoto1984chemical}, which we have shown in Section 8.7 recovers a semiclassical form of the Margolus-Levitin limit for a closed cycle of many coupled, highly degenerate oscillators. It is also important to consider how drive and dissipation may modify the corresponding  bounds on computation, whether using classical or quantum degrees of freedom. Quantum speed limits for open systems undergoing non-unitary evolution have been developed in several works \cite{del2013quantum, taddei2013quantum, brody2012mixed}, and studies mapping QSLs into classical phase space using Wigner quasi-probability distributions have concluded that similar limits can hold when considering classical and semiclassical systems, or those at the quantum-to-classical transition~\cite{shanahan_quantum_2018, okuyama_quantum_2018}.}

\textcolor{black}{Here, following the framework of del Campo et al.~\cite{del2013quantum}, we consider coupled-oscillator models incorporating drive and dissipation in order to estimate corrections to the corresponding computational bounds. The open-system QSL formulation in Ref. \cite{del2013quantum} extends Mandelstam--Tamm-type bounds for non-unitary systems governed by non-Hermitian ($H_\text{eff}=H-i\Gamma$) dynamics, relating the minimum evolution time not only to the energy uncertainty ($\Delta H$) of the underlying system, but also to non-unitary contributions encoded through the non-Hermitian component ($\Gamma$) of the effective Hamiltonian.} 

\textcolor{black}{Within this framework, we incorporate dissipative and drive channels into the corresponding Lindbladian dynamics in order to estimate computational bounds for driven-dissipative coupled-oscillator systems.
To quantify the resulting enhancement in computational bounds, we compare the driven-dissipative computational bound with the corresponding closed-system Margolus-Levitin bound by computing their ratio, $\widetilde{\mathcal{N}}=N_{\text{open}}/N_{ML}$. For a simple system of two coupled oscillators, this ratio takes the form
\begin{equation}\label{Enhancement_comp}
\small
\begin{aligned}
\widetilde{\mathcal N}
=&\, \frac{\pi}{\sqrt{3/2}}
\Bigg[\left(N_{+}^{2}-1\right)+(1+2\tilde{k})
\left(N_{-}^{2}-1\right)+(\tilde{\gamma}+\tilde{\kappa})^{2}
\Big(7(N_{+}^{2}+N_{-}^{2})+24(N_{+}+N_{-})+12N_{+}N_{-}
+12\Big)\\
&+48\tilde{\kappa}(\tilde{\gamma}+\tilde{\kappa})(N_{+}+N_{-}+2)+96\tilde{\kappa}^{2}\Bigg]^{1/2}
\times
\Big[(N_{+}+1)+\sqrt{1+2\tilde{k}}(N_{-}+1)\Big]^{-1}.
\end{aligned}
\end{equation}
Here, $N_{+}$ and $N_{-}$ denote the normal-mode energy cutoffs, $\tilde{\gamma}=\gamma/\omega$ and $\tilde{\kappa}=\kappa/\omega$ are the dimensionless dissipation and gain rates, respectively, and $\tilde{k}=k/(m\omega^2)$ is the dimensionless coupling strength. Detailed derivations of the corresponding enhancement ratios for both the single harmonic oscillator and coupled harmonic oscillator systems are presented in the Supplementary Material. In the single-oscillator limit corresponding to $N_{-}=0$ and $\tilde{k}\rightarrow0$, and in the weak-dissipation and weak-drive regime $\tilde{\gamma},\tilde{\kappa}\rightarrow 0$, Eq.~\ref{Enhancement_comp} reduces to $\widetilde{\mathcal{N}}\rightarrow \frac{\pi} {\sqrt{3/2}}\frac{\sqrt{N_{+}^{2}-2}}{N_{+}+2}$, which approaches an enhancement factor of order $\pi/\sqrt{3/2}$ for large $N_{+}$.
This enhancement may be interpreted as arising from two contributions: a characteristic open-system correction factor of order $\pi/\sqrt{2}$ associated with the driven-dissipative extension of Mandelstam--Tamm-type quantum speed limits, together with an additional factor associated with the ratio between the Margolus--Levitin (ML) and Mandelstam--Tamm (MT) bounds for the corresponding closed-system oscillator. More generally for an arbitrary state, within the weak-coupling regime consistent with the Markovian approximation, the enhancement of the driven-dissipative computational bound relative to this ratio may be simply expressed as
\begin{equation}
\widetilde{\mathcal N}=\frac{\pi}{\sqrt2}
\frac{\mathcal N_{\mathrm{MT}}}{\mathcal N_{\mathrm{ML}}}.
\end{equation}
When the MT and ML computational rates remain comparable, i.e.,
\[
\mathcal N_{\mathrm{MT}}
\approx
\mathcal N_{\mathrm{ML}},
\]
this enhancement reduces to a characteristic factor of order $\pi/\sqrt2$. We therefore adopt $\pi/\sqrt2$ as a representative correction factor to the closed-system ML and MT bounds in order to estimate the corresponding driven-dissipative computational bounds considered here.
}

\subsection{Calculation of group-aggregate mean bounds on computation}

The area and perimeter time series of all samples were fitted with the appropriate models: a sigmoid for the Japanese groups (young Japanese, old Japanese, and young Japanese-starved groups) and a bi-sigmoid for the Carolina strain. The extracted fit parameters for each sample were then substituted into the analytical expressions for the various bounds to calculate the upper bounds on computation for that individual trial over a given time interval.

Group-aggregate mean values of the bounds at a time point $t$ were computed as follows. First, the extracted fit parameters from each sample were applied to the analytical expressions, and the bound values for that sample were calculated from the start of the experiment up to time $t$. This process was repeated across all samples within a group to obtain the distribution of bound values at $t$. Next, the geometric mean of these values was computed to represent the group-aggregate mean, since the bounds can span several orders of magnitude. This procedure was repeated for successive time points in 0.5-hour increments (the resolution of our experiments), continuing throughout the experimental window and thereby producing a time series of group-aggregate bounds over the entire duration. Error estimates were obtained by calculating a multiplicative standard error factor across all trials within a group at each time point, and the results are shown as shaded ± error bars in the figures. All bounds were calculated from the first time point giving nonzero values for the bounds, in 0.5-hour increments, over time windows ranging from 24 to 72 hours. 

\bibliography{aapmsamp} 

@article{muller2017morphological,
  title={What is morphological computation? On how the body contributes to cognition and control},
  author={M{\"u}ller, Vincent C and Hoffmann, Matej},
  journal={Artificial Life},
  volume={23},
  number={1},
  pages={1--24},
  year={2017},
  publisher={MIT Press One Rogers Street, Cambridge, MA 02142-1209, USA journals-info~…}
}

@article{nowakowski2017bodily,
  title={Bodily processing: the role of morphological computation},
  author={Nowakowski, Przemys{\l}aw Robert},
  journal={Entropy},
  volume={19},
  number={7},
  pages={295},
  year={2017},
  publisher={MDPI}
}

@article{zahedi2013quantifying,
  title={Quantifying morphological computation},
  author={Zahedi, Keyan and Ay, Nihat},
  journal={Entropy},
  volume={15},
  number={5},
  pages={1887--1915},
  year={2013},
  publisher={MDPI}
}

@article{nakagaki2000maze,
  title={Maze-solving by an amoeboid organism},
  author={Nakagaki, Toshiyuki and Yamada, Hiroyasu and T{\'o}th, {\'A}gota},
  journal={Nature},
  volume={407},
  number={6803},
  pages={470--470},
  year={2000},
  publisher={Nature Publishing Group UK London}
}

@article{nakagaki2007minimum,
  title={Minimum-risk path finding by an adaptive amoebal network},
  author={Nakagaki, Toshiyuki and Iima, Makoto and Ueda, Tetsuo and Nishiura, Yasumasa and Saigusa, Tetsu and Tero, Atsushi and Kobayashi, Ryo and Showalter, Kenneth},
  journal={Physical Review Letters},
  volume={99},
  number={6},
  pages={068104},
  year={2007},
  publisher={APS}
}

@article{tero2010rules,
  title={Rules for biologically inspired adaptive network design},
  author={Tero, Atsushi and Takagi, Seiji and Saigusa, Tetsu and Ito, Kentaro and Bebber, Dan P and Fricker, Mark D and Yumiki, Kenji and Kobayashi, Ryo and Nakagaki, Toshiyuki},
  journal={Science},
  volume={327},
  number={5964},
  pages={439--442},
  year={2010},
  publisher={American Association for the Advancement of Science}
}

@article{dussutour2010amoeboid,
  title={Amoeboid organism solves complex nutritional challenges},
  author={Dussutour, Audrey and Latty, Tanya and Beekman, Madeleine and Simpson, Stephen J},
  journal={Proceedings of the National Academy of Sciences},
  volume={107},
  number={10},
  pages={4607--4611},
  year={2010},
  publisher={National Acad Sciences}
}

@article{jones2014computation,
  title={Computation of the travelling salesman problem by a shrinking blob},
  author={Jones, Jeff and Adamatzky, Andrew},
  journal={Natural Computing},
  volume={13},
  pages={1--16},
  year={2014},
  publisher={Springer}
}

@article{baumgarten2010plasmodial,
  title={Plasmodial vein networks of the slime mold \textit{Physarum polycephalum} form regular graphs},
  author={Baumgarten, Werner and Ueda, Tetsuo and Hauser, Marcus JB},
  journal={Physical Review E},
  volume={82},
  number={4},
  pages={046113},
  year={2010},
  publisher={APS}
}

@article{saigusa2008amoebae,
  title={Amoebae anticipate periodic events},
  author={Saigusa, Tetsu and Tero, Atsushi and Nakagaki, Toshiyuki and Kuramoto, Yoshiki},
  journal={Physical Review Letters},
  volume={100},
  number={1},
  pages={018101},
  year={2008},
  publisher={APS}
}

@article{aono2007amoeba,
author = {Aono, Masashi and Hara, Masahiko and Aihara, Kazuyuki},
title = {Amoeba-based neurocomputing with chaotic dynamics},
year = {2007},
issue_date = {September 2007},
publisher = {Association for Computing Machinery},
address = {New York, NY, USA},
volume = {50},
number = {9},
issn = {0001-0782},
journal = {Communications of the ACM},
month = sep,
pages = {69–72},
numpages = {4}
}

@article{aono2009amoeba,
  title={Amoeba-based chaotic neurocomputing: combinatorial optimization by coupled biological oscillators},
  author={Aono, Masashi and Hirata, Yoshito and Hara, Masahiko and Aihara, Kazuyuki},
  journal={New Generation Computing},
  volume={27},
  pages={129--157},
  year={2009},
  publisher={Springer}
}

@article{zhu2013amoeba,
  title={Amoeba-based computing for traveling salesman problem: Long-term correlations between spatially separated individual cells of \textit{Physarum polycephalum}},
  author={Zhu, Liping and Aono, Masashi and Kim, Song-Ju and Hara, Masahiko},
  journal={Biosystems},
  volume={112},
  number={1},
  pages={1--10},
  year={2013},
  publisher={Elsevier}
}

@article{zhu2018remarkable,
  title={Remarkable problem-solving ability of unicellular amoeboid organism and its mechanism},
  author={Zhu, Liping and Kim, Song-Ju and Hara, Masahiko and Aono, Masashi},
  journal={Royal Society Open Science},
  volume={5},
  number={12},
  pages={180396},
  year={2018},
  publisher={The Royal Society}
}

@article{babcock2024ultraviolet,
  title={Ultraviolet superradiance from mega-networks of tryptophan in biological architectures},
  author={Babcock, NS and Montes-Cabrera, G and Oberhofer, KE and Chergui, M and Celardo, GL and Kurian, P},
  journal={The Journal of Physical Chemistry B},
  volume={128},
  number={17},
  pages={4035--4046},
  year={2024},
  publisher={ACS Publications}
}

@article{patwa2024quantum,
  title={Quantum-enhanced photoprotection in neuroprotein architectures emerges from collective light-matter interactions},
  author={Patwa, Hamza and Babcock, Nathan S and Kurian, Philip},
  journal={Frontiers in Physics},
  volume={12},
  pages={1387271},
  year={2024},
  publisher={Frontiers Media SA}
}

@article{bajpai2025tracking,
  title={Tracking and distinguishing slime mold solutions to the traveling salesman problem through synchronized amplification in the non-equilibrium steady state},
  author={Bajpai, Suyash and Aono, Masashi and Kurian, Philip},
  journal={arXiv:2504.03492\ \textbf{[physics.bio-ph]}},
  year={2025}
}

@article{kurian2025computational,
  title={Computational capacity of life in relation to the universe},
  author={Kurian, Philip},
  journal={Science Advances},
  volume={11},
  number={13},
  pages={eadt4623},
  year={2025},
  publisher={American Association for the Advancement of Science}
}

@article{murugan2021mechanosensation,
  title={Mechanosensation mediates long-range spatial decision-making in an aneural organism},
  author={Murugan, Nirosha J and Kaltman, Daniel H and Jin, Paul H and Chien, Melanie and Martinez, Ramses and Nguyen, Cuong Q and Kane, Anna and Novak, Richard and Ingber, Donald E and Levin, Michael},
  journal={Advanced Materials},
  volume={33},
  number={34},
  pages={2008161},
  year={2021},
  publisher={Wiley Online Library}
}

@inproceedings{paul2004morphology,
  title={Morphology and computation},
  author={Paul, Chandana},
  booktitle={Proceedings of the International Conference on the Simulation of Adaptive Behaviour},
  year={2004},
pages={33--38}
}

@article{wootton1992functional,
  title={Functional morphology of insect wings},
  author={Wootton, RJ},
  journal={Annual Review of Entomology},
  volume={37},
  number={1},
  pages={113--140},
  year={1992}
}

@article{lloyd2002computational,
  title={Computational capacity of the universe},
  author={Lloyd, Seth},
  journal={Physical Review Letters},
  volume={88},
  number={23},
  pages={237901},
  year={2002},
  publisher={APS}
}

@article{ueda1987patterns,
  title={Patterns in the distribution of intracellular ATP concentration in relation to coordination of amoeboid cell behavior in \textit{Physarum polycephalum}},
  author={Ueda, Tetsuo and Mori, Yoshihito and Kobatake, Yonosuke},
  journal={Experimental Cell Research},
  volume={169},
  number={1},
  pages={191--201},
  year={1987},
  publisher={Elsevier}
}

@article{shanahan_quantum_2018,
	title = {Quantum {speed} {limits} across the {quantum}-to-{classical} {transition}},
	volume = {120},
	issn = {0031-9007, 1079-7114},
	number = {7},
	urldate = {2024-08-16},
	journal = {Physical Review Letters},
	author = {Shanahan, B. and Chenu, A. and Margolus, N. and Del Campo, A.},
	month = feb,
	year = {2018},
	pages = {070401},
}

@article{okuyama_quantum_2018,
	title = {Quantum {speed} {limit} is {not} {quantum}},
	volume = {120},
	issn = {0031-9007, 1079-7114},
	number = {7},
	urldate = {2024-08-16},
	journal = {Physical Review Letters},
	author = {Okuyama, Manaka and Ohzeki, Masayuki},
	month = feb,
	year = {2018},
	pages = {070402},
}

@article{ben2009learning,
  title={Learning from bacteria about natural information processing},
  author={Ben-Jacob, Eshel},
  journal={Annals of the New York Academy of Sciences},
  volume={1178},
  number={1},
  pages={78--90},
  year={2009},
  publisher={Wiley Online Library}
}

@article{adleman1994molecular,
  title={Molecular computation of solutions to combinatorial problems},
  author={Adleman, Leonard M},
  journal={Science},
  volume={266},
  number={5187},
  pages={1021--1024},
  year={1994},
  publisher={American Association for the Advancement of Science}
}

@incollection{feynman2018there,
  title={There’s plenty of room at the bottom},
  author={Feynman, Richard},
  booktitle={Feynman and Computation},
  pages={63--76},
  year={2018},
  publisher={CRC Press}
}

@article{winfree1998design,
  title={Design and self-assembly of two-dimensional DNA crystals},
  author={Winfree, Erik and Liu, Furong and Wenzler, Lisa A and Seeman, Nadrian C},
  journal={Nature},
  volume={394},
  number={6693},
  pages={539--544},
  year={1998},
  publisher={Nature Publishing Group UK London}
}

@inproceedings{rothemund2000program,
author = {Rothemund, Paul W. K. and Winfree, Erik},
title = {The program-size complexity of self-assembled squares},
year = {2000},
isbn = {1581131844},
publisher = {Association for Computing Machinery},
booktitle = {Proceedings of the Thirty-Second Annual ACM Symposium on Theory of Computing},
pages = {459–468},
numpages = {10},
location = {Portland, Oregon, USA},
}

@article{barry2010entropy,
author = {Edward Barry  and Zvonimir Dogic },
title = {Entropy driven self-assembly of nonamphiphilic colloidal membranes},
journal = {Proceedings of the National Academy of Sciences},
volume = {107},
number = {23},
pages = {10348-10353},
year = {2010},
eprint = {https://www.pnas.org/doi/pdf/10.1073/pnas.1000406107}}

@article{gibaud2012reconfigurable,
  title={\ Reconfigurable self-assembly through chiral control of interfacial tension},
  author={Gibaud, Thomas and Barry, Edward and Zakhary, Mark J and Henglin, Mir and Ward, Andrew and Yang, Yasheng and Berciu, Cristina and Oldenbourg, Rudolf and Hagan, Michael F and Nicastro, Daniela},
  journal={Nature},
  volume={481},
  number={7381},
  pages={348--351},
  year={2012},
  publisher={Nature Publishing Group UK London}
}

@article{kari1997dna,
  title={DNA computing: arrival of biological mathematics},
  author={Kari, Lila},
  journal={Mathematical Intelligencer},
  volume={19},
  number={2},
  pages={9--22},
  year={1997},
  publisher={Citeseer}
}

@article{tekle2022new,
  title={New insights on the evolutionary relationships between the major lineages of Amoebozoa},
  author={Tekle, Yonas I and Wang, Fang and Wood, Fiona C and Anderson, O Roger and Smirnov, Alexey},
  journal={Scientific Reports},
  volume={12},
  number={1},
  pages={11173},
  year={2022},
  publisher={Nature Publishing Group UK London}
}

@article{nakajima2020physical,
  title={Physical reservoir computing—an introductory perspective},
  author={Nakajima, Kohei},
  journal={Japanese Journal of Applied Physics},
  volume={59},
  number={6},
  pages={060501},
  year={2020},
  publisher={IOP Publishing}
}

@article{cai2023brain,
  title={Brain organoid reservoir computing for artificial intelligence},
  author={Cai, Hongwei and Ao, Zheng and Tian, Chunhui and Wu, Zhuhao and Liu, Hongcheng and Tchieu, Jason and Gu, Mingxia and Mackie, Ken and Guo, Feng},
  journal={Nature Electronics},
  volume={6},
  number={12},
  pages={1032--1039},
  year={2023},
  publisher={Nature Publishing Group UK London}
}

@article{nikolic2023computational,
  title={Computational capabilities of a multicellular reservoir computing system},
  author={Nikoli{\'c}, Vladimir and Echlin, Moriah and Aguilar, Boris and Shmulevich, Ilya},
  journal={PLOS One},
  volume={18},
  number={4},
  pages={e0282122},
  year={2023},
  publisher={Public Library of Science San Francisco, CA USA}
}

@inproceedings{fernando2003pattern,
  title={Pattern recognition in a bucket},
  author={Fernando, Chrisantha and Sojakka, Sampsa},
  booktitle={European Conference on Artificial Life},
  pages={588--597},
  year={2003},
  organization={Springer}
}

@article{coulombe2017computing,
  title={Computing with networks of nonlinear mechanical oscillators},
  author={Coulombe, Jean C and York, Mark CA and Sylvestre, Julien},
  journal={PLOS One},
  volume={12},
  number={6},
  pages={e0178663},
  year={2017},
  publisher={Public Library of Science San Francisco, CA USA}
}

@article{cucchi2021reservoir,
  title={\ Reservoir computing with biocompatible organic electrochemical networks for brain-inspired biosignal classification},
  author={Cucchi, Matteo and Gruener, Christopher and Petrauskas, Lautaro and Steiner, Peter and Tseng, Hsin and Fischer, Axel and Penkovsky, Bogdan and Matthus, Christian and Birkholz, Peter and Kleemann, Hans},
  journal={Science Advances},
  volume={7},
  number={34},
  pages={eabh0693},
  year={2021},
  publisher={American Association for the Advancement of Science}
}

@article{kleiber1947body,
  title={Body size and metabolic rate},
  author={Kleiber, Max},
  journal={Physiological Reviews},
  volume={27},
  number={4},
  pages={511--541},
  year={1947}
}

@article{west2002allometric,
  title={Allometric scaling of metabolic rate from molecules and mitochondria to cells and mammals},
  author={West, Geoffrey B and Woodruff, William H and Brown, James H},
  journal={Proceedings of the National Academy of Sciences},
  volume={99},
  number={suppl\_1},
  pages={2473--2478},
  year={2002},
  publisher={National Academy of Sciences}
}

@article{boussard2021adaptive,
  title={Adaptive behaviour and learning in slime moulds: the role of oscillations},
  author={Boussard, Aur{\`e}le and Fessel, Adrian and Oettmeier, Christina and Briard, L{\'e}a and D{\"o}bereiner, Hans-G{\"u}nther and Dussutour, Audrey},
  journal={Philosophical Transactions of the Royal Society B},
  volume={376},
  number={1820},
  pages={20190757},
  year={2021},
  publisher={The Royal Society}
}

@article{reid2016decision,
  title={Decision-making without a brain: how an amoeboid organism solves the two-armed bandit},
  author={Reid, Chris R and MacDonald, Hannelore and Mann, Richard P and Marshall, James AR and Latty, Tanya and Garnier, Simon},
  journal={Journal of The Royal Society Interface},
  volume={13},
  number={119},
  pages={20160030},
  year={2016},
  publisher={The Royal Society}
}

@inproceedings{siriwardana2012fast,
  title={Fast shortest path optimization inspired by shuttle streaming of \textit{Physarum polycephalum}},
  author={Siriwardana, Jayantha and Halgamuge, Saman K},
  booktitle={2012 IEEE Congress on Evolutionary Computation},
  pages={1--8},
  year={2012},
  organization={IEEE}
}

@article{quinlan1981correlation,
  title={A correlation between in vivo and in vitro effects of the microtubule inhibitors colchicine, parbendazole and nocodazole on myxamoebae of \textit{Physarum polycephalum}},
  author={Quinlan, RA and Roobol, A and Pogson, CI and Gull, K},
  journal={Microbiology},
  volume={122},
  number={1},
  pages={1--6},
  year={1981},
  publisher={Microbiology Society}
}

@article{salles1991physarum,
  title={\textit{Physarum} plasmodia do contain cytoplasmic microtubules!},
  author={Salles-Passador, Isabelle and Moisand, Andr{\'e} and Planques, Viviane and Wright, Michel},
  journal={Journal of Cell Science},
  volume={100},
  number={3},
  pages={509--520},
  year={1991},
  publisher={The Company of Biologists Ltd}
}

@article{salles1992intranuclear,
  title={The intranuclear microtubule organizing centre in the plasmodium of the myxomycete \textit{Physarum polycephalum}},
  author={Salles-Passador, I and Lajoie-Mazenc, I and Tollon, Y and Ahkavan-Niaki, H and Oustrin, ML and Moisand, A and Raynaud-Messina, B and Wright, M},
  journal={Cell Biology International Reports},
  volume={16},
  number={11},
  pages={1193--1204},
  year={1992},
  publisher={Elsevier}
}

@article{pfeifer2005new,
  title={New robotics: Design principles for intelligent systems},
  author={Pfeifer, Rolf and Iida, Fumiya and Bongard, Josh},
  journal={Artificial Life},
  volume={11},
  number={1-2},
  pages={99--120},
  year={2005},
  publisher={MIT Press One Rogers Street, Cambridge, MA 02142-1209, USA journals-info~…}
}

@article{vallverdu2018slime,
  title={Slime mould: the fundamental mechanisms of biological cognition},
  author={Vallverd{\'u}, Jordi and Castro, Oscar and Mayne, Richard and Talanov, Max and Levin, Michael and Balu{\v{s}}ka, Frantisek and Gunji, Yukio and Dussutour, Audrey and Zenil, Hector and Adamatzky, Andrew},
  journal={Biosystems},
  volume={165},
  pages={57--70},
  year={2018},
  publisher={Elsevier}
}

@article{beekman2015brainless,
  title={Brainless but multi-headed: decision making by the acellular slime mould \textit{Physarum polycephalum}},
  author={Beekman, Madeleine and Latty, Tanya},
  journal={Journal of Molecular Biology},
  volume={427},
  number={23},
  pages={3734--3743},
  year={2015},
  publisher={Elsevier}
}

@article{nakagaki2008intelligent,
  title={Intelligent behaviors of amoeboid movement based on complex dynamics of soft matter},
  author={Nakagaki, Toshiyuki and Guy, Robert D},
  journal={Soft Matter},
  volume={4},
  number={1},
  pages={57--67},
  year={2008},
  publisher={Royal Society of Chemistry}
}

@article{burland1983cell,
  title={Cell type-dependent expression of tubulins in \textit{Physarum}.},
  author={Burland, Timothy G and Gull, Keith and Schedl, Tim and Boston, Rebecca S and Dove, William F},
  journal={The Journal of Cell Biology},
  volume={97},
  number={6},
  pages={1852--1859},
  year={1983}
}

@article{burland1988gene,
  title={A gene encoding the major beta tubulin of the mitotic spindle in \textit{Physarum polycephalum} plasmodia},
  author={Burland, Timothy G and Paul, Eileen CA and Oetliker, Michael and Dove, William F},
  journal={Molecular and Cellular Biology},
  volume={8},
  number={3},
  pages={1275--1281},
  year={1988},
  publisher={Taylor \& Francis}
}

@article{goodman1969plasmodial,
  title={Plasmodial mitosis in \textit{\textit{Physarum polycephalum}}},
  author={Goodman, EM and Ritter, H},
  journal={Arch. Protistenk},
  volume={111},
  pages={161--169},
  year={1969}
}

@article{green1987developmental,
  title={Developmental regulation and identification of an isotype encoded by altB, an alpha-tubulin locus in \textit{Physarum polycephalum}},
  author={Green, Larry L and Schroeder, Marianne M and Diggins, Maureen A and Dove, William F},
  journal={Molecular and Cellular Biology},
  volume={7},
  number={9},
  pages={3337--3340},
  year={1987},
  publisher={Taylor \& Francis}
}

@article{gull1974ultrastructural,
  title={Ultrastructural effects of griseofulvin on the myxomycete \textit{Physarum polycephalum}: inhibition of mitosis and the production of microtubule crystals},
  author={Gull, K and Trinci, APJ},
  journal={Protoplasma},
  volume={81},
  number={1},
  pages={37--48},
  year={1974},
  publisher={Springer}
}

@article{guttes1968electron,
  title={Electron microscope study of mitosis in \textit{\textit{Physarum polycephalum}}},
  author={Guttes, Sophie and Guttes, Edmund and Ellis, Richard A},
  journal={Journal of Ultrastructure Research},
  volume={22},
  number={5-6},
  pages={508--529},
  year={1968},
  publisher={Elsevier}
}

@article{havercroft1983demonstration,
  title={Demonstration of different patterns of microtubule organization in \textit{Physarum polycephalum} myxamoebae and plasmodia using immunofluorescence microscopy.},
  author={Havercroft, JC and Gull, K},
  journal={European Journal of Cell Biology},
  volume={32},
  number={1},
  pages={67--74},
  year={1983}
}

@article{paul1987patterns,
  title={Patterns of tubulin isotype synthesis and usage during mitotic spindle morphogenesis in \textit{Physarum}},
  author={Paul, Eileen CA and Roobol, Anne and Foster, Kay E and Gull, Keith},
  journal={Cell Motility and the Cytoskeleton},
  volume={7},
  number={3},
  pages={272--281},
  year={1987},
  publisher={Wiley Online Library}
}

@article{roobol1984identification,
  title={Identification of tubulin isoforms in the plasmodium of \textit{Physarum polycephalum} by in vitro microtubule assembly.},
  author={Roobol, A and Wilcox, M and Paul, EC and Gull, K},
  journal={European Journal of Cell Biology},
  volume={33},
  number={1},
  pages={24--28},
  year={1984}
}

@article{sakai1972electron,
  title={Electron microscopy of dividing cells: IV. behaviour of spindle microtubules during nuclear division in the plasmodium of the myxomycete, \textit{\textit{Physarum polycephalum}}},
  author={Sakai, Aiko and Shigenaga, Michio},
  journal={Chromosoma},
  volume={37},
  number={1},
  pages={101--116},
  year={1972},
  publisher={Springer}
}

@book{sauer1982developmental,
  title={Developmental biology of Physarum},
  author={Sauer, Helmut W},
  volume={11},
  year={1982},
  publisher={Cambridge University Press}
}

@article{solnica1990variable,
  title={Variable pathways for developmental changes in composition and organization of microtubules in \textit{Physarum polycephalum}},
  author={Solnica-Krezel, Lilianna and Diggins-Gilicinski, Maureen and Burland, Timothy G and Dove, William F},
  journal={Journal of Cell Science},
  volume={96},
  number={3},
  pages={383--393},
  year={1990},
  publisher={The Company of Biologists Ltd}
}

@article{wille1979fine,
  title={Fine structure of plasmodial nuclei in the slime mold \textit{Physarum polycephalum} I. comparison of diploid and haploid vegetative mitosis},
  author={Wille, JJ and Steffens, WL},
  journal={Protoplasma},
  volume={101},
  number={3},
  pages={165--180},
  year={1979},
  publisher={Springer}
}

@article{dugas1961electron,
  title={Electron microscopy of the slime mold \textit{Physarum polycephalum}},
  author={Dugas, DJ and Bath, Jean D},
  journal={Protoplasma},
  volume={54},
  number={3},
  pages={421--431},
  year={1961},
  publisher={Springer}
}

@article{rhea1966electron,
  title={Electron microscopic observations on the slime mold \textit{Physarum polycephalum} with specific reference to fibrillar structures},
  author={Rhea, Robert P},
  journal={Journal of Ultrastructure Research},
  volume={15},
  number={3-4},
  pages={349--379},
  year={1966},
  publisher={Elsevier}
}

@article{margolus1998maximum,
  title={The maximum speed of dynamical evolution},
  author={Margolus, Norman and Levitin, Lev B},
  journal={Physica D: Nonlinear Phenomena},
  volume={120},
  number={1-2},
  pages={188--195},
  year={1998},
  publisher={Elsevier}
}

@article{ohl1991studies,
  title={Studies on microplasmodia of \textit{Physarum polycephalum}: Quantitative analysis of contractile activities and microfilament organization},
  author={Ohl, Christiane and Brix, Klaudia and Stockem, Wilhelm},
  journal={Cell and Tissue Research},
  volume={264},
  number={2},
  pages={283--291},
  year={1991},
  publisher={Springer}
}

@article{hirose1980changes,
  title={Changes in ATP concentration triggered by chemoreception in the plasmodia of the myxomycete \textit{Physarum polycephalum}},
  author={Hirose, Tatsumi and Ueda, Tetsuo and Kobatake, Yonosuke},
  journal={Microbiology},
  volume={121},
  number={1},
  pages={175--180},
  year={1980},
  publisher={Microbiology Society}
}

@article{stephanopoulos2020hybrid,
  title={Hybrid nanostructures from the self-assembly of proteins and DNA},
  author={Stephanopoulos, Nicholas},
  journal={Chem},
  volume={6},
  number={2},
  pages={364--405},
  year={2020},
  publisher={Elsevier}
}

@article{han2011dna,
  title={DNA origami with complex curvatures in three-dimensional space},
  author={Han, Dongran and Pal, Suchetan and Nangreave, Jeanette and Deng, Zhengtao and Liu, Yan and Yan, Hao},
  journal={Science},
  volume={332},
  number={6027},
  pages={342--346},
  year={2011},
  publisher={American Association for the Advancement of Science}
}

@article{mccullough1976defined,
  title={Defined and semi-defined media for the growth of amoebae of \textit{Physarum polycephalum}},
  author={Mccullough, Clare HR and Dee, Jennifer},
  journal={Microbiology},
  volume={95},
  number={1},
  pages={151--158},
  year={1976},
  publisher={Microbiology Society}
}

@article{rosina2025mathematical,
  title={A mathematical model to predict network growth in \textit{Physarum polycephalum} as a function of extracellular matrix viscosity, measured by a novel viscometer},
  author={Rosina, Philipp and Grube, Martin},
  journal={Journal of the Royal Society Interface},
  volume={22},
  number={224},
  pages={20240720},
  year={2025},
  publisher={The Royal Society}
}

@article{rolland2023behavioural,
  title={Behavioural changes in slime moulds over time},
  author={Rolland, Ang{\`e}le and Pasquier, Emilie and Malvezin, Paul and Cassandra, Craig and Dumas, Mathilde and Dussutour, Audrey},
  journal={Philosophical Transactions of the Royal Society B},
  volume={378},
  number={1874},
  pages={20220063},
  year={2023},
  publisher={The Royal Society}
}

@article{vogel2015phenotypic,
  title={Phenotypic variability in unicellular organisms: from calcium signalling to social behaviour},
  author={Vogel, David and Nicolis, Stamatios C and Perez-Escudero, Alfonso and Nanjundiah, Vidyanand and Sumpter, David JT and Dussutour, Audrey},
  journal={Proceedings of the Royal Society B: Biological Sciences},
  volume={282},
  number={1819},
  pages={20152322},
  year={2015},
  publisher={The Royal Society}
}

@article{nakagaki2000interaction,
  title={Interaction between cell shape and contraction pattern in the \textit{Physarum} plasmodium},
  author={Nakagaki, Toshiyuki and Yamada, Hiroyasu and Ueda, Tetsuo},
  journal={Biophysical Chemistry},
  volume={84},
  number={3},
  pages={195--204},
  year={2000},
  publisher={Elsevier}
}

@article{azizi2023examining,
  title={Examining the origins of observed terahertz modes from an optically pumped atomistic model protein in aqueous solution},
  author={Azizi, Khatereh and Gori, Matteo and Morzan, Uriel and Hassanali, Ali and Kurian, Philip},
  journal={PNAS Nexus},
  volume={2},
  number={8},
  pages={pgad257},
  year={2023},
  publisher={Oxford University Press US}
}

@article{niklas2006phyletic,
  title={A phyletic perspective on the allometry of plant biomass-partitioning patterns and functionally equivalent organ-categories},
  author={Niklas, Karl J},
  journal={New Phytologist},
  volume={171},
  number={1},
  pages={27--40},
  year={2006},
  publisher={Wiley Online Library}
}

@article{shephard1975efficiency,
  title={Efficiency of muscular work: some clinical implications},
  author={Shephard, Roy J},
  journal={Physical Therapy},
  volume={55},
  number={5},
  pages={476--481},
  year={1975},
  publisher={Oxford University Press}
}

@article{yoshimoto1981atp,
  title={ATP oscillation in Physarum plasmodium},
  author={Yoshimoto, Y and Sakai, T and Kamiya, N},
  journal={Protoplasma},
  volume={109},
  number={1},
  pages={159--168},
  year={1981},
  publisher={Springer}
}

@article{del2013quantum,
  title={Quantum speed limits in open system dynamics},
  author={del Campo, Adolfo and Egusquiza, Inigo L and Plenio, Martin B and Huelga, Susana F},
  journal={Physical Review Letters},
  volume={110},
  number={5},
  pages={050403},
  year={2013},
  publisher={APS}
}

@incollection{kuramoto1984chemical,
  title={Chemical turbulence},
  author={Kuramoto, Yoshiki},
  booktitle={Chemical oscillations, waves, and turbulence},
  pages={111--140},
  year={1984},
  publisher={Springer}
}

@article{taddei2013quantum,
  title={Quantum speed limit for physical processes},
  author={Taddei, M{\'a}rcio M and Escher, Bruno M and Davidovich, Luiz and de Matos Filho, Ruynet L},
  journal={Physical Review Letters},
  volume={110},
  number={5},
  pages={050402},
  year={2013},
  publisher={APS}
}

@article{brody2012mixed,
  title={Mixed-state evolution in the presence of gain and loss},
  author={Brody, Dorje C and Graefe, Eva-Maria},
  journal={Physical Review Letters},
  volume={109},
  number={23},
  pages={230405},
  year={2012},
  publisher={APS}
}

@article{ronellenfitsch2019phenotypes,
  title={Phenotypes of vascular flow networks},
  author={Ronellenfitsch, Henrik and Katifori, Eleni},
  journal={Physical Review Letters},
  volume={123},
  number={24},
  pages={248101},
  year={2019},
  publisher={APS}
}

@article{katifori2010damage,
  title={Damage and fluctuations induce loops in optimal transport networks},
  author={Katifori, Eleni and Sz{\"o}ll{\H{o}}si, Gergely J and Magnasco, Marcelo O},
  journal={Physical Review Letters},
  volume={104},
  number={4},
  pages={048704},
  year={2010},
  publisher={APS}
}

@article{bhattacharyya2022memory,
  title={Memory formation in adaptive networks},
  author={Bhattacharyya, Komal and Zwicker, David and Alim, Karen},
  journal={Physical Review Letters},
  volume={129},
  number={2},
  pages={028101},
  year={2022},
  publisher={APS}
}

@article{bhattacharyya2023memory,
  title={Memory capacity of adaptive flow networks},
  author={Bhattacharyya, Komal and Zwicker, David and Alim, Karen},
  journal={Physical Review E},
  volume={107},
  number={3},
  pages={034407},
  year={2023},
  publisher={APS}
}

@article{fessel2012physarum,
  title={Physarum polycephalum Percolation as a Paradigm for Topological Phase Transitions<? format?> in Transportation Networks},
  author={Fessel, Adrian and Oettmeier, Christina and Bernitt, Erik and Gauthier, Nils C and D{\"o}bereiner, Hans-G{\"u}nther},
  journal={Physical Review Letters},
  volume={109},
  number={7},
  pages={078103},
  year={2012},
  publisher={APS}
}
\bibliographystyle{sciencemag}
\newpage

\textcolor{black}{\section*{Acknowledgments}}

\textcolor{black}{
We would like to thank Dr. Yukinori Nishigami and Prof. Toshiyuki Nakagaki for kindly providing the Japanese (Sonobe) strain of \textit{Physarum polycephalum}. We thank Dr. Sanghita Sengupta for developing the PyPETANA 1.0 scripts, which were modified and customized for the analysis in this work. Portions of this work were presented at the Sloan Foundation Matter-to-Life meetings in New York and Germany, the Active Solids Program at the Kavli Institute for Theoretical Physics (KITP) in California, the Quantum Thermodynamics Conference in Maryland, the Princeton–Texas A\&M-Casper Quantum Summer School in Wyoming, the Molecular Biophysics Workshops in France and Japan, and several American Physics Society meetings.} 
\subsection*{\textcolor{black}{Funding}}
\textcolor{black}{We acknowledge financial support from the Alfred P. Sloan Foundation's Matter-to-Life Program (P.K.), the Howard University Office of Research, and the Howard University Graduate School (P.K.). KITP is supported in part by a grant from the National Science Foundation.}
\subsection*{\textcolor{black}{Author contributions}}
\textcolor{black}{\textbf{S.B.:} Conceptualization, methodology, formal analysis, software development, data curation, visualization, writing---original draft, writing---review and editing.\\
\textbf{A.L.-D.:} Experimental investigation.\\
\textbf{N.M.:} Experimental investigation, writing---review and editing.\\
\textbf{M.L.:} Investigation, writing---review and editing.\\
\textbf{P.K.:} Conceptualization, methodology, formal analysis, data curation, visualization, writing---original draft, writing---review and editing, funding acquisition, project administration, supervision.}
\subsection*{\textcolor{black}{Competing interests}}
\textcolor{black}{All authors declare that they have no competing interests.}
\subsection*{\textcolor{black}{Data, code, and materials availability}}
\textcolor{black}
{The data and code needed to evaluate and reproduce the results presented in this paper are available in the PyPETANA 2.0 GitHub repository (\url{https://github.com/suyashb4/PyPETANA2.0.git}), which is also linked in the Methods section.
The repository contains images of \textit{Physarum} growth for representative Carolina and Japanese (Sonobe) samples, time-series datasets of the mean morphological quantities for the four experimental groups, and the PyPETANA 2.0 code used for image processing and extraction of morphological indices. This study did not generate any new materials.
}
\enlargethispage{\baselineskip}
\clearpage 

\end{document}